\documentclass[11pt]{article}
\pdfoutput=1
\usepackage{jheppub}
\usepackage[dvipsnames]{xcolor}
\usepackage{mathrsfs} % for mathscr
\usepackage[T1]{fontenc}

\usepackage{tikz}
\usetikzlibrary{decorations.pathmorphing,decorations.pathreplacing,calligraphy,calc,cd,external,arrows.meta,patterns}

\usepackage{svg}
\usepackage[export]{adjustbox}
\usepackage{hyperref}
\usepackage{bbold}
\usepackage{mathtools}
\usepackage{empheq}
\usepackage{nccmath}
\usepackage{blkarray}
\usepackage{esint}
\usepackage{microtype}
\usepackage{slashed}
\usepackage{amsmath}
\usepackage{dsfont}
\usepackage{amsfonts}
\usepackage{amssymb}
\usepackage{mathtools}
\usepackage{microtype}
\usepackage{bm}
\usepackage[utf8]{inputenc}
\usepackage{blkarray}
\usepackage{bigstrut}
\usepackage{nccmath}
\usepackage{enumitem}
\usepackage{float}
\usepackage{braket}
\usepackage{dsfont}
\usepackage[export]{adjustbox}
\usepackage[english]{babel}
\usepackage{stmaryrd}
\usepackage{orcidlink}
\usepackage{tikz}
\usepackage{bbm}
\usepackage{lipsum}
\usepackage{blindtext}
\usepackage{fontawesome}
\usepackage[]{notes2bib}
\usepackage{enumitem}
\usepackage{cleveref}
\usepackage[mathscr]{euscript}
\usepackage{stackengine}
\usepackage{esint}
\usepackage{lmodern}
\usepackage{textcomp}
\let\oldFootnote\footnote
\newcommand\nextToken\relax

\renewcommand\footnote[1]{%
    \oldFootnote{#1}\futurelet\nextToken\isFootnote}

\newcommand\isFootnote{%
    \ifx\footnote\nextToken\textsuperscript{,}\fi}
    
\hypersetup{
    pdfencoding=unicode,
	colorlinks=true,
	urlcolor=sidsMaroon,
	linkcolor=sidsblue,
	citecolor=sidsMaroon,
	pdftitle={High-energy evolution in planar QCD to three loops: the non-conformal contribution},
	pdfauthor={Giacomo Brunello, Simon Caron-Huot, Giulio Crisanti, Mathieu Giroux, Sid Smith},
	pdfdisplaydoctitle=true,
	pdfstartview=FitH,
	linktocpage=true
}

\def\e{\mathrm{e}}
\def\D{\mathrm{D}}
\def\d{\mathrm{d}}
\def\xb{x_{\rm B}}
\def\Dperp{\D_{\perp}}

\newcommand{\dbar}{\mathchar'26\mkern-12mu\raisebox{0.8pt}{$\d$}}

\def\Uu{\mathcal{U}}
\def\Aa{\mathcal{A}}
\def\Dperp{{\D_\perp}}
\def\mubar{\bar{\mu}}
\def\kslash{k\!\!\!/}

\def\mubar{\overline{\mu}}
\def\Li{{\rm Li}}
\def\cd{{C_\varepsilon}}
\def\aeff{a}
\def\T{\top}
\def\ns{{n_{\rm adj}^s}}
\def\nf{{n_{\rm adj}^f}}
\def\Vv{\mathcal{V}}
\definecolor{mgGreen}{rgb}{0.0, 0.29, 0.33}

\definecolor{orcidlogocol}{named}{Maroon}
\makeatletter
\newenvironment{sqcases}{%
  \matrix@check\sqcases\env@sqcases
}{%
  \endarray\right.%
}
\def\env@sqcases{%
  \let\@ifnextchar\new@ifnextchar
  \left\lbrack
  \def\arraystretch{1.2}%
  \array{@{}l@{\quad}l@{}}%
}
\makeatother

\DeclareFontFamily{U}{mathx}{}
\DeclareFontShape{U}{mathx}{m}{n}{<-> mathx10}{}
\DeclareSymbolFont{mathx}{U}{mathx}{m}{n}
\DeclareMathAccent{\widehat}{0}{mathx}{"70}
\DeclareMathAccent{\widecheck}{0}{mathx}{"71}

\definecolor{sidsblue}{RGB}{44,90,160}
\definecolor{sidsMaroon}{RGB}{128,0,0}
\definecolor{sidsOrange}{RGB}{204,85,0}
\definecolor{darkmagenta}{rgb}{0.55, 0.0, 0.55}

\usepackage[titles]{tocloft}
\usepackage{parskip}
\usepackage{dirtytalk}
\usepackage{changepage}
\usepackage{subcaption}
\usepackage{float}
\usepackage{enumitem}  
\usepackage{stackengine}
\usepackage{bm}

\title{High-energy evolution in planar QCD to three loops:
\\ the non-conformal contribution}
\usepackage{orcidlink}
\author[\!a,b,c,d,\orcidlink{0009-0004-4788-738X}]{Giacomo Brunello,}\emailAdd{giacomo.brunello@phd.unipd.it}
\author[\!e,\orcidlink{0000-0002-7005-9652}]{Simon Caron-Huot,}\emailAdd{schuot@physics.mcgill.ca}
\author[\!f,\orcidlink{0009-0009-3053-2394}]{Giulio Crisanti,}\emailAdd{g.crisanti@ed.ac.uk}
\author[\!e,\orcidlink{0000-0002-2672-634X}]{Mathieu Giroux,}\emailAdd{mathieu.giroux2@mail.mcgill.ca}
\author[\,a,b,f,\orcidlink{0009-0007-7799-0136}]{\\ Sid Smith}\emailAdd{sid.smith@ed.ac.uk}

\affiliation{$^a$Dipartimento di Fisica e Astronomia, Universita di Padova, Via Marzolo 8, 35131 Padova, Italy}
\affiliation{$^b$INFN, Sezione di Padova,
Via Marzolo 8, I-35131 Padova, Italy.}
\affiliation{$^c$Institut de Physique Théorique, CEA, CNRS, Université Paris-Saclay, F–91191 Gif-sur-Yvette cedex, France}
\affiliation{$^d$Scuola Normale Superiore, Piazza dei Cavalieri 7, 56126, Pisa, Italy and INFN Sezione di Pisa, Largo
Pontecorvo 3, 56127 Pisa, Italy}
\affiliation{$^e$Department of Physics, McGill University, 3600 Rue University, Montr\'eal, H3A 2T8, QC Canada}
\affiliation{$^f$Higgs Centre for Theoretical Physics, University of Edinburgh, James Clerk Maxwell Building,Peter Guthrie Tait Road, Edinburgh, EH9 3FD, United Kingdom}

\abstract{
The Balitsky–Kovchegov (BK) equation offers a tractable description of the high‑energy growth of gauge‑theory scattering amplitudes and the nonlinear saturation effects that eventually tame it. Motivated by the upcoming Electron-Ion Collider (EIC), whose extended kinematic reach promises more decisive tests of saturation at high energies, we present a framework based on the spacelike-timelike correspondence that streamlines the computation of multi-loop corrections to the BK equation. We explicitly verify the correspondence at three loops in the large-flavor limit and predict the full non-conformal component of the three-loop BK Hamiltonian in the planar limit of a generic gauge theory, treating the numbers of fermions and scalars as free parameters. 
}

\begin{document} 

\addtocontents{toc}{\protect\thispagestyle{empty}}

\maketitle

\thispagestyle{empty}

\setcounter{page}{3}
\flushbottom
\raggedbottom
\allowdisplaybreaks
\newpage

\section{Introduction}

The discovery of quarks and the advent of deep-inelastic scattering (DIS) have profoundly reshaped our view of the fundamental constituents of matter in the second half of the last century. Although these breakthroughs have clarified many aspects of nucleon structure, they have also exposed significant puzzles. One of them is the complex role that gluons play in the emergent properties of composite matter, such as mass and spin, and their behavior at high energies (small $\xb$) where gluon densities become large.  The planned Electron-Ion Collider (EIC) promises to transform our understanding of hadronic structure in general, and of the properties of dense partonic systems in particular \cite{Accardi:2012qut,AbdulKhalek:2022hcn}.

On the theoretical side, there is a timely need to understand better the high-energy growth of QCD---and more broadly, gauge theory---scattering amplitudes. This involves understanding the \emph{nonlinear} effects that are eventually expected to saturate this growth, in order to ensure the preservation of unitarity. Mathematically, saturation is characterized by scattering amplitudes reaching opacity at a certain impact parameter or projectile size, where the projectile is effectively absorbed by the target. In the saturation regime, the (gluon-dominated) densities of partons become so high that their otherwise rapid growth is curbed by recombination effects. 

Over the years, various collider experiments, notably at HERA, have reported hints of gluon saturation \cite{Morreale:2021pnn}.
However, although observations are compatible with the saturation picture, they are also compatible with a linear description based on collinear resummation (DGLAP) \cite{Caola:2010cy,Alekhin:2012ig};  direct comparison of linear and nonlinear fits does not suggest significant nonlinearities \cite{Mantysaari:2018nng}.
In this context, an essential aspect of EIC physics is the enhancement of nonlinear effects for nuclear targets.
The corresponding increases in the saturation scale $Q_s^2 \sim A^{1/3}$
suggests
the possibility of a perturbative treatment of saturation, the so-called ``color glass condensate'' (see \cite{Iancu:2003xm,Gelis:2010nm,Kovchegov:2012mbw,Albacete:2014fwa,Morreale:2021pnn} for reviews).
Assessing the stability of such a perturbative treatment, notably with respect to higher order corrections in $\alpha_s$, is a main motivation for this paper.

The most general framework describing the high-energy (small-$\xb$) evolution of fundamental QCD degrees of freedom in perturbative QCD is the Balitsky--JIMWLK\footnote{The acronym is a shorthand for Jalilian-Marian--Iancu--McLerran--Weigert--Leonidov--Kovner.} hierarchy of equations. This hierarchy involves correlators of arbitrarily many (light-like) Wilson lines and retains full information about color flow, which makes its solution difficult in practice. For applications to phenomenology, it often suffices to consider its mean-field version that involves only color dipoles, 
known as the Balitsky--Kovchegov (BK) equation, which is theoretically justified in the planar (large-$N$) limit. At leading order (LO/“one-loop”), the BK and Balitsky--JIMWLK equations were derived respectively in \cite{Balitsky:1995ub,Kovchegov:1999yj} and \cite{Jalilian-Marian:1997qno,Jalilian-Marian:1997ubg}, and at the next order in \cite{Balitsky:2006wa,Balitsky:2007feb} and \cite{Balitsky:2013fea,Kovner:2013ona,Kovner:2014lca}.

The two-loop corrections were found to be sizeable.  As was previously observed for the linearized equation of rapidity evolution (``BFKL'') at small $\xb$, this can be attributed to large collinear logarithms independently predicted by the DGLAP equation \cite{Fadin:1998py,Ciafaloni:1998gs,Salam:1998tj}.
In this way a ``collinearly improved’’ BK equation (see, e.g., \cite{Iancu:2015joa,Ducloue:2019jmy}) has been obtained.
This refined approach appears to be numerically stable \cite{Lappi:2016fmu}, thus restoring the predictive power lost in the first numerical NLO study \cite{Lappi:2015fma}, which revealed instabilities and significant negative contributions.
At the same time, while the improvement is unique in the collinear limit where the BK and DGLAP equations overlap, it is an interesting theoretical question to determine its best extension to generic configurations
where nonlinear effects are important. To confirm that the perturbative expansion is really under control, a NNLO (“three-loop”) correction seems needed.

The main goal of this paper is to initiate the calculation of the three-loop rapidity evolution equation in planar QCD and obtain a specific set of (``non-conformal'') terms.

One principal obstacle which we overcome is the substantial technical challenge posed by the increasingly complex integrals (most notably impact-parameter space Fourier transforms) required at NNLO, which is a feature already made apparent in the NLO foundational studies \cite{Balitsky:2006wa,Balitsky:2007feb,Balitsky:2009xg}.
As we will see, the relevant integrals are all within the reach of modern Feynman-calculus techniques developed for scattering amplitudes, such as integration-by-parts \cite{Chetyrkin:1981qh,Laporta:2000dsw}, 
canonical differential equations \cite{Kotikov:1990kg,Kotikov:1991pm,Remiddi:1997ny,Gehrmann:1999as,Argeri:2007up,Smirnov:2012gma,Henn:2013pwa,Argeri:2014qva} and intersection theory \cite{Mastrolia:2018uzb,Frellesvig:2019kgj,Frellesvig:2019uqt,Frellesvig:2020qot,Chestnov:2022xsy,Fontana:2023amt,Brunello:2023fef,Brunello:2023rpq,Brunello:2024tqf}.

The ``non-conformal'' terms which we will calculate are precisely the  $\mathcal{O}(\varepsilon)$ part of the \emph{two-loop} BK equation in $\D=4-2\varepsilon$ dimensions.
Their relevance to three-loop QCD stems from the spacelike-timelike correspondence, which relates the  evolution equations for non-global logarithms (NGLs) in jet physics \cite{Banfi:2002hw} with the high-energy (BK/JIMWLK) equations mentioned above \cite{Hatta:2008st,Caron-Huot:2015bja}.
This correspondence relies on conformal symmetry, which is broken by the running of the QCD coupling, but can be effectively restored if one knows the equation in the conformal dimension $\varepsilon\approx -\alpha_sb_0$ \cite{Vladimirov:2016dll}, which results in a mixing of loop orders. The technical benefit of this approach is that the remaining ``conformal'' terms can be obtained from a technically simpler calculation of three-loop NGLs---whose gluonic
contribution is already contained in the $\mathcal{N}=4$ supersymmetric Yang--Mills result of \cite{Caron-Huot:2016tzz}.

We will work in the so-called 't Hooft--Veneziano planar limit and take the number of colors to be large ($N\to\infty$ with $g_s^2N_c$ fixed) while also formally keeping the number of quark flavors large, $n_f\sim N$.  As in the standard 't Hooft planar limit, this grants us the simplification of treating all color connections as products of dipoles, but without dropping the quarks.

\paragraph{Organization of the paper.} 
In Sec.~\ref{sec:setup} we review the necessary background to introduce the BK equation. In Sec.~\ref{ex:BKmain} we present our main results: the detailed computation of the relevant $\varepsilon$-corrections to the LO and NLO BK kernels. In Sec.~\ref{sec:correspondence} we review the spacelike-timelike correspondence and explain how these $\varepsilon$-corrections give rise to the conformal-symmetry-breaking contributions in the three-loop BK kernel, which we explicitly check at three loops in the large number of flavors limit.
Sec.~\ref{sec:conclusion} presents our conclusions and outlines directions for future work. In the appendices we then briefly review lightcone Feynman rules and give additional technical details on the calculation of real-virtual amplitudes. An ancillary file accompanying this manuscript lists all new results from \eqref{eq:resultthree-loop} in Wolfram syntax.

\paragraph{Conventions.} 
In this work, we use the mostly-\emph{plus} metric and decompose the $\D$-dimensional coordinates as
$x^\mu = (x^+,x^-,z)$ where $x^\pm=x^0\pm x^1$ and $z^i$ denote the
$\D_\perp$-dimensional transverse components,
with $\D = 2 + \D_\perp$ and $\D_\perp \equiv 2-2\varepsilon$. In this notation, $x^2=-x^+x^-+z^2$.
The transverse separation between vectors $x_a^\mu$ and $x_b^\mu$ is denoted $z_{ab}^i\equiv z_a^i-z_b^i$.  Greek indices run over all $\D$ dimensions, while Latin ones run over the $\Dperp$-dimensional Euclidean transverse subspace.
In these coordinates, the $\D$-dimensional measure splits as
\begin{equation}
    \d^\D x=\frac12 \d x^+\d x^-\d^{\Dperp} z\,.
\end{equation}
Finally, we often write transverse position and momentum integrals using $\,\dbar^{\Dperp}\equiv \frac{\d^{\Dperp}}{(2\pi)^{\Dperp}}$ and
\begin{equation}
    \int\limits_{z_{0},z_{0'},...}\equiv{}\int \d^\Dperp z_{0}\,\d^\Dperp z_{0'}\,... \qquad \text{and} \qquad \int\limits_{\ell_{1},\ell_{2},...}\equiv{}\int \dbar^\Dperp \ell_{1}\,\dbar^\Dperp \ell_{2}\,...
\end{equation}

{\section{Background and setting the stage}\label{sec:setup}}

In this section, we briefly review how the Wilson-line amplitudes we study relate to physical observables. After fixing our notation, we introduce a renormalization scheme in which the renormalized quantity gives rise to the Balitsky--Kovchegov (BK) equation. 

\subsection{Dipoles as probes of nuclei at high energies} 
We consider the high-energy and near-forward scattering of a dilute (``simple'') projectile against a dense (``complex'') target ``$\rm t$.'' The paradigmatic example is the deep inelastic scattering of an electron against
a hadron or heavy ion. As far as QCD is concerned, we can treat the electron probe as a virtual photon with four-momentum $Q^\mu$.
In forward kinematics,
$Q^\mu=(Q^+,-Q^2/Q^+,0)$ is highly boosted, corresponding to $\xb\equiv \frac{Q^2}{-2Q{\cdot}P_{\rm t}}\ll 1$ (``small-$\xb$'').
Geometrically, this means that the lifetime of typical fluctuations of the probe greatly exceeds the longitudinal extent of the Lorentz-contracted target.
This enables a factorized treatment of the probe and target.

\begin{figure}[t]
  \centering
  \begin{subfigure}[c]{1\textwidth}
    \centering
    \includegraphics[scale=1]{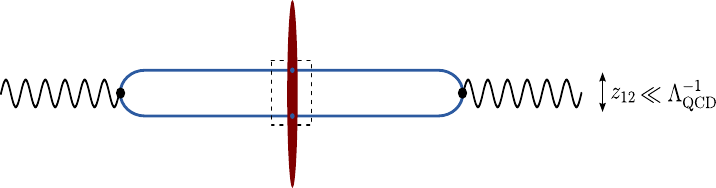}
    \caption{}
    \label{fig:correlatorA}
  \end{subfigure}
  \\
  \begin{subfigure}[c]{0.9\textwidth}
    \centering
    \adjustbox{valign=c}{%
      \includegraphics[scale=0.6]{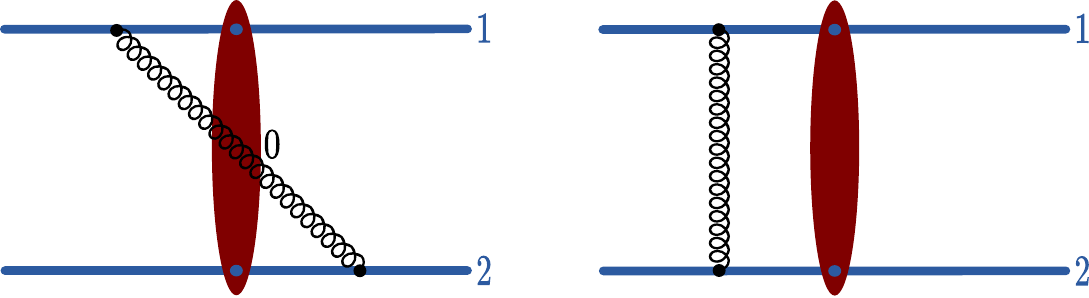}%
    }
    \caption{}
    \label{fig:correlatorB}
  \end{subfigure}
  \caption{%
    (a) At small $\xb$ and leading order in $\alpha_s(Q)$, a virtual photon probes a hadron as a quark-antiquark dipole. (b) Examples of real (left) and virtual (right) perturbative corrections in $\alpha_s(Q)$ generating more complicated color correlations.
  }
  \label{fig:correlator}
\end{figure}

At leading order in the strong coupling $\alpha_s(Q)\equiv \frac{g_s^2(Q)}{4\pi}$, the total $\gamma^* \rm{t}$ cross section (for example) factorizes into the lightcone wavefunction for the $\gamma^*\to q\bar q$ splitting and the subsequent interaction of that color dipole with the target \cite{Mueller:1993rr,Mueller:1994gb,Gelis:2012ri,Kovchegov:2012mbw}:
\begin{equation}\label{eq:obs}
\sigma^{\gamma^* \rm{t}}(\xb,Q^2)\approx \sum_f\int\!\d^{\Dperp} z_{12} \int_0^1\!\d\xi\;
\bigl|\Psi^{(f)}(z_{12},\xi;Q^2)\bigr|^2\;
\sigma_{\rm dip.}(
\xb
,z_{12})\,.
\end{equation}
Here, the squared lightcone wavefunction $\Psi^{(f)}$ is known in practice (see, e.g., \cite{Kovchegov:2012mbw}), but it is left implicit as it will not play an explicit role in what follows. Physically, it gives the probability for a photon to split into a quark-antiquark dipole of (transverse) size $z_{12}$ carrying “+” momentum fractions $\xi$ and $(1-\xi)$ respectively.  The sum in \eqref{eq:obs} runs over the quark flavors, all experiencing the same
dipole-target cross section $\sigma_{\rm dip.}$, which is the expectation value of a dipole Wilson loop in the target state:
\begin{equation} \label{dipole}
   \sigma_{\rm dip.}(\xb,z_{12})=
1-\langle \text{t}| \mathcal{U}_{12}|\text{t}\rangle\,,\qquad
{\cal U}_{12}=\frac{1}{N}\mathrm{Tr}[U_1 U_2^\dagger]\,.
\end{equation}
The dipole operator ${\cal U}_{12}$ is shown schematically in Fig.~\ref{fig:correlatorA} and is defined more precisely as the color trace
of a product of fundamental and antifundamental Wilson lines
$U_1=U(z_1)$ and $U_2^\dagger=U^\dagger(z_2)$,\footnote{Details of how the Wilson lines join far from the target are suppressed from the notation as they will not affect our calculations.} where
\begin{equation}\label{eq:WL0}
U(z)
=\mathbb{P}\exp\!\Bigg[i g_s\!\int_{-\infty}^{\infty}\!\d x^+\,A_+^a(x^+,0,z)\,t^a\Bigg]\;\in\;\mathrm{SU}(N)\,.
\end{equation}
Thus, the dipole ${\cal U}_{12}$ accumulates the eikonal phases that encode the color rotations experienced along the paths of the partons that cross the target.
A value for $\mathcal{U}_{12}$ close to unity indicates that the probe retains its coherent color structure (i.e., a transparent/dilute target), while deviations from unity signal significant interactions (opaque/dense target), leading to color decoherence. Roughly speaking, $\langle \text{t}| \mathcal{U}_{12}|\text{t}\rangle$ is therefore a measure of the color charge present in the target within a surrounding disc of radius $\sim z_{12}$.
Thus, a target probed by a dipole of sufficiently small transverse size automatically appears transparent. The dipole size $z_{12}\sim 1/Q_s$ at which the scattering becomes opaque is called \emph{saturation scale}.

Naively, the task of computing observables that probe color coherence within a nucleon near saturation would seem inherently non-perturbative: the only natural scale, on dimensional grounds, is $Q_s\sim \Lambda_{\rm QCD}$.
However, two effects can increase the value of $Q_s$: large logarithms $\eta\simeq\log(1/\xb)$ caused by a large rapidity\footnote{ 
The variable $\eta$ can be related to the usual rapidity of a particle defined as $\frac12\log \frac{p^+}{p^-}$, but its only essential feature here is that it increases by $\Delta\eta$ when the projectile is boosted and its energy increases by a factor $\e^{\Delta\eta}$.} difference
between target and projectile, or the use of heavy nuclei where charge density is enhanced by the thickness $\sim A^{1/3}$ \cite{Gelis:2010nm}.
Thus, there exists the potential to observe gluon saturation in a regime where $\alpha_s(Q)$ is sufficiently small to justify perturbation theory.

Operating at a larger $Q^2$ also increases sensitivity to DGLAP‐type collinear logarithms, so a careful treatment of both small-$\xb$ saturation dynamics and conventional $Q^2$ evolution is typically required (see, for example, \cite{Iancu:2015joa,Ducloue:2019jmy}).
A controlled understanding of both collinear and soft physics and of their interplay will likely be essential to a successful saturation phenomenology at the EIC.

It is important to understand what form the perturbative corrections to \eqref{eq:obs} can take. Certainly, the wavefunctions $\Psi$ receive $\mathcal{O}(\alpha_s)$ perturbative corrections.
This is not unlike the familiar situation in collinear factorization, where the large-$Q$ expansion of $\sigma^{\gamma^* \rm{t}}$ 
is described in terms of parton distribution functions multiplied by operator product expansion (OPE) coefficients that admit calculable series in $\alpha_s$.
These coefficients are also scale dependent, in accordance with the $Q^2$ dependence of parton distribution functions that is described by the DGLAP equation. In contrast, the main complicating feature of rapidity factorization and evolution
(i.e., in the limit $\xb\to 0$ with $Q$ fixed) is that the corresponding equations are \emph{nonlinear}: 
at each order in the evolution, new operator structures (i.e., higher-point Wilson-line correlators) are generated. This nonlinear nature is what enables novel phenomena like saturation.

As a specific example, the presence of the gluon crossing the shock in Fig.~\ref{fig:correlatorB} (left) adds an adjoint Wilson line to the quark and antiquark ones in \eqref{dipole}. This creates an operator with three Wilson lines at different transverse positions.
Keeping full-color information makes the bookkeeping considerably more challenging---this leads to the Balitsky-JIMWLK hierarchy mentioned in introduction.
In this work, we settle on the large-$N$ (and large fermion degrees of freedom $n_F$) compromise, where a complete basis of operators over which we can expand the dilute projectile consists of \emph{products of color dipoles}, which evolve independently of each other.

We note that, in addition to the DIS cross section in \eqref{eq:obs}, a variety of other physically interesting observables are also mediated by dipole scattering. Examples include back-to-back dijet correlators \cite{Caucal:2022ulg}, energy flow correlators \cite{Kang:2023oqj,Yang:2023dwc}, and gluon transverse momentum distributions (TMDs) \cite{Angeles-Martinez:2015sea,Moult:2018jzp}.

{\subsection{Small-\texorpdfstring{$\xb$}{dum0} evolution: intuitive picture}\label{sec:heuristic}}

As mentioned, the high-energy limit features a large hierarchy between the Lorentz-dilated lifetime of projectile fluctuations (in the center-of-mass frame) and the extent of the Lorentz-contracted target.
This means that each vertex in a Feynman diagram (i.e., any perturbative correction to Fig.~\ref{fig:correlatorA}) can be effectively assigned to one of three $x^+$-regions: large $x^+\sim z_{12}^2Q^+$ comparable to that of the projectile, small $x^+\sim 1/P_\text{t}^-$ comparable to the target's thickness,
and an intermediate ``scaling or boost-invariant'' regime between them.

Contributions from large $x^+$ correct the wavefunction $\Psi$ in \eqref{eq:obs} (and its nonlinear generalizations), those from small $x^+$ correct the nonperturbative target
in a way that is automatically included in the Wilson-loop expectation value \eqref{dipole}, while the scaling region produces the large logarithms $\sim \log(1/\xb)\gg 1$ that will be our main focus (more on that below \eqref{eq:bareDipole}).  Physics in the scaling region can be understood by replacing the projectile by infinite Wilson lines and the target by an infinitesimal shockwave or ``pancake.'' The large logarithms then appear as divergences, which signal that null infinite Wilson lines implicitly depend on a rapidity regulator.

At leading logarithmic order, large logarithms are resummed by the celebrated Balitsky--Kovchegov (BK) equation \cite{Balitsky:1995ub,Kovchegov:1999yj}:
\begin{align}
    \partial_\eta\mathcal{U}_{12} 
    &= \adjustbox{valign=c,scale=0.90}{\tikzset{every picture/.style={line width=0.75pt}}   
\begin{tikzpicture}[x=0.75pt,y=0.75pt,yscale=-1,xscale=1]
%Straight Lines 
\draw[->,color=gray]    (95.46,206.91) -- (166.76,135.38) ;
%Straight Lines 
\draw[<-,color=gray]    (95.35,135.85) -- (166.88,206.45) ;
%Straight Lines 
\draw[very thick, color=sidsblue, line cap =round]    (151,120.08) -- (79.69,191.61) node [right] {$1$};
%Straight Lines 
\draw[very thick, color=sidsblue, line cap =round]    (181,150.08) -- (109.69,221.61) node [right] {$2$};
%Shape: Ellipse
\draw  [draw opacity=0][fill=sidsMaroon  ,fill opacity=1 ] (110.11,150) .. controls (112.05,148.07) and (122.79,155.68) .. (134.11,167) .. controls (145.43,178.32) and (153.05,189.07) .. (151.11,191) .. controls (149.18,192.93) and (138.43,185.32) .. (127.11,174) .. controls (115.79,162.68) and (108.18,151.93) .. (110.11,150) -- cycle ;
%Shape: Arc 
\draw  [draw opacity=0] (127.63,143.19) .. controls (131.04,146.37) and (133.17,150.87) .. (133.17,155.85) .. controls (133.17,165.5) and (125.19,173.32) .. (115.35,173.32) .. controls (110.25,173.32) and (105.66,171.23) .. (102.41,167.87) -- (115.35,155.85) -- cycle ; \draw[decorate, decoration={coil, segment length=3pt, amplitude=2pt}]   (127.63,143.19) .. controls (131.04,146.37) and (133.17,150.87) .. (133.17,155.85) .. controls (133.17,165.5) and (125.19,173.32) .. (115.35,173.32) .. controls (110.25,173.32) and (105.66,171.23) .. (102.41,167.87) ;  
% Text Node
\draw (123.11,176) node [anchor=north west][inner sep=0.75pt]   [align=left] {0};
\end{tikzpicture}}
+
\adjustbox{valign=c,scale=0.90}{\tikzset{every picture/.style={line width=0.75pt}}   
\begin{tikzpicture}[x=0.75pt,y=0.75pt,yscale=-1,xscale=1]
%Straight Lines 
\draw[->,color=gray]    (216.46,206.91) -- (287.76,135.38) ;
%Straight Lines 
\draw[<-,color=gray]    (216.35,135.85) -- (287.88,206.45) ;
%Straight Lines 
\draw[very thick, color=sidsblue, line cap =round]    (272,120.08) -- (200.69,191.61) ;
%Straight Lines 
\draw[very thick, color=sidsblue, line cap =round]    (302,150.08) -- (230.69,221.61) ;
%Shape: Ellipse
\draw  [draw opacity=0][fill=sidsMaroon  ,fill opacity=1 ] (231.11,150) .. controls (233.05,148.07) and (243.79,155.68) .. (255.11,167) .. controls (266.43,178.32) and (274.05,189.07) .. (272.11,191) .. controls (270.18,192.93) and (259.43,185.32) .. (248.11,174) .. controls (236.79,162.68) and (229.18,151.93) .. (231.11,150) -- cycle ;
%Straight Lines 
\draw[decorate, decoration={coil, segment length=3pt, amplitude=2pt}]   (218.54,172.99) -- (291.68,161.01) ;
% Text Node
\draw (250.11,151) node [anchor=north west][inner sep=0.75pt]   [align=left] {0};
\end{tikzpicture}}
+
\adjustbox{valign=c,scale=0.90}{\tikzset{every picture/.style={line width=0.75pt}}   
\begin{tikzpicture}[x=0.75pt,y=0.75pt,yscale=-1,xscale=1]
%Straight Lines 
\draw[->,color=gray]    (335.46,206.91) -- (406.76,135.38) ;
%Shape: Arc 
\draw  [draw opacity=0] (329.77,181.49) .. controls (326.74,179.39) and (324.76,175.88) .. (324.76,171.92) .. controls (324.76,165.48) and (329.97,160.27) .. (336.41,160.27) .. controls (340.14,160.27) and (343.47,162.02) .. (345.6,164.76) -- (336.41,171.92) -- cycle ; \draw[decorate, decoration={coil, segment length=3pt, amplitude=2pt}]    (329.77,181.49) .. controls (326.74,179.39) and (324.76,175.88) .. (324.76,171.92) .. controls (324.76,165.48) and (329.97,160.27) .. (336.41,160.27) .. controls (340.14,160.27) and (343.47,162.02) .. (345.6,164.76) ;  
%Straight Lines 
\draw[<-,color=gray]    (335.35,135.85) -- (406.88,206.45) ;
%Straight Lines 
\draw[very thick, color=sidsblue, line cap =round]    (391,120.08) -- (319.69,191.61) ;
%Straight Lines 
\draw[very thick, color=sidsblue, line cap =round]    (421,150.08) -- (349.69,221.61) ;
%Shape: Ellipse 
\draw  [draw opacity=0][fill=sidsMaroon  ,fill opacity=1 ] (350.11,150) .. controls (352.05,148.07) and (362.79,155.68) .. (374.11,167) .. controls (385.43,178.32) and (393.05,189.07) .. (391.11,191) .. controls (389.18,192.93) and (378.43,185.32) .. (367.11,174) .. controls (355.79,162.68) and (348.18,151.93) .. (350.11,150) -- cycle ;
\end{tikzpicture}}
+
\adjustbox{valign=c,scale=0.90}{\tikzset{every picture/.style={line width=0.75pt}}   
\begin{tikzpicture}[x=0.75pt,y=0.75pt,yscale=-1,xscale=1]
%Straight Lines 
\draw[->,color=gray]    (454.46,205.91) -- (525.76,134.38) ;
%Straight Lines 
\draw[<-,color=gray]    (454.35,134.85) -- (525.88,205.45) ;
%Straight Lines 
\draw[very thick, color=sidsblue, line cap =round]    (510,119.08) -- (438.69,190.61) ;
%Straight Lines 
\draw[very thick, color=sidsblue, line cap =round]    (540,149.08) -- (468.69,220.61) ;
%Shape: Ellipse 
\draw  [draw opacity=0][fill=sidsMaroon  ,fill opacity=1 ] (469.11,149) .. controls (471.05,147.07) and (481.79,154.68) .. (493.11,166) .. controls (504.43,177.32) and (512.05,188.07) .. (510.11,190) .. controls (508.18,191.93) and (497.43,184.32) .. (486.11,173) .. controls (474.79,161.68) and (467.18,150.93) .. (469.11,149) -- cycle ;
%Straight Lines 
\draw[decorate, decoration={coil, segment length=3pt, amplitude=2pt}]    (455.35,173.85) -- (485.35,203.85) ;
\end{tikzpicture}}+\text{perms}\notag
\\&
\equiv [\hat{\mathcal{H}}_{\rm BK}\cdot \Uu_{12}]|_{\rm LO }\notag
\\&
\xrightarrow{\D=4} \frac{\alpha_s N}{2\pi}\int \frac{\d^2z_0}{\pi} \frac{z_{12}^2}{z_{10}^2z_{02}^2} \left( \Uu_{10} \Uu_{02}-\Uu_{12}\right) + \mathcal{O}[(\alpha_s N)^2] \,.
    \label{eq:BK0}
\end{align}
Here, $\hat{\mathcal{H}}_{\mathrm{BK}}$ denotes the so-called ``BK Hamiltonian,'' which will be discussed in greater detail later, and $\partial_\eta\simeq -\xb\partial_{\xb}$ is the evolution under boost. For now, we emphasize that its leading‑order (LO) action is proportional to $\alpha_s$. 
Eq.~\eqref{eq:BK0} thus describes the fact that, as one increases the projectile's energy, the radiation of gluons with smaller and smaller energy fraction $\ell^+/Q^+$ will occur at times
$x^+\sim \ell^+/\ell_\perp^2$ that are parametrically outside the target.  Since the gluon carries the same color charge as a $q\bar{q}$ pair inserted at a new transverse point $z_0$, the net effect is to probe the target by a product of two daughter dipoles, $\Uu_{10}$ and $\Uu_{02}$, rather than the original dipole.

For finite $N$, the evolution is governed by the Balitsky--JIMWLK hierarchy of equations \cite{Balitsky:1995ub,Balitsky:1997mk,Balitsky:2001mr,Jalilian-Marian:1997qno}. In the large-$N$ limit two distinct
simplifications arise. First, \eqref{eq:BK0} can be iterated without tracking an ever‐growing set of color connections: each daughter dipole can split further into grand-daughter dipoles such that the color structure reduces to simple products of elementary dipoles (see, e.g., Fig.~\ref{fig:largeN}). Second, target expectation values factorize naturally: $\langle \text{t}|\,\mathcal{U}_{10}\,\mathcal{U}_{02}\,|\text{t}\rangle
= \langle \text{t}|\,\mathcal{U}_{10}\,|\text{t}\rangle\;\langle \text{t}|\,\mathcal{U}_{02}\,|\text{t}\rangle.$
This property reduces the BK equation to an ordinary integro-differential equation for the classical function $\langle \text{t}|\,\mathcal{U}_{ij}\,|\text{t}\rangle(z_i,z_j;\eta)$.
Since we do not discuss expectation values here, we will rely solely on the first simplification, which follows directly from large-$N$ counting rules.

The $\mathcal{O}(\alpha_s^2)$ correction to $\partial_\eta\,\mathcal{U}_{12}$ was computed in \cite{Balitsky:2007feb,Balitsky:2009xg}.  Incorporating finite-$N$ non-planar corrections demands iterating the evolution with a general color-charge distribution on the left-hand side, which was worked out at two loops in \cite{Kovner:2013ona,Grabovsky:2013mba,Caron-Huot:2015bja}.  In this paper, we focus on deriving the planar contribution to the $\mathcal{O}([\alpha_s N]^3)$ small-$\xb$ evolution equation.

Physically, one can interpret $\langle\Uu_{ij}\rangle \equiv \langle \text{t}|\,\Uu_{ij}\,|\text{t}\rangle$ as the probability that a dipole of transverse size $z_{ij}$ escapes the target unscathed \cite{Marchesini:2015ica}.  Hence, the configuration $\langle {\cal U}_{ij}\rangle = 1$ is an (unstable) fixed point, realized only in the absence of any scattering (i.e., in the absence of a target). In general, the evolution equation must be solved with an initial condition at some initial rapidity scale $\eta_0=\log(1/x_0)$, and generically one finds that $\langle 1-{\cal U}\rangle$ increases with rapidity: the scattering region gradually ``eats'' into the transparent region and the opaque fixed point ${\langle \cal U}\rangle=0$ is approached pointwise.
For an in-depth review of the leading-order BK equation and related topics and a concise review of the initial conditions, we point the reader to \cite{Gelis:2012ri,Kovchegov:2012mbw,AbdulKhalek:2022hcn}.

On physical grounds, scattering cannot grow without bounds, and the nonlinear term in \eqref{eq:BK0} ensures that unitarity is preserved throughout the evolution and the solution remains bounded as $0 < \langle\Uu_{ij}(\eta)\rangle < 1$. Since $\langle 1-\Uu\rangle\sim \xb G(\xb,q^2)$ in some limits, this can be interpreted physically as the saturation of parton distributions at large occupation number \cite{Kovchegov:2012mbw,Iancu2019High}. 

\subsection{Setup: planar gauge theory in \texorpdfstring{$\D=4-2\varepsilon$}{dum1} dimensions}\label{ssec:setup}

We study SU($N$) gauge theory (whose Lagrangian density is given in App.~\ref{App:lcFR}) in $\D=4-2\varepsilon$ spacetime dimensions in the large-$N$ limit.
Instead of the standard 't Hooft large-$N$ limit, where fermions are subleading \cite{tHooft:1973alw}, we work in the so-called Veneziano limit
where $n_F,n_S\sim N$ \cite{Veneziano:1976wm} so that matter loops are retained; we allow both scalar and fermion matter.
Actually, for bookkeeping purposes it will be more convenient to consider a theory with an ${\cal O}(1)$ number $\nf$, $\ns$ of adjoint fields.
For adjoint matter, the $\beta$-function ($\mu\partial_\mu g_s=\beta(g_s)$)
\begin{equation}
    \frac{1}{g_s} \beta(g_s) = -\varepsilon -
\bar\beta(\aeff)=-\varepsilon-
(\aeff b_0+\aeff^2 b_1+\ldots)\,,\qquad \aeff\equiv \frac{\alpha_s N}{4\pi}=\frac{g_s^2 N}{16 \pi^2}\,,
\end{equation}
is given by the coefficient $b_0=\frac{11}{3}-\frac{2}{3}\nf-\frac{1}{6}\ns$ (see, e.g., \cite[Eq. (30)]{Luo:2002ti} for $b_1$),
where $\nf$ and $\ns$
denote respectively the number of adjoint Majorana fermions and adjoint real scalar fields.

For most of the paper, we will only record formulas for \emph{adjoint} matter. Equivalent formulas for $n_F$ fundamental Dirac fermions and $n_S$ fundamental complex scalars can be readily obtained using the substitutions $\ns\mapsto \frac{n_S}{N}$, $\nf\mapsto \frac{n_F}{N}$ (together with the erasure of certain Wilson lines as noted below \eqref{eq:summary_res}).

As will be further discussed in the context of the spacelike-timelike correspondence in Sec.~\ref{sec:correspondence}, special simplifications occur when the theory happens to be conformal.
The \emph{Wilson--Fisher fixed point} is picked out by the requirement that the $\beta$‑function vanishes for a particular choice of the dimensional‑regularization parameter.  In other words, for any (sufficiently small) value of $\alpha_s$ there exists a special dimension $\varepsilon=\varepsilon_{*}$ where the theory is conformal:
\begin{equation}\label{eq:WF_point_def}
     \varepsilon_{*}\equiv-\bar{\beta}(\aeff)\qquad\mbox{(conformal dimension)}\,.
\end{equation}
Motivated by treating $\varepsilon\sim \aeff$, we derive the ${\cal O}(\varepsilon)$ contributions to the two-loop BK evolution equation below. Via the spacelike-timelike correspondence (see Sec.~\ref{sec:correspondence}), this then predicts a nontrivial part of the three-loop evolution \emph{in the physical, non-conformal theory} (here, a planar approximation to QCD)---specifically, all three-loop terms that are \emph{not} conformally invariant.

\paragraph{Dimensional regularization conventions.}
Since we are interested in ${\cal O}(\varepsilon)$ corrections to the $\mathcal{O}(\alpha_s^2)$ BK kernel, let us clearly spell out the conventions that we follow.
As customary, we treat the physical coupling $g_s^2=4\pi\alpha_s$ as dimensionless. The bare coupling $g_0$ that appears in the Lagrangian thus comes with an explicit scale $\mu$ related to the $\overline{\rm MS}$ scale $\mubar$:
\begin{equation}
    g_0= \mu^{\varepsilon}g_s Z_{g_s},
    \qquad Z_{g_s} = 1-\frac{b_0}{2\varepsilon}a 
    +\left(\frac{3b_0^2}{8\varepsilon^2}-\frac{b_1}{4\varepsilon}\right)a^2
    +\ldots,\qquad \mu^2\equiv\mubar^2 \frac{\e^{\gamma_{\rm E}}}{4\pi} \,.
\label{bare coupling}
\end{equation}
$Z_{g_s}$ is a series of pure $\varepsilon$ poles fixed by the renormalization group equation $(\mu\partial_\mu+\beta(g_s)\partial_{g_s})g_0=0$.

We will always use the renormalized coupling $\aeff=\frac{\alpha_s N}{4\pi}$ as a loop counting parameter.  The renormalization scale and volume factors will
often appear through the following combinations:
\begin{equation} \label{def cd}
\mathcal{N}_\varepsilon
\equiv 2\pi \mu^{2\varepsilon}, \qquad \cd\equiv\frac{\Gamma\big(1{-}\varepsilon\big)}{\pi^{1{-}\varepsilon}}\,.
\end{equation}
We use conventional \emph{dimensional regularization} (DREG) where the number of
physical gluon states running inside loops
is $g^\mu_{\phantom{a}\mu}-2=2-2\varepsilon$.
For fermions, the Clifford algebra $\{\gamma_\mu,\gamma_\nu\}=2g_{\mu\nu}\mathbb{1}$ is assumed to hold for all $\D$-dimensional components, and fermionic traces are kept normalized to ${\rm tr}_f[1]=4$ for a Dirac fermion (2 for a Majorana fermion);  $\gamma_5$ and complications attached to this prescription will not appear in our calculations.

Sometimes, especially (but not exclusively) if supersymmetry is present, it is convenient to consider the \emph{dimensional reduction} (DRED) scheme \cite{Siegel:1979wq,Stockinger:2005gx} (equivalent to ``four-dimensional helicity'' (FDH) regularization scheme for our purposes \cite{Bern:1991aq}) where $+2\varepsilon$ real adjoint scalars are added to the theory to keep the number of bosonic degrees of freedom constant.  Our results will thus feature an additional parameter $\delta$, defined as:
\begin{equation}\label{eq:DREGvsDRED}
\delta = 
\begin{sqcases}
0 & \text{in DREG}\\
1 & \text{in DRED}
\end{sqcases}\,.
\end{equation}

\subsection{Bare dipole: integrands and amplitudes}

To formalize the high-energy evolution of the dipole amplitude in \eqref{dipole}, it is helpful to make explicit the rapidity factorization underlying \eqref{eq:obs}. The core idea is that, after a sufficiently large boost, the product of projectile operators (specifically the two currents $J^\mu$ that create and absorb the virtual photon in DIS) can be replaced by Wilson lines \cite{Balitsky:2013fea}. In schematic form reminiscent of OPEs \cite{Cheng:1984vwu}, with Lorentz indices on $J$ and $C$ omitted, we write:
\begin{equation} \label{factorization JJ}
    J(Q) J(-Q) \simeq \int\limits_{z_1,z_2} C^{{\rm ren}}(\nu,Q;\{z\}) \Uu_{12}^{{\rm ren}}(\nu)+\!\!
    \int\limits_{z_1,z_2,z_3} \!\!\!C^{{\rm ren}}(\nu,Q;\{z\}) \Uu_{12}^{{\rm ren}}(\nu)\Uu_{23}^{{\rm ren}}(\nu) +\ldots
\end{equation}
where the ``impact factors'' $C^{{\rm ren}}(\nu)$, which encode, for instance, details on how the photon decays into correlated Wilson lines (color singlet dipoles), can be computed perturbatively in $\alpha_s$ and are the same for any target.
The \emph{rapidity factorization scale} $\underline{\Lambda}\ll \nu\ll \overline{\Lambda}$ is chosen to lie between the two longitudinal energy scales in the problem: the projectile’s total energy, $\overline{\Lambda}=Q^+$, and the typical longitudinal energy probed in the target, $\underline{\Lambda}\sim
Q^2/P_{t}^-$.

Thus, \eqref{factorization JJ} expresses the correlator $\langle \text{t}|J J|\text{t}\rangle$ in terms of 
(simpler)
target expectation values of Wilson‑line operators.\footnote{%
Keeping the factorization scale $\nu$ arbitrary is useful, even though it ultimately cancels in \eqref{factorization JJ}: choosing $\nu\sim\underline{\Lambda}$ removes large logarithms from the target expectation value, while choosing $\nu\sim\overline{\Lambda}$ removes them from $C^{\mathrm{ren}}(\nu)$. Integrating the BK equation
$\nu\,\partial_\nu\,\mathcal{U}^{\mathrm{ren}}(\nu)$ over the range $\nu\in [\underline{\Lambda},\overline{\Lambda}]$
thus resums all large logarithms.%
}
Their perturbative calculation can be organized as integrals over the $+$‑components of loop momenta:
\begin{align}\label{eq:bareDipole}
\Bigl\langle \text{t}\Bigl|\mathcal{U}_{12}^{\mathrm{bare}}\Bigr|\text{t}\Bigr\rangle&= 
{\cal I}^{(0)}
+ \aeff\,\mathcal{N}_\varepsilon
\int\limits^{\infty}_{\sim\underline{\Lambda}}\frac{\d \ell_{0}^+}{\ell_{0}^+}\mathcal{I}^{(1)}(\ell_0^+)
+(\aeff\,\mathcal{N}_\varepsilon)^2
\int\limits^{\infty}_{\sim\underline{\Lambda}}\frac{\d \ell_{0}^+}{\ell_{0}^+}\frac{\d \ell_{0'}^+}{\ell_{0'}^+}\mathcal{I}^{(2)}(\ell_0^+,\ell_{0'}^+)+\ldots
\end{align}
where the superscript $(L)$ on the integrand $\mathcal{I}^{(L)}$ indicates the loop order and the ellipsis represents similar corrections from three loops onward. 

Next, we explain how to obtain $\mathcal{I}^{(L)}$ directly from the corresponding amplitudes.

As the integration ranges suggest, the integrals in \eqref{eq:bareDipole} converge at small $\ell^+$ due to the finite energy of the target, but they diverge at infinity. This is the origin of the rapidity divergences. Every $\ell^+$ integration can be split into three regions:
\begin{enumerate}[label=(\roman*)]
    \item Modes with $\ell^+\gtrsim \nu$: these belong to the impact factors $C^{{\rm ren}}(\nu)$ and should be subtracted from the Wilson lines (see Sec.~\ref{ssec:renDipole}).
    \item Modes with $\ell^+\lesssim \underline{\Lambda}$: these are part of the target and therefore are not explicitly included in the calculations after taking the target expectation value.
    \item Intermediate modes with $\underline{\Lambda}\ll \ell^+ \lesssim \nu$ contribute to the aforementioned \emph{scaling regime}: these produce perturbatively calculable corrections and give rise to large logarithms, which will be our focus.
\end{enumerate}
In the scaling regime, the Wilson‑line integrand (identical to that of the original current correlator for
$\underline{\Lambda}\ll\ell^+\ll\overline{\Lambda}$) is universal, i.e., independent of the fine‑grained details of both projectile and target, and can be computed straightforwardly. Since the target is highly Lorentz‑contracted for modes in this kinematic window, its sole effect is to color‑rotate the wavefunction of each parton that crosses it \cite{Balitsky:1995ub}. Consequently, the $L$‑loop integrand $\mathcal{I}^{(L)}$ decomposes into contributions with varying numbers of “real” partons crossing the target, each associated with a transverse coordinate (e.g., $z_{0},\,z_{0'},\,z_{0''},\dots$) to be integrated over:

\begin{equation}\label{eq:loopExpansion}
\hspace{-0.3cm}
\mathcal{I}^{(L)}(\ell_1^+,\ldots,\ell_{0^{|L|'}}^+)=\,
\hat{\mathcal{I}}^{(L)}_{[12]}+\hat{\mathcal{I}}^{(L)}_{[102]}+\hat{\mathcal{I}}^{(L)}_{[100'2]}+\hat{\mathcal{I}}^{(L)}_{[100'0''2]}+\ldots\qquad(\mbox{when all }\ell_{0^{i'}}^+\gg \underline{\Lambda})\,,
\end{equation}
where 
\begin{subequations}\label{eq:Ihats}
    \begin{align}
    \hat{\mathcal{I}}^{(L)}_{[12]}&=\mathcal{I}^{(L)}_{[12]}\,\Uu_{12}\,,\\
        \hat{\mathcal{I}}^{(L)}_{[102]}&=\int\limits_{z_0}\,\mathcal{I}^{(L)}_{[102]}\,\Uu_{10}\Uu_{02}\,,\label{eq:Ihat1}\\
    \hat{\mathcal{I}}^{(L)}_{[100'2]}&=\!\!\!\int \limits_{z_0,z_{0'}}\,\mathcal{I}^{(L)}_{[100'2]}\,\Uu_{10}\Uu_{00'}\Uu_{0'2}\,,
    \end{align}
\end{subequations}
and so on. To lighten notation, from \eqref{eq:Ihats} onward we omit the target expectation values, which are understood to accompany each term passively along the way.
The functions ${\cal I}^{(L)}_{[\cdots]}$ that will appear in our actual calculations are target-independent.
(Conceptually, the dipoles in \eqref{eq:Ihats} are evaluated at the low scale $\sim \underline{\Lambda}$, since our calculations only include harder modes.) 

\begin{figure}
    \centering
    \includegraphics[scale=0.5]{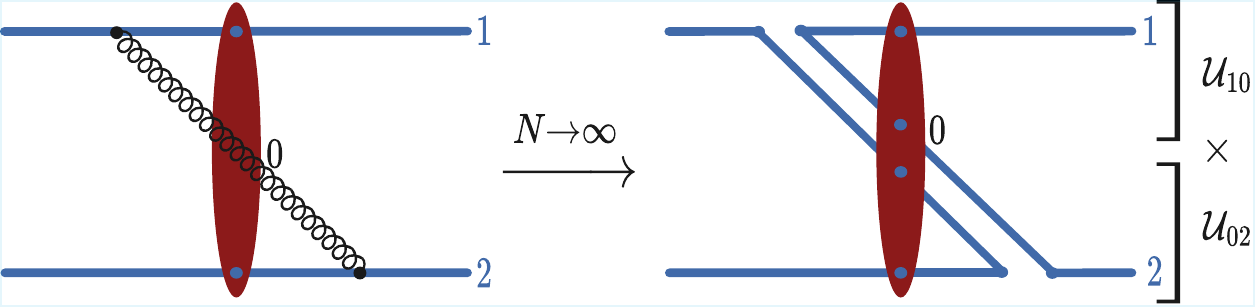}
    \caption{In the planar limit ($N\to \infty$), the dipole configuration on the left reduces to a product of two color dipoles $\Uu_{10}\Uu_{02}$.}
    \label{fig:largeN}
\end{figure}

In writing \eqref{eq:Ihats}, we have specialized to the planar limit, where all relevant color structures are products of elementary dipoles $\mathcal{U}_{ij} \equiv\frac{1}{N} \mathrm{Tr}[U(z_i) U^\dagger(z_j)]$ (see Fig.\,\ref{fig:largeN}).
As already mentioned, this simplification constitutes the main use of the planar limit in this work. 

As a bookkeeping device, we use the following labeling scheme.  For \emph{real emissions} (i.e., partons crossing the target), we use the momenta 
\begin{equation}
    \ell_\bullet^+\in\{k_{0}^+, k_{0'}^+, k_{0''}^+, \ldots\}\,,
\end{equation}
while for \emph{virtual emissions}  (i.e., those \emph{not} crossing the target), we use
\begin{equation}
    \ell_\bullet^+\in\{l_{0}^+, l_{0'}^+, l_{0''}^+, \ldots\}\,,
\end{equation}
at each loop order. The integrand where a (possibly empty) set $X$ of real partons crosses the target naturally factorizes into a product of amplitudes that capture the interactions on either side of the target:
\begin{equation} \label{I from A}
    \mathcal{I}_{[100'\ldots 0^{|X|'}2]}^{(L)}(\ell_0^+,\ldots, \ell_{0^{|L|'}}^+)= \sum_{X}\sum_{L_1+L_2=L-|X|}\, \mathcal{A}^{(X,L_1)}_{[100'\ldots0^{|X|'}2]}\, \mathcal{A}_{[100'\ldots0^{|X|'}2]}^{(X,L_2)} \,.
\end{equation}
Therefore, the amplitudes depend on $|X|$ transverse positions $z_{0}, z_{0'}, z_{0''}, \ldots, z_{0^{|X|'}}$
and at most
$L$
energies $\ell_0^+,\ldots, \ell_{L}^+$ associated to $|X|$ real emissions ($k_{0}^+, k_{0'}^+, k_{0''}^+, \ldots, k_{0^{|X|'}}^+$) and $L-|X|$ virtual ones ($l_{0}^+, l_{0'}^+, l_{0''}^+, \ldots, l_{0^{(L-|X|)'}}^+$) sprinkled around both sides of the target in all possible ways allowed by the interactions. In two loops ($L=2$), for example, we could have $X\in\{gg,ff,ss\}$ and $|X|=2$ for double emissions at tree level.

\paragraph{One-loop example.} As a simple example to illustrate the notation introduced above, let us briefly revisit the well-known result for the one-loop ($L=1$) calculation (see App.~\ref{App:lcFR} for a derivation from the standard (covariant) Feynman rules). The amplitude for a single real gluon emission ($X=g$) is a function of its transverse position $z_0$ and polarization index $i$ (but does not depend on the energy $k_0^+$):
\begin{equation} \label{amplitude single real}
\begin{split}
\Aa_{[102]}^{(g,0)i}(z_0)&=\adjustbox{valign=c}{\includegraphics[scale=0.5]{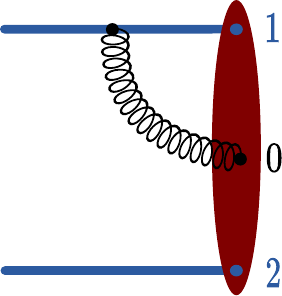}}\quad+\quad\adjustbox{valign=c}{\includegraphics[scale=0.5]{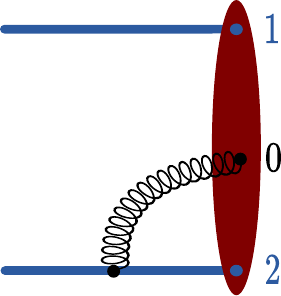}}\\
&=
\int\limits_{k_0}
\frac{2i k_0^i}{k_0^2} \left(\e^{ik_0\cdot z_{01}}-\e^{ik_0\cdot z_{02}}\right) \xrightarrow{\varepsilon\mapsto 0}
 \frac{1}{\pi}\left(\frac{z_{01}^i}{z_{01}^2}-\frac{z_{02}^i}{z_{02}^2}\right)\,,
\end{split}
\end{equation}
while the amplitude for a single virtual gluon emission is 
\begin{equation}\label{amplitude single virtual}
\begin{split}
\Aa_{[12]}^{(\varnothing,1)}&=\adjustbox{valign=c}{\includegraphics[scale=0.5]{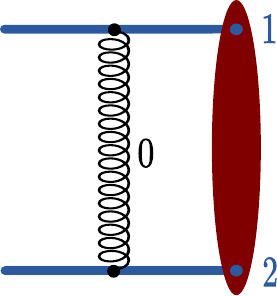}}\quad+\quad\adjustbox{valign=c}{\includegraphics[scale=0.5]{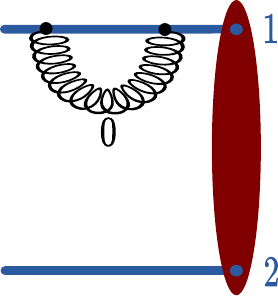}}\quad+\quad\adjustbox{valign=c}{\includegraphics[scale=0.5]{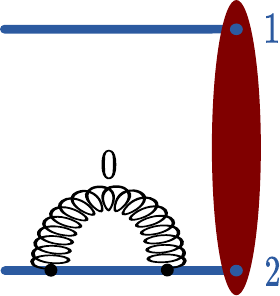}}
    \\
&=  \int\limits_{l_0} \frac{4}{l_0^2} \left(\e^{il_0\cdot z_{21}}-1\right)\,,
\end{split}
\end{equation}
where $\varnothing$ is there to remind us that no emitted partons cross the target (i.e., they are all virtual). Both of these amplitudes contribute to the leading-order term in \eqref{eq:bareDipole}, which we write in full as (see \eqref{I from A}):
\begin{subequations}\label{eq:I1sq}
    \begin{align}
\mathcal{I}_{[102]}^{(1)}&=\Aa_{[102]}^{(g,0)i}\,\Aa_{[102]}^{(g,0)i}=\cd^2\left[\frac{1}{(z_{01}^2)^{1-2\varepsilon}}+\frac{1}{(z_{02}^2)^{1-2\varepsilon}}-\frac{2z_{01}\cdot z_{02}}{(z_{01}^2z_{02}^2)^{1-\varepsilon}}\right]\xrightarrow{\varepsilon\mapsto 0} \frac{z_{12}^2}{\pi^2 z_{01}^2z_{02}^2}\,,\label{eq:I1_102}
        \\
        \mathcal{I}_{[12]}^{(1)}&=\left( \Aa_{[12]}^{(\varnothing,1)}\, \Aa_{[12]}^{(\varnothing,0)}+\Aa_{[12]}^{(\varnothing,0)}\, \Aa_{[12]}^{(\varnothing,1)}\right)=2\Aa_{[12]}^{(\varnothing,1)}\,,
   \end{align}
\end{subequations}
where $\cd$ is the constant in \eqref{def cd} and $\Aa_{[12]}^{(\varnothing,0)}=1$ is the ``integrand'' where nothing happens to the parent dipole.

A useful fact is that the real and virtual corrections are not independent; it is easy to see from the momentum space expressions in \eqref{amplitude single real} and \eqref{amplitude single virtual} that, in any $\Dperp$,
\begin{equation} \label{real-virtual one-loop}
    2\Aa_{[12]}^{(\varnothing,1)}=-
    \int\limits_{z_0}
    \Aa_{[102]}^{(g,0)i}\Aa_{[102]}^{(g,0)i}\,.
\end{equation}
This enables us to write the full integrand \eqref{eq:loopExpansion} as
\begin{equation}\label{eq:I1}
    \mathcal{I}^{(1)}(\ell_0^+)=
   \int\limits_{z_0}
   \,\mathcal{I}_{[102]}^{(1)}(z_0)\, (\Uu_{10} \Uu_{02}-\Uu_{12})\,.
\end{equation}
Here, $\mathcal{I}^{(1)}(\ell_0^+)$ is independent of the longitudinal momentum ($\mathcal{I}^{(1)}(\ell_0^+)=\mathcal{I}^{(1)}$), which leads to a rapidity divergence in the bare dipole \eqref{eq:bareDipole} (we will revisit this later in this section). 
Consequently, \eqref{eq:I1} is what controls the leading‑order small-$\xb$ evolution equation as shown in detail later in \eqref{eq:LOBKnu}: the $\D$-dimensional version of the one-loop BK equation \eqref{eq:BK0}.
Note that the $z_0$ integral in the two terms in \eqref{eq:I1} separately diverge as $\varepsilon\to 0$ but the combination is finite since a sufficiently small dipole is transparent:  $\lim_{z_0\to z_1}\Uu_{10}=1$.

The elegant way in which virtual and real contributions combine in \eqref{eq:I1} is not specific to one loop.  Physically, it ensures that the evolution has a fixed point for a trivial target (one where $\Uu_{ij}=1$ for all coordinates $z_i$, $z_j$) and for a fully opaque target where $\Uu_{ij}=0$, but these properties must hold to all loops. We will use them to avoid the direct calculation of purely virtual diagrams at two loops. That is, the coefficient of $\Uu_{12}$ must be minus the sum of all other coefficients, and we can generate it automatically by substituting in \eqref{eq:Ihats}:
\begin{equation}\label{eq:noFullyVirtual}
\Uu_{10}\Uu_{00'}\ldots\Uu_{0^{|L|'}2}\mapsto \Uu_{10}\Uu_{00'}\ldots\Uu_{0^{|L|'}2}-\Uu_{12}\,.
\end{equation}

In the next subsection, we describe how to obtain the renormalized, cutoff‑independent (and thus physical) dipole from the ultraviolet cutoff-dependent definition in \eqref{eq:bareDipole}.

{\subsection{Renormalized dipole, high-energy evolution and conformal symmetry}\label{ssec:renDipole}}

Since the integrand in \eqref{eq:I1} is independent of $\ell^+$, integrating it in \eqref{eq:bareDipole} up to $\ell^+=\infty$ generates logarithmic divergences, which precisely drive the BK evolution in \eqref{eq:BK0}. The region $\ell^+>\overline\Lambda$ is unphysical, arising solely from our Wilson‑line approximation of the projectile. We now show how to formally subtract all contributions with $\ell^+>\nu$ (that is, above the factorization scale) to define a renormalized dipole operator, in a way that maintains conformal symmetry if it is present.

Similarly to the usual renormalization theory of local operators, the subtraction can be understood in terms of a suitable $Z$-factor
\begin{equation}\label{eq:renSeed}
    \mathcal{U}_{12}^{\text{ren}}(\nu)\equiv \hat{Z}(\nu)\cdot\mathcal{U}_{12}^{\text{bare}}\,.
\end{equation}
The expectation values of this operator are finite and depend only on the energy scale $\nu$ through the BK equation:\footnote{
In deep inelastic scattering, evolution to larger $\nu\sim Q^+$ plays the same role as evolution toward small $\xb$:
$\nu\partial_\nu = \partial_\eta = -x\partial_x$.
Using an energy scale $\nu$ instead of ``rapidity'' is frequently done in the context of soft-collinear effective field theory; see, e.g., \cite{Chiu:2011qc}.
Thus, in analogy with how the Hamiltonian generates time evolution in quantum mechanics, the ``BK Hamiltonian'' $\hat{{\cal H}}_{\rm BK}$ generates rapidity evolution.
}
\begin{equation}\label{eq:Zevol}
    \nu \partial_{\nu} \hat{Z}(\nu)\equiv
    \hat{\mathcal{H}}_{\text{BK}}\cdot \hat{Z}(\nu) 
    \qquad \stackrel{\eqref{eq:renSeed}}{\implies} \qquad
    \nu \partial_{\nu}\, \mathcal{U}_{12}^{\text{ren}}(\nu)=
    \hat{\mathcal{H}}_{\text{BK}}\cdot \mathcal{U}_{12}^{\text{ren}}(\nu)\,.
\end{equation}
To be fully precise, one should first regulate the bare dipole $\Uu^{\rm bare}_{12}$ before renormalization (e.g., by imposing the hard cutoff $\ell^+<\overline{\Lambda}$ with $\overline{\Lambda}\gg\nu$, or by tilting the associated Wilson lines slightly off the lightcone; both regularization schemes are discussed in \cite{Balitsky:2007feb}). This ensures that the renormalization factor $\hat Z$ depends on the chosen regulator, while the renormalized dipole $\Uu^{\rm ren}_{12}$ remains finite and regulator‑independent. Since our procedure is entirely formulated in terms of $\Uu^{\rm ren}_{12}$ and does not rely on any auxiliary regulator, we suppress it in our notation.

Our starting point is thus the (formal) solution of \eqref{eq:Zevol}, which can be written in terms of the path-ordered exponential:  
\begin{equation}\label{eq:Uren1}
\begin{split}
        \mathcal{U}_{12}^{\mathrm{ren}}(\nu)
    &\equiv \mathbb{P}\exp\Bigg[-\int\limits_{\nu}^{\infty} \frac{\d \nu'}{\nu'} \hat{\mathcal{H}}_{\text{BK}}(\nu')\Bigg]\cdot\mathcal{U}_{12}^{\text{bare}}
    \\&=\Bigg[1
        -\int\limits_{\nu}^{\infty} \frac{\d \nu_0}{\nu_0}\hat{\mathcal{H}}_{\text{BK}}(\nu_0)
        {+}
        \int\limits_{\nu}^{\infty} \frac{\d \nu_0}{\nu_0}\int\limits_{\nu_0}^{\infty} \frac{\d \nu_{0'}}{\nu_{0'}}\hat{\mathcal{H}}_{\text{BK}}(\nu_{0'})\cdot\hat{\mathcal{H}}_{\text{BK}}(\nu_{0})+
        \ldots
        \Bigg]\cdot\mathcal{U}_{12}^{\text{bare}}\,.
\end{split}
\end{equation}
From the perturbative expansion of \eqref{eq:Uren1}, together with the definitions
\begin{equation}\label{eq:expBareRen}
  \mathcal{U}_{12}^{\text{bare}}
    \equiv \sum_{L=0}^\infty \aeff^L\,
      \mathcal{U}_{12}^{(L),\text{bare}}\,, \qquad
  \mathcal{U}_{12}^{\mathrm{ren}}(\nu)
    \equiv \sum_{L=0}^\infty \aeff^L\,
      \mathcal{U}_{12}^{(L),\mathrm{ren}}(\nu)\,,
      \qquad
      \hat{\mathcal{H}}_{\rm BK}
    \equiv \sum_{L=0}^\infty \aeff^L\,
      \hat{\mathcal{H}}_{\rm BK}^{(L)}\,,
\end{equation}
where $\mathcal{U}_{12}^{(0),\text{bare}}=\mathcal{U}_{12}^{(0),\text{ren}}\equiv\mathcal{U}_{12}^{(0)}\equiv \mathcal{I}^{(0)}$ and $\hat{\mathcal{H}}_{\rm BK}^{(0)}\equiv0$, one obtains the BK equation at one- and two-loop orders. For clarity, we discuss next these two cases separately.

\paragraph{One-loop evolution equation.} At one loop, we see from \eqref{eq:Uren1} that
\begin{equation}
    \mathcal{U}_{12}^{(1),\text{ren}}
=\mathcal{U}_{12}^{(1),\text{bare}}
-
\int\limits_{\nu}^{\infty} \frac{\d \nu_0}{\nu_0}\hat{\mathcal{H}}_{\text{BK}}^{(1)}\cdot \mathcal{U}_{12}^{(0)}
\,.\label{eq:expUren2}
\end{equation}
By definition, $\hat{\mathcal{H}}_{\text{BK}}^{(1)}$ must be such that \eqref{eq:expUren2} is finite. This is achieved provided that
\begin{equation}\label{eq:1loopAction}
    \hat{\mathcal{H}}_{\text{BK}}\cdot \mathcal{U}_{12}^{(0)}\equiv\mathcal{N}_\varepsilon\,\mathcal{I}^{(1)}(\ell_{0}^+)=\mathcal{N}_\varepsilon\,\mathcal{I}^{(1)}\,,
\end{equation}
with $\mathcal{I}^{(1)}$ given in \eqref{eq:I1}.
Indeed, after introducing the ordering variable $\nu_{[102]}=\nu_0\propto\ell_0^+$, which will be defined more formally below (the precise proportionality constant will only affect the Hamiltonian starting from two loops), we can immediately rewrite \eqref{eq:expUren2} as
\begin{equation}
\begin{split}
    \mathcal{U}_{12}^{(1),\mathrm{ren}}(\nu)&
= \mathcal{N}_\varepsilon\,\int\limits_{\sim \underline{\Lambda}}
\frac{\d \ell_{0}^+}{\ell_{0}^+}\,
\mathcal{I}^{(1)}\,\theta(\nu_{[102]}<\nu)
\,.\label{eq:expUren2_1}
\end{split}
\end{equation}
The support $\ell_0^+\in[\underline{\Lambda},\propto \nu]$ of \eqref{eq:expUren2_1} covers the scaling region mentioned earlier and is depicted in Fig.~\ref{fig:support1loop}. Recalling \eqref{eq:I1}, the one-loop evolution equation can thus be written as:
\begin{subequations}
    \begin{align}
    \label{eq:LOBKnu}
     & 
     \hat{\mathcal{H}}_{\rm BK}^{(1)}\cdot \mathcal{U}_{12}^{\mathrm{ren}}(\nu)
     =\mathcal{I}^{(1)}\equiv C_\varepsilon\int\limits_{z_0}H_{[102]}^{(1)}(z_0)\,(\Uu_{10}\Uu_{02}-\Uu_{12})\,,
     \\&
      \hspace{-0.3cm} 
      \text{where:} \quad      H^{(1)}_{[102]} \equiv \frac{{\cal N}_\varepsilon}{\cd}\Aa_{[102]}^{(g,0)i}\,\Aa_{[102]}^{(g,0)i}=
    2\pi \mu^{2\varepsilon}\cd\left[
    \frac{1}{(z_{01}^2)^{1-2\varepsilon}}+\frac{1}{(z_{02}^2)^{1-2\varepsilon}}-\frac{2z_{01}\cdot z_{02}}{(z_{01}^2z_{02}^2)^{1-\varepsilon}}\right]\,.
\label{eq:H1_102}
    \end{align}
\end{subequations}
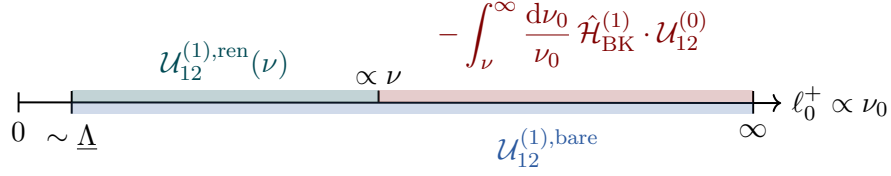
\begin{figure}
    \centering
    \tikzset{every picture/.style={line width=0.75pt}}
\begin{tikzpicture}[x=0.75pt,y=0.75pt,yscale=-1,xscale=2]
%Straight Lines
\draw[->]    (170,130) -- (361.38,130.34) node[right]{$\ell_0^+\propto\nu_0$};
\draw [shift={(170,130)}, rotate = 180.1] [color=black  ][line width=0.75]    (0,5.03) -- (0,-5.03)   node[below]{0};
\draw    (200.33-17,130-5.7) -- (200.33-17,135.67) node[below]{$\sim\underline{\Lambda}$};
\begin{scope}[yshift=-0.155cm]
    \draw    (260.33,130) -- (260.33,135.67) node[yshift=0.1cm,above]{$\propto\nu$};
\end{scope}
\draw    (329.33+25,130-5.7) -- (329.33+25,135.67) node[below]{$\infty$};
%Shape: Rectangle 
    \draw  [draw opacity=0][fill=sidsblue  ,fill opacity=0.22 ] (200.33-17,124.07+5.9) -- (329.33+25,124.07+5.9) node[pos=0.70,yshift=-0.2cm,below,color=sidsblue, fill opacity=1]{$\mathcal{U}_{12}^{(1),\text{bare}}$} -- (329.33+25,130+5.9) -- (200.33-17,130+5.9) -- cycle ;
%Shape: Rectangles 
\draw  [draw opacity=0][fill=sidsMaroon  ,fill opacity=0.2 ] (260.33,124.07) -- (329.33+25,124.07)  -- (329.33+25,130) -- (260.33,130) -- cycle;
\node[above, text=sidsMaroon] 
  at ($(265.33+15,118.43)!.45!(329.33+15,118.43)$) {%
  $\displaystyle
    -\int_{\nu}^{\infty}\frac{\mathrm{d}\nu_0}{\nu_0}
    \,\hat{\mathcal{H}}_{\mathrm{BK}}^{(1)}\cdot
    \mathcal{U}_{12}^{(0)}
  $};
\draw  [draw opacity=0][fill=mgGreen  ,fill opacity=0.24 ] (200.33-17,130-5.9) -- (260.33,130-5.9) node[yshift=0cm,above,midway,color=mgGreen, fill opacity=1]{$\mathcal{U}_{12}^{(1),\mathrm{ren}}(\nu)$} -- (260.33,135.67-5.9) -- (200.33-17,135.67-5.9) -- cycle ;
\end{tikzpicture}
    \caption{
    The support of the one‑loop renormalized dipole integral over $\ell_0^+$ in \eqref{eq:expUren2_1} is shown (upper green rectangle).  Also displayed are the supports of the individual terms on the right-hand side of \eqref{eq:expUren2}: the lower blue rectangle corresponds to the bare dipole support (target expectation values left implicit as explained in the text), while the upper red rectangle represents the support of the subtraction term.
    }
    \label{fig:support1loop}
\end{figure}
In $\D=4$, we manifestly recover the leading-order BK equation quoted earlier in \eqref{eq:BK0}.

\paragraph{Two-loop evolution equation.} At two loops, we see from \eqref{eq:Uren1} that
\begin{align}\label{eq:2l1}
\mathcal{U}_{12}^{(2),\mathrm{ren}}(\nu)
&=\mathcal{U}_{12,(2)}^{\text{bare}}
-\int\limits_{\nu}^{\infty} \frac{\d \nu_0}{\nu_0}[\hat{\mathcal{H}}_{\text{BK}}^{(1)}\cdot \mathcal{U}_{12}^{\text{ren}}]^{(1)}-\int\limits_{\nu}^{\infty} \frac{\d \nu_0}{\nu_0}\hat{\mathcal{H}}_{\text{BK}}^{(2)}\cdot \mathcal{U}_{12}^{(0)}\,,
\end{align}
where the contributions from one-loop iterations are grouped as
\begin{equation}\label{eq:tauDivs}
    \int\limits_{\nu}^{\infty} \frac{\d \nu_0}{\nu_0}[\hat{\mathcal{H}}_{\text{BK}}^{(1)}\cdot \mathcal{U}_{12}^{\text{ren}}]^{(1)}\equiv \int\limits_{\nu}^{\infty} \frac{\d \nu_0}{\nu_0}\hat{\mathcal{H}}_{\text{BK}}^{(1)}\cdot \mathcal{U}_{12}^{(1),\text{bare}}
-\int\limits_{\nu}^{\infty}\frac{\d \nu_0}{\nu_0}\int\limits\limits_{\nu_0}^{\infty}\frac{\d \nu_{0'}}{\nu_{0'}}\hat{\mathcal{H}}_{\text{BK}}^{(1)}(\nu_{0'})\cdot\hat{\mathcal{H}}_{\text{BK}}^{(1)}(\nu_{0})\cdot \mathcal{U}_{12}^{(0)}\,,
\end{equation}
and $(\nu_0,\nu_{0'})\propto(\ell_0^+,\ell_{0'}^+)$.

By construction, $\mathcal{U}_{12}^{(2),\mathrm{ren}}(\nu)$ in \eqref{eq:2l1} is finite. As for the one-loop case, this finiteness results from nontrivial cancellations of divergences on the right‐hand side of the equation. The new feature at two loops is that the bare dipole $\mathcal{U}_{12}^{(2),\mathrm{bare}}$ now contains rapidity \emph{subdivergences}, 
which occur when one daughter emission forms over a timescale much longer or shorter than the other 
(i.e., as $\tau \equiv \ell_0^+ / \ell_{0'}^+ \to 0$ or $\infty$), in addition to an overall rapidity divergence which will be associated with $\hat{\mathcal{H}}^{(2)}_{\rm BK}$.

Recall that, in the scaling (boost-invariant) regime, the integrand $\mathcal{I}^{(2)}(\ell_0^+,\ell_{0'}^+)=\mathcal{I}^{(2)}(\tau)$ is homogeneous and thus depends only on the energy ratio $\tau$.
Furthermore, in singular limits, it universally factorizes into products of lower-loop evolution. At the level of the integrands discussed earlier, this means that
\begin{equation}\label{eq:facto}
  \lim_{\tau\to 0\,{\rm or}\,\infty}\mathcal{I}_{[\dots]}^{(2)}(\tau)\;=\;\mathcal{I}_{[\dots]}^{(1)}\,\mathcal{I}_{[\dots]}^{(1)}\,,
\end{equation}
(see, e.g., \eqref{amp factorization} and \eqref{virtual factorization} for explicit examples). \begin{figure}
    \centering
\includegraphics[scale=0.5]{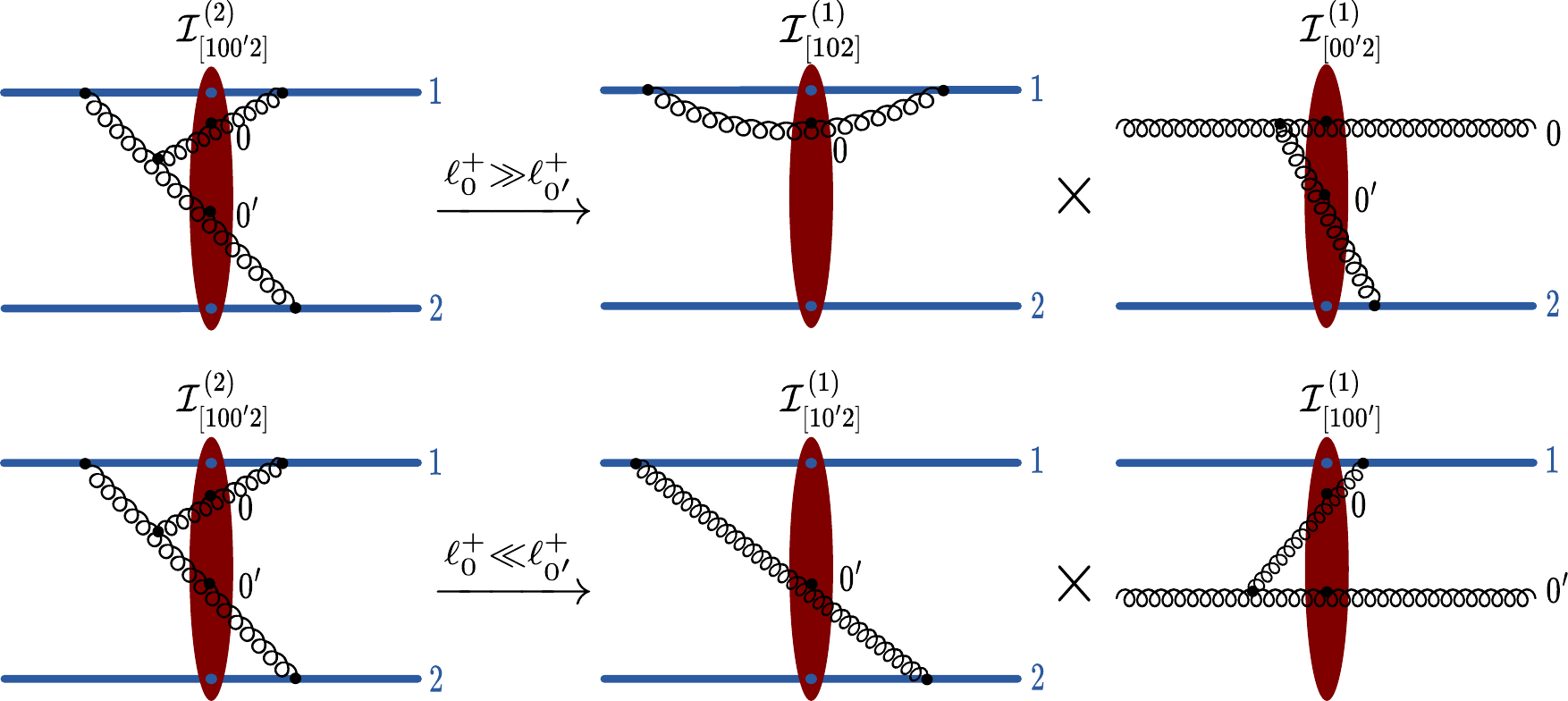}
    \caption{Example of diagram factorizations in small and large $\tau$ integration regions. The formation time of each quantum is $t_i \propto \ell_{0^{i'}}^+$.}
    \label{fig:facto1}
\end{figure}
Note that \eqref{eq:facto} admits a simple diagrammatic interpretation (see Fig.~\ref{fig:facto1} and Fig.~\ref{fig:facto2}). Since \eqref{eq:tauDivs} exactly encodes the one‑loop evolution of the one‑loop dipole and the successive one‑loop evolutions of the parent dipole, this factorization captures (and ends up canceling) \emph{all} rapidity subdivergences, as will be manifest in the explicit computations of Sec.\,\ref{ex:BKmain}. 

Following \cite{Caron-Huot:2015bja}, we thus ``slice'' the two-loop phase space by using, in addition to $\tau$, an ordering variable proportional to the geometric mean of energies $\nu_{[100'2]}$:
\begin{equation}\label{eq:slicing}
  (\tau,\nu_{[100'2]})
  =\Bigl(\tfrac{\ell_0^+}{\ell_{0'}^+},\;\propto\sqrt{\ell_0^+\,\ell_{0'}^+}\Bigr)\,.
\end{equation}
In these variables, the two-loop renormalized dipole in \eqref{eq:2l1} becomes
\begin{align} \label{Uren two-loop}
\mathcal{U}_{12}^{(2),\mathrm{ren}}(\nu)
&=\mathcal{N}_\varepsilon^2\int\limits^{\infty}_{\sim\underline{\Lambda}}\frac{\d \nu_{[100'2]}}{ \nu_{[100'2]}}\int\limits^{\infty}_{0}\frac{\d \tau}{\tau}
\Bigl[\mathcal{I}^{(2)}(\tau)-\mathcal{S}^{(2)}(\tau)\Bigr] -
\int\limits_{\nu}^{\infty} \frac{\d \nu_0}{\nu_0}\hat{\mathcal{H}}_{\text{BK}}^{(2)}\cdot \mathcal{U}_{12}^{(0)}\,,
\end{align}
where ${\cal S}$ collects all the one-loop iterations from \eqref{eq:2l1}:
\begin{equation}
    \mathcal{S}^{(2)}(\tau)=\!\!\!
          \int\limits_{z_{0},z_{0'}}
          \left[\begin{array}{rll}
          &\phantom{a}(\Uu_{12}{-}\Uu_{10}\Uu_{02})\mathcal{I}_{[10'2]}^{(1)}\mathcal{I}_{[102]}^{(1)}\theta(\nu_{[10'2]}{>}\nu_{[102]})
          \\
          &+(\Uu_{10}\Uu_{00'}\Uu_{0'2}{-}\Uu_{10}\Uu_{02})\mathcal{I}_{[102]}^{(1)}\mathcal{I}_{[00'2]}^{(1)}\theta(\nu_{[102]}{>}\nu_{[00'2]})
          \\
          &+(\Uu_{10}\Uu_{00'}\Uu_{0'2}{-}\Uu_{10'}\Uu_{0'2})\mathcal{I}_{[10'2]}^{(1)}\mathcal{I}_{[100']}^{(1)}\theta(\nu_{[10'2]}{>}\nu_{[100']})
          \end{array}\right]\,.\label{eq:Sdef} 
\end{equation}
Hence, because ${\cal S}$ includes all factorization channels,
the integrand in \eqref{Uren two-loop} has all subdivergences canceled, that is, 
\begin{equation}
      \lim_{\tau\to 0\,{\rm or}\,\infty}\Bigl[\mathcal{I}^{(2)}(\tau)-\mathcal{S}^{(2)}(\tau)\Bigr]=0\,.
\end{equation}
The first term in \eqref{Uren two-loop} thus only has an overall (single-logarithmic) rapidity divergence, which is canceled by choosing:
\begin{equation}
\hat{\mathcal{H}}_{\mathrm{BK}}^{(2)}\!\cdot\mathcal{U}_{12}^{(0)}
  \equiv\mathcal{N}_\varepsilon^2\int_{0}^{\infty}\frac{\mathrm{d}\tau}{\tau}
   \bigl[\mathcal{I}^{(2)}(\tau)-\mathcal{S}^{(2)}(\tau)\bigr]\,.
\end{equation}
Organizing the different dipoles into double-real and single-real emissions, we write
\begin{equation}\label{eq:evolEqU2}
\begin{split}
    \hat{\mathcal{H}}_{\rm BK}^{(2)}\cdot \mathcal{U}_{12}^{\mathrm{ren}}(\nu)&
        \equiv 
        C_\varepsilon \!\int\limits_{z_{0}}\,\big(\Uu_{10}\Uu_{02}-\Uu_{12}\big)\,H_{[102]}^{(2)}
        +
        C_\varepsilon^2
        \int\limits_{z_0,z_{0'}}\big(\Uu_{10}\Uu_{00'}\Uu_{0'2}-\Uu_{12}\big)\,H_{[100'2]}^{(2)}\,,
\end{split}
\end{equation}
from which the two-loop kernels can be calculated as
\begin{subequations}\label{eq:evolEqU2Hs}
    \begin{align}
        H_{[102]}^{(2)}&\equiv
\frac{{\cal N}_\varepsilon^2}{\cd}
    \int_0^{\infty} 
    \frac{\d\tau}{\tau}
\left[{\cal I}^{(2)}_{[102]}(\tau)
- 
\int\limits_{z_{0'}} {\cal I}_{[102]}^{(1)}\times \left[\begin{array}{lll}
& \phantom{+}\mathcal{I}_{[10'2]}^{(1)}\theta\Big(1>\tfrac{\nu_{[102]}}{\nu_{[10'2]}}\Big)
\\&+ \mathcal{I}_{[00'2]}^{(1)}\theta\Big(\tfrac{\nu_{[102]}}{\nu_{[00'2]}}>1\Big)
\\&+ \mathcal{I}_{[10'0]}^{(1)}\theta\Big(\tfrac{\nu_{[102]}}{\nu_{[10'0]}}>1\Big)
\end{array}\right]
\right]\,,\label{eq:L2RV}
\\
H^{(2)}_{[100'2]}&\equiv \frac{{\cal N}_\varepsilon^2}{\cd^2}
\int_0^\infty \frac{\d\tau}{\tau}
\left[\begin{array}{ll}
{\cal I}^{(2)}_{[100'2]}(\tau)
&-{\cal I}^{(1)}_{[10'2]}{\cal I}^{(1)}_{[100']}\theta(\nu_{[10'2]}{>}\nu_{[100']})
\\&-{\cal I}^{(1)}_{[102]}{\cal I}^{(1)}_{[00'2]}\theta(\nu_{[102]}{>}\nu_{00'2})
\end{array}\right]\,.\label{eq:L2DR}
    \end{align}
\end{subequations}

The evaluation of \eqref{eq:evolEqU2Hs} in $\D = 4 - 2\varepsilon$ will be the main result of this paper (see Sec.~\ref{ex:BKmain}).
As we will discuss subsequently in Sec.~\ref{sec:correspondence}, this is then used to predict, for the first time, the non-conformal contribution to the three-loop BK Hamiltonian in the planar limit of a generic gauge theory.
Note that while \eqref{eq:evolEqU2Hs} manifests the cancellation of rapidity divergences, ultraviolet and collinear divergences as $\varepsilon\to 0$ still require separate care (see \eqref{eq:H102 def}).
The final result (see \eqref{eq:summary_res} or, equivalently, \eqref{eq:resultthree-loop}) is obtained in the ``conformal subtraction scheme,'' which we discuss next. 

\paragraph{Choice of ordering variables and the conformal dipole.} 
Before proceeding to the explicit computation of \eqref{eq:evolEqU2Hs}, we briefly comment on the “geometric‐mean” or \emph{ordering variables}
\begin{equation}\label{eq:orderingVarProp}
\nu_{[i\,a_1\ldots a_L\,j]}\equiv\lambda_L\,\times\,(\ell_{a_1}^+\cdots\ell_{a_L}^+)^{1/L}\,,
\end{equation}
introduced around \eqref{eq:1loopAction} (for $L=1$) and \eqref{eq:slicing} (for $L=2$), where $L$ denotes the loop order and $\lambda_L$ is a positive proportionality constant.

These variables are interesting for two reasons we did not yet emphasize: first, they naturally enforce a strict decrease in the geometric mean of longitudinal momenta of each cluster at each emission step: by requiring 
$\ell^+_{a_{L+1}}<\nu_{[i\,a_1\cdots a_L\,j]}$, one has 
$\nu_{[i\,a_1\cdots a_{L+1}\,j]}<\nu_{[i\,a_1\cdots a_L\,j]}$; and second, by making the specific choice
\begin{equation}\label{eq:conformalScheme}
\lambda_L\equiv\biggl[\frac{z_{i a_1}^2\,z_{a_1 a_2}^2\cdots z_{a_L j}^2}{z_{i j}^2}\biggr]^{\frac{1}{2L}}  \qquad (\text{conformal scheme})\,,
\end{equation}
we can make the regulator manifestly conformal from the outset. Indeed, it is straightforward to verify that $\nu_{[i\,a_1\cdots a_L\,j]}$, as defined in \eqref{eq:orderingVarProp} with \eqref{eq:conformalScheme}, is invariant under inversion 
$(z_k,\ell_k^+)\mapsto(z_k/z_k^2,\;z_k^2\ell_k^+)$ for all $k$ 
and under translations of the transverse coordinates.
(Through the spacelike-timelike correspondence discussed in Sec.\,\ref{ssec:correspondence}, $\nu_{[i\,a_1\cdots a_L\,j]}$ is equivalent to the ordering variable which was used for non-global logarithms in \cite{Caron-Huot:2016tzz}.)
Hence, in this “conformal scheme,” we refer to $\mathcal{U}_{12}^{\mathrm{ren}}(\nu)$ as the \emph{conformal dipole} (see also \cite{Balitsky:2009xg}), and any breaking of conformal symmetry will be due to the physical running of the coupling.

As a closing remark we note that in the BK literature (see, e.g., \cite{Balitsky:2007feb}), $\lambda_L$ in \eqref{eq:orderingVarProp} is often set to 1, which gives rise to the so-called ``$+$-prescription scheme'' (see \eqref{plus example}). Although this choice is mathematically correct and even natural due to its simplicity, it is not the one we adopt to present our results in \eqref{eq:resultthree-loop}, as such a rapidity regulator would break conformal invariance, which would then have to be restored by hand \cite{Balitsky:2009xg}.

\section{Dimensionally regularized NLO evolution of color dipoles}\label{ex:BKmain}

In this section, we assemble the two-loop BK kernel in impact parameter space, expanded to first order in $\varepsilon$. Let us first briefly review the one-loop BK Hamiltonian in \eqref{eq:H1_102}.
In practice we will need the expansion of this expression in $\varepsilon$, where in the context of the spacelike-timelike correspondence (see Sec.\,\ref{sec:correspondence}) we treat $\varepsilon\sim \alpha_s$.
Anticipating application to three loops, we will thus need to expand the one-loop kernel up to and including the order $\varepsilon^2$.
It is convenient to pull out a conformal-invariant measure factor which makes the kernel dimensionless:
\begin{equation} \label{eq:H1_102 pre}
     H^{(1)}_{[102]}= \left(\frac{z_{12}^2}{z_{10}^2z_{02}^2}\right)^{1-\varepsilon}\tilde{H}^{(1)}_{[102]}\,.
\end{equation}
Expanding \eqref{eq:H1_102} then gives:
\begin{equation}\label{eq:H1_102exp}\begin{aligned}
\tilde{H}^{(1)}_{[102]} &= 
2+2\varepsilon\left(L_\mu-\frac{z_{01}^2-z_{02}^2}{z_{12}^2}\log\frac{z_{01}^2}{z_{02}^2}\right)
\\ &\quad+\varepsilon^2\left(\frac{\pi^2}{6}+L_\mu^2 -2L_\mu\frac{z_{01}^2-z_{02}^2}{z_{12}^2}\log\frac{z_{01}^2}{z_{02}^2}+\frac{z_{01}^2+z_{02}^2}{z_{12}^2}\log^2\frac{z_{01}^2}{z_{02}^2}\right)+{\cal O}(\varepsilon^3)\,,
\end{aligned}\end{equation}
with $L_\mu\equiv\log\Big[\frac{\mubar^2z_{12}^2}{4\e^{-2\gamma_\text{E}}}\Big]$ where
$\mubar^2\equiv 4\pi\mu^2\e^{-\gamma_\text{E}}$ is the standard $\overline{\rm MS}$ scale.
A similar organization will be used at two loops.

\begin{figure}[t]
    \centering
    \adjustbox{valign=c,scale=0.8}{\includegraphics[scale=0.8]{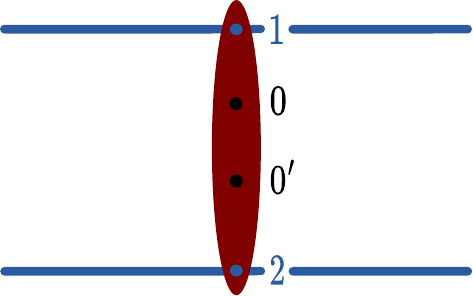}
    }   
    \qquad \qquad \qquad
    \adjustbox{valign=c,scale=0.8}{\includegraphics[scale=0.8]{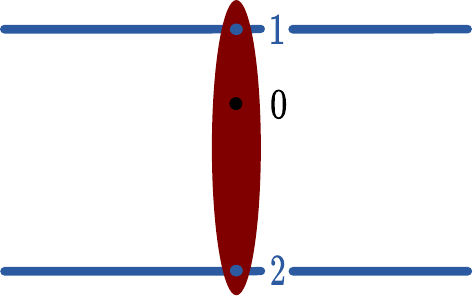} }
    \caption{Transverse coordinate conventions for the double-real (left) and real-virtual (right) NLO color-dipole. The labels $i=0,0',1$ and $2$ denote the positions $z_i$ at which partons cross the shock and experience the color potential of the target hadron.}
    \label{fig:coords}
\end{figure}

We now detail our two-loop calculation where we will keep the $\mathcal{O}(\varepsilon)$ terms. There will be double-real and real-virtual contributions to the two-loop Hamiltonian, namely $H^{(2)}_{[100'2]}$ and $H^{(2)}_{[102]}$ in \eqref{eq:evolEqU2Hs}.
The transverse coordinates conventions are depicted in Fig.\,\ref{fig:coords}: the double-real contribution is characterized by two particles crossing the shock at positions $z_0$ and $z_{0'}$, while the real-virtual contribution includes only one crossing at position $z_0$.
We will consider these contributions separately, as they include different classes of integrals.

Our method is heavily based on \cite{Balitsky:2006wa,Balitsky:2007feb,Balitsky:2009xg}.  We start from the same momentum-space amplitudes from the Feynman rules, Fourier-transform to transverse coordinate space, and integrate over relative energies. The only change is that we keep all $\mathcal{O}(\varepsilon)$ terms.

{\subsection{Starting point: splitting amplitudes in momentum space}\label{ssec:amplitudes}}

The building blocks for evolution are the splitting amplitudes that depend on the transverse position of the partons crossing the shock. However, the Feynman rules in lightcone gauge naturally produce expressions in momentum space.  Here, we record these initial expressions, whose Fourier transforms will be calculated in the next subsection and in App.~\ref{app:detailsHRV}.

We start by recalling the leading-order amplitude for the emission of one (color-ordered) gluon, given earlier in \eqref{amplitude single real}:
\begin{align} \label{LO amplitude mom}
    \Aa_{[102]}^{(g,0)i}
    &=\int\limits_{k}
\frac{2i k^i}{k^2} \left(\e^{ik{\cdot}z_{01}}-\e^{ik{\cdot}z_{02}}\right)\,.
\end{align}
For the two-loop calculation, we need the color‑ordered amplitudes for:\footnote{Recall from the argument above \eqref{eq:noFullyVirtual} that the double-virtual contributions need not be computed explicitly, since they are determined by the fixed points of the evolution equation.}
\begin{enumerate}
    \item \textbf{Tree-level double-real contribution:} emission of two soft partons crossing the shock, which can be either two gluons, two scalars or two fermions.
    \item \textbf{One-loop real-virtual contribution:} emission of a single soft gluon crossing the shock, accompanied by a loop correction entirely on \emph{one} side of the shock.
\end{enumerate}

\begin{figure}
    \centering
    \begin{equation*}
        \begin{split}
        \adjustbox{valign=c}{\includegraphics[scale=0.6]{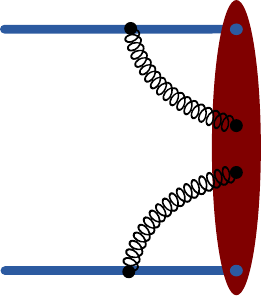}}\quad+\adjustbox{valign=c}{\includegraphics[scale=0.6]{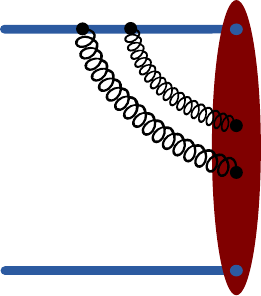}}\quad+\adjustbox{valign=c}{\includegraphics[scale=0.6]{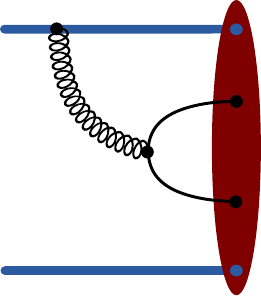}}\quad+\text{perms}
        \end{split}
    \end{equation*}
    \caption{
       The set of Feynman diagrams contributing to the double-real (left) amplitude. Non-spiral black lines accounts for all possible pair of intermediate states: $(g,g), (f,\bar{f}), (s,s)$.
        }\label{fig:dr_amp}
\end{figure}

The two soft-gluon double-real amplitude is given by the sum of diagrams shown in Fig.~\ref{fig:dr_amp}, which are shown in the order in which they appear in the square bracket below. Their momentum space expressions are directly obtained from the Feynman rules in lightcone gauge reviewed in App.~\ref{App:lcFR}. Since this starting point is entirely dimension-independent and identical to that of \cite{Balitsky:2007feb}, which we reproduced, here we simply quote the result: 
\begin{align} \label{gluon amplitude def}
 &\Aa_{[100'2]}^{(gg,0)ij}(\tau)
= 4\int\limits_{k_1,k_2}\,
\e^{ik_1{\cdot}z_0+ik_2{\cdot}z_{0'}}
\nonumber\\&
\quad\times\Bigg[\e^{-ik_1{\cdot}z_1-i k_2{\cdot}z_2} \frac{k_1^ik_2^j}{k_1^2k_2^2}-
\frac{k_1^ik_2^j}{k_1^2+\tau k_2^2}
\left(
\frac{\e^{-i(k_1{+}k_2){\cdot}z_1}}{k_2^2}
+\tau\frac{\e^{-i(k_1{+}k_2){\cdot}z_2}}{k_1^2}
\right)
\nonumber\\
&
\qquad{+}\left(\e^{-i(k_1+k_2){\cdot}z_1}{-}\e^{-i(k_1+k_2){\cdot}z_2}\right)
\frac{k_1^i(k_1{+}k_2)^j-\tau (k_1{+}k_2)^ik_2^j+\frac{\tau}{2(1+\tau)}(k_2^2-k_1^2)\delta^{ij}}{(k_1^2+\tau k_2^2)(k_1+k_2)^2}
\Bigg]\,,
\end{align}
where $\tau=k_0^+/k_{0'}^+$ denotes the ratio of the energies of the two gluons. 

We will also need the amplitudes controlling the emission of a pair of soft (real adjoint) scalars or (adjoint Majorana)
fermions crossing the shock:
\begin{subequations}\label{eq:dr_sf_amp}
\begin{align}\label{scalar amplitude def}
 \Aa_{[100'2]}^{(ss,0)}(\tau) &= \!
\int\limits_{k_1,k_2}
\left(\e^{ik_1{\cdot}z_{01}+ik_2{\cdot}z_{0'1}}
-(z_1{\mapsto}z_2)\right)
\frac{2\tau}{(1+\tau)}\frac{k_2^2-k_1^2}{(k_1^2+\tau k_2^2)(k_1+k_2)^2}\,,
\\
\label{fermion amplitude def}
 \Aa_{[100'2]}^{(ff,0)}(\tau) &= \!
\int\limits_{k_1,k_2}
\left(\e^{ik_1{\cdot}z_{01}+ik_2{\cdot}z_{0'1}}
-(z_1{\mapsto}z_2)\right)
\frac{2\sqrt{\tau}\, \kslash_1\kslash_2}{(k_1^2+\tau k_2^2)(k_1+k_2)^2}\,.
\end{align}
\end{subequations}

Diagrammatically, \eqref{eq:dr_sf_amp} originate from the same diagram as the last one in Fig.~\ref{fig:dr_amp}, with the gluons crossing the shock replaced by scalars and fermions, respectively. The fermion amplitude involves Dirac $\gamma$-matrices, which act on the fermion polarizations. After we square the amplitude, the polarization sum will yield a sum of four $\gamma$ matrices.
Note that the scalar amplitude coincides with the coefficient of $\delta_{ij}$ in \eqref{gluon amplitude def}.

Using simple shifts such $k_2\mapsto k_2-k_1$ where appropriate, the following limits of the gluon amplitudes can be verified:
\begin{equation}\begin{aligned} \label{amp factorization}
 \lim_{\tau\to 0} \Aa_{[100'2]}^{(gg,0)ij}(\tau)
 &=\Aa_{[100']}^{(g,0)i}\Aa_{[10'2]}^{(g,0)j}\,,
 \\
 \lim_{\tau\to \infty} \Aa_{[100'2]}^{(gg,0)ij}(\tau)
 &=\Aa_{[00'2]}^{(g,0)j}\Aa_{[102]}^{(g,0)i}\,.
\end{aligned}\end{equation}
The right-hand sides correspond to the amplitudes for the sequential emission of two gluons: the gluon with the larger formation time (i.e., higher energy) splits off first from $\Uu_{12}$, in agreement with the factorization in Fig.\,\ref{fig:facto1}. This factorization property will be crucial to obtain a finite and conformal result for the NLO kernel in the calculation below. In the analogous limits, the scalar and fermion amplitudes vanish.

\subsection{Splitting amplitudes in impact parameter space}\label{ssec:splitting amps}

We are now ready to discuss the Fourier transform of the splitting amplitudes just described.  

\paragraph{Tree-level single-real.} At leading order, \eqref{LO amplitude mom} is given by the straightforward transform of a power law (recorded in \eqref{eq:powerLawFT}):
\begin{align}\label{eq:LOamp}
    \Aa_{[102]}^{(g,0)i}(z_0)&=
    \cd\left[\frac{z_{10}^i}{(z_{10}^2)^{1{-}\varepsilon}}- \frac{z_{20}^i}{(z_{20}^2)^{1{-}\varepsilon}}\right]\,.
\end{align}
\paragraph{Tree-level double-real.} At the next order, the Fourier transforms defined in the preceding subsection can be calculated using the basis of Fourier integrals evaluated in \cite{Brunello:2023fef}.
It is shown that these organize into families of integrals that each contain a pair of master integrals. The latter evaluate to the following pair of pure transcendental functions:\footnote{This basis are related to those quoted in \cite{Brunello:2023fef} through simple ${}_{2}F_1$ identities.}
\begin{equation}\label{eq:LOFT}
 F_1[x]=\frac{\Gamma (1{-}2 \varepsilon)}{\Gamma^2(1{-}\varepsilon)}(1+x)^{2\varepsilon}\,,\qquad
 F_2[x]=\frac{\varepsilon x \Gamma (1{-}2 \varepsilon)}{(1{-}\varepsilon) \Gamma^2(1{-}\varepsilon)} \,{}_2F_1(1{-}2
   \varepsilon,1{-}\varepsilon;2{-}\varepsilon;-x)\,.
\end{equation}
These are pure transcendental functions satisfying the following canonical-form differential equation \cite{Henn:2013pwa}, which will be useful below:
\begin{align} \label{F12 diff eq}
 \frac{\d}{\d x} \begin{bmatrix}
F_1 \\ F_2
 \end{bmatrix} = \varepsilon\begin{bmatrix}\frac{2}{1+x} & 0 \\ \frac{1}{1+x} &\frac{1}{x}\end{bmatrix}\cdot\begin{bmatrix}
F_1 \\ F_2
 \end{bmatrix}\,,
\end{align}
Using the simple boundary condition at $x=0$ one then finds, for example, the expansions
\begin{subequations}
    \begin{align}
   F_1[x]&= 1+2\varepsilon\log(1{+}x)+\varepsilon^2(2\log^2(1{+}x)+\tfrac{\pi^2}{6})+\mathcal{O}(\varepsilon^3)\,,\\
   F_2[x]&= \varepsilon\log(1{+}x)+\varepsilon^2(\log^2(1{+}x)-{\rm Li}_2(-x))+\mathcal{O}(\varepsilon^3)\,.
\end{align}
\end{subequations}
Below we will only need up to the $\mathcal{O}(\varepsilon)$ terms.
In terms of these transcendental functions, we then have
\begin{subequations}\label{eq:forDer}
    \begin{align}
\frac{1}{\cd^2}\int\limits_{k_1,k_2} \,
\frac{\e^{ik_1{\cdot}z_{01}+ik_2{\cdot}z_{0'1}}}{(k_1^2+\tau k_2^2)k_2^2}&=
\frac{\tau^{\varepsilon} (z_{10}^2)^{2\varepsilon}}{32\varepsilon^2}
\tilde{F}\left[\frac{z^2_{10'}}{\tau z^2_{10}}\right], \\
\frac{1}{\cd^2}\int\limits_{k_1,k_2}\,
\frac{\e^{ik_1{\cdot}z_{01}+ik_2{\cdot}z_{0'1}}}{(k_1^2+\tau k_2^2)(k_1+k_2)^2}&=
\frac{\tau^{\varepsilon} (z_{00'}^2)^{2\varepsilon}}{32\varepsilon^2(1{+}\tau)^{1+2\varepsilon}}
\tilde{F}\left[\frac{(\tau z_{10}+z_{10'})^2}{\tau z^2_{00'}}\right]\,,
\end{align}
\end{subequations}
where we abbreviated $\tilde{F}[x]=F_1[x]-2F_2[x]$.
Note that these integrals are not independent: their equivalence can be seen by applying the substitution $(k_1,k_2)\mapsto (k_1-\tau k_2,k_2+k_1)$ to the first \cite{Brunello:2023fef}.

By writing the momenta in the numerator as derivatives, i.e., $k_0^i\rightarrow-i\frac{\partial}{\partial z_0^i}$ and using \eqref{F12 diff eq} to simplify derivatives as they appear, 
the Fourier integrals \eqref{gluon amplitude def}-\eqref{fermion amplitude def} can be expressed in terms of the $F_1$ and $F_2$ transcendental functions in a relatively automated manner.
After the dust settles, we obtain the following result for the amplitude of two real gluons \eqref{gluon amplitude def} in $\Dperp=2-2\varepsilon$ transverse dimensions:
\begin{align} \label{amp gluons}
\frac{1}{\cd^2}\Aa_{[100'2]}^{(gg,0)ij}(\tau)
&=(1+\mathcal{P})\Bigg\{
-\frac{z_{10}^i z_{20'}^j}{2
(z_{10}^2 z_{20'}^2)^{1{-}\varepsilon}}
+\frac{\tau^\varepsilon z_{10}^i z_{10'}^j}{(\tau z_{10}^2+ z_{10'}^2)(z^2_{10})^{1-2\varepsilon}}G\left[\frac{z_{10'}^2}{\tau z_{10}^2}\right]
\nonumber\\
&\quad+
\frac{\tau^\varepsilon}{(1+\tau)^{2\varepsilon}}
\frac{z_{00'}^i z_{10'}^j+\tau z_{10}^i z_{00'}^j+ \frac{\tau}{2(1+\tau)}(z_{10}^2{-}z_{10'}^2)\delta^{ij}}{(\tau z^2_{10}+z^2_{10'})(z_{00'}^2)^{1-2\varepsilon}}
G\left[\frac{(\tau z_{10}+z_{10'})^2}{\tau z_{00'}^2}\right]
\nonumber\\ &\quad+ \frac{(1-\tau)\tau^{1+\varepsilon}\delta^{ij}}{2(1+\tau)^{1+2\varepsilon}} 
\frac{(z^2_{00'})^{2\varepsilon}}{(\tau z_{10}+z_{10'})^2}
F_2\left[\frac{(\tau z_{10}+z_{10'})^2}{\tau z_{00'}^2}\right]\Bigg\}
\,,
\end{align}
where ${\cal P}$ is the permutation $(z_1,z_0,i,\tau){\leftrightarrow}(z_2,z_{0'},j,\tau^{-1})$ and $G$ is the combination
\begin{equation}
    G[x]=F_1[x]-\frac{1+x}{x}F_2[x]\,.
\end{equation}
For the scalar and fermion amplitudes \eqref{scalar amplitude def} and \eqref{fermion amplitude def}, using the same method, we find, respectively
\def\zslash{z\!\!\!/}
\begin{subequations}
\begin{align}
\frac{1}{\cd^2}\Aa_{[100'2]}^{(ss,0)}(\tau)
&= 
 \frac{\tau^{1+\varepsilon}(z^2_{00'})^{2\varepsilon-1}}{2(1+\tau)^{1+2\varepsilon}}
\Bigg(
       \frac{(z_{10}^2{-}z_{10'}^2)}{\tau z_{10}^2+z_{10'}^2}
G\left[\frac{(\tau z_{10}+z_{10'})^2}{\tau z_{00'}^2}\right]
\nonumber\\&\hspace{30mm}
+ \frac{(1-\tau)z_{00'}^2}{(\tau z_{10}+z_{10'})^2}F_2\left[\frac{(\tau z_{10}+z_{10'})^2}{\tau z_{00'}^2}\right]
-(z_1{\mapsto}z_2)\Bigg),
\\
 \frac{1}{\cd^2}\Aa_{[100'2]}^{(ff,0)}(\tau)
        &=
       \frac{\tau^{\frac12+\varepsilon}(z^2_{00'})^{2\varepsilon-1}}{(1+\tau)^{1+2\varepsilon}}
\Bigg(
\left(1-\zslash_{10}\zslash_{10'} \frac{1+\tau}{\tau z_{10}^2+z_{10'}^2}\right)
G\left[\frac{(\tau z_{10}+z_{10'})^2}{\tau z_{00'}^2}\right]
\nonumber\\ &\hspace{30mm}
-\frac{2\tau z_{00'}^2}{(\tau z_{10}+z_{10'})^2} F_2\left[\frac{(\tau z_{10}+z_{10'})^2}{\tau z_{00'}^2}\right]-(z_1{\mapsto}z_2)\Bigg).
\label{amp ff position}
\end{align}
\end{subequations}

A simple consistency check on the gluon amplitude is its factorization property \eqref{amp factorization}, which can be verified in any $\Dperp$ using $\lim_{x\to\infty}G[x]=x^\varepsilon+\mathcal{O}(x^{-1})$.
Another check is the limit as $\D\to 4$, which can be easily obtained by setting $G\mapsto 1$ and $F_2\mapsto 0$.
This reproduces, for example, the gluon amplitude given in \cite[Eq. (43)]{Balitsky:2007feb} after the color traces there are reduced using \cite[Eq. (55)]{Balitsky:2007feb}:
\begin{equation}
\begin{split}
\hspace{-0.4cm}
\pi^2\Aa_{[100'2]}^{(gg,0)ij}(\tau)
 &\!\stackrel{\D=4}{=}\!(1{+}{\cal P})\Bigg\{
 \frac{z_{00'}^i z_{10'}^j{+}\tau z_{10}^i z_{00'}^j{+}\frac{\tau (z_{10}^2{-}z_{10'}^2)}{2(1{+}\tau)}\delta^{ij}}{(\tau z^2_{10}+z^2_{10'})z^2_{00'}}
{+}\frac{z_{10}^i z_{10'}^j}{z^2_{10}(\tau z^2_{10}{+}z^2_{10'})}\Bigg\}\!
-\frac{z_{10}^i z_{20'}^j}{z^2_{10}z^2_{20'}}\,.
\end{split}
\end{equation}
Finally, note that the two-scalar amplitude is simply the coefficient of $\delta^{ij}$ in the two-gluon amplitude.

\subsection{Double-real evolution kernel}

Squaring the preceding two-parton amplitudes and summing over intermediate states yields the double-real integrand:
\begin{equation} \label{I from A squared}
  {\cal I}^{(2)}_{[100'2]}(\tau) =
\Aa_{[100'2]}^{(gg,0)ij}(\tau)
\Aa_{[100'2]}^{(gg,0)ij}(\tau)
+
n^s_{\rm adj}
\Aa_{[100'2]}^{(ss,0)}(\tau)^2
+
n^f_{\rm adj}
{\rm tr}_f[\Aa_{[100'2]}^{(ff,0)}(\tau)
\Aa_{[100'2]}^{(ff,0)}(\tau)^\dagger]\,.
\end{equation}
Note that while the energies $k_i^+$ are conserved across the shock, the transverse momentum is not. But in transverse coordinate space, the amplitudes simply multiply.

As explained in Sec.\,\ref{ssec:setup}, we perform all intermediate calculations in the large $N$ limit of a theory with $n^s_{\rm adj}$ real adjoint scalars and $n^f_{\rm adj}$ adjoint Majorana fermions. For Majorana fermions, the $\dagger$ operation in \eqref{I from A squared} reverses all $\gamma$-matrices in the amplitude \eqref{amp ff position}.
The sum over Majorana fermion states is then computed using standard trace identities:
\begin{equation}
\hspace{-0.25cm}
    {\rm tr}_f[\zslash_1\zslash_2]=2z_1{\cdot}z_2\,{\rm tr}_f[1]\,, \, \,
{\rm tr}_f[\zslash_1\zslash_2\zslash_3\zslash_4]=2(
z_1{\cdot}z_2\,z_3{\cdot}z_4+z_1{\cdot}z_4\,z_2{\cdot}z_3-z_1{\cdot}z_3\,z_2{\cdot}z_4){\rm tr}_f[1]\,,
\end{equation}
where we take ${\rm tr}_f[1]=2$ following the standard
DRED
convention (see, e.g., \cite{Stockinger:2005gx}) to not continue the dimension of fermion representations. (For a Dirac fermion we would have ${\rm tr}_F[1]=4$; as far as our calculation is concerned, an overall factor of two is the only difference between a Dirac and Majorana fermion.)
For the gluons in \eqref{I from A squared} we sum over indices in $\Dperp=2{-}2\varepsilon$ dimensions.  Note that these conventions imply that $\gamma_i\gamma_i= \D_\perp$.

This yields the integrand \eqref{I from A squared}, which is still a function of the (relative) energies of the partons crossing the shock, which must be integrated over.
Before integrating, we must subtract iterations of leading-order emissions, as explained in the previous subsection (see \eqref{eq:L2DR}):
\begin{equation} \label{double-real subtraction}
    H^{(2)}_{[100'2]} = \frac{{\cal N}_\varepsilon^2}{\cd^2}
\int_0^\infty \frac{\d\tau}{\tau}
\left[\begin{array}{ll}
{\cal I}^{(2)}_{[100'2]}(\tau)
&-{\cal I}^{(1)}_{[10'2]}{\cal I}^{(1)}_{[100']}\theta(\nu_{[10'2]}{>}\nu_{[100']})
\\&-{\cal I}^{(1)}_{[102]}{\cal I}^{(1)}_{[00'2]}\theta(\nu_{[102]}{>}\nu_{[00'2]})
\end{array}\right]\,.
\end{equation}
The integral converges due to the factorization limits \eqref{amp factorization}.  We are thus allowed to expand the integrand in $\varepsilon$, which produces a series in logarithms (and polylogarithms should one want the $\mathcal{O}(\varepsilon^2)$ term); the $\tau$ integration then increases the transcendental weight of the functions by (at most) one.

The $\tau$ integrals can be performed using standard integration methods, applied in a similar context in \cite{Caron-Huot:2016tzz}. In particular, at the order of interest, we can use partial fractions to reduce the integrand to a sum of terms of the following forms:
\begin{equation}\label{eq:tauIntegrals}
    \int \d\log(\tau+a)\log(\tau+b)\quad\mbox{or}
\quad \int \frac{\d\tau}{(\tau+a)^2}\log(\tau+b)\,.
\end{equation}
For the latter, we apply integration-by-parts to obtain elementary integrals. Note that the integrals in \eqref{eq:tauIntegrals} can be written as iterated integrals over rational functions and are, in fact, \emph{linearly reducible} \cite{Brown:2008um},\footnote{Meaning one can order the integrations so that at each step the remaining denominator polynomials factor with roots linear in the next integration variable, precluding any increase in algebraic complexity, such as the appearance of elliptic curves or more complicated geometries.} ensuring that \eqref{eq:tauIntegrals} evaluates to (multiple)polylogarithms. This allows us to use general integration packages such as \texttt{HyperInt} \cite{Panzer:2014caa} for independent cross-checks.

The main technical difficulty of this calculation is that the partial-fraction step can produce complicated rational or algebraic denominators.
It proves convenient to organize the final result in terms of these denominators.

\paragraph{Basis of transcendental functions and denominators.}

The simplest example of a nontrivial denominator generated by the calculation is the following weight-one function:
\begin{equation} \label{short integral}
\int_0^\infty \frac{\d\tau}{(z_{10}^2\tau+z_{10'}^2)(z_{20}^2\tau+z_{20'}^2)}
= \frac{1}{z_{10}^2z_{20'}^2-z_{10'}^2z_{20}^2}\log \frac{z_{10}^2z_{20'}^2}{z_{10'}^2z_{20}^2}\,.
\end{equation}
While this result can be derived using an elementary partial-fraction step, this procedure produces a nontrivial denominator which can vanish for some physical configurations.
However, when this happens, the logarithm automatically vanishes and the integral is, in fact, regular throughout the physical region where all $z_{ij}^2$ are positive. This was, indeed, manifest from the integral representation.

A weight-two example of this phenomenon is the following function:
\begin{equation}\begin{aligned}
&\int_0^\infty \frac{\d\tau}{(z_{10}^2\tau+z_{10'}^2)(z_{20}^2\tau+z_{20'}^2)}
\log\frac{\tau z_{10'}^2z_{20}^2}{(z_{10}^2\tau+z_{10'}^2)(z_{20}^2\tau+z_{20'}^2)}\\
&= \frac{2}{z_{10}^2z_{20'}^2-z_{10'}^2z_{20}^2}\Li_2\left(1-\frac{z_{10}^2z_{20'}^2}{z_{10'}^2z_{20}^2}\right)\,.
\end{aligned}\end{equation}
These two examples exhaust all the ways that this particular denominator will appear in the calculation.

Some other integrals produce denominators with a square root:
\begin{align}
\int_0^\infty \frac{z_{10'}^2\,\d\tau}{(\tau z_{10}+z_{10'})^2}\log \frac{(1+\tau)(\tau z_{10}^2+z_{10'}^2)}{\tau z_{00'}^2}=
B_2\left(\frac{z_{10}^2}{z_{10'}^2},\frac{z_{00'}^2}{z_{10'}^2}\right)\,,
\end{align}
where $B_2$ is the so-called Bloch--Wigner function
\begin{equation} \label{def BW}
\hspace{-0.4cm}
    B_2(u,v)\equiv \frac{2[\Li_2(z){-}\Li_2(\bar{z})]{+}(\log(1{-}z){-}\log(1{-}\bar{z}))\log z\bar{z}}{z-\bar{z}} \quad\mbox{with}\,\,\,
    \begin{sqcases}
        u=z\bar{z}\,,\\ 
        v=(1{-}z)(1{-}\bar{z})\,.
    \end{sqcases}
\end{equation}
For $(u,v)=\Big(\frac{z_{10}^2}{z_{10'}^2},\frac{z_{00'}^2}{z_{10'}^2}\Big)$, the denominator can be written more explicitly as
\begin{equation}
    z-\bar{z} = \frac{1}{z_{10'}^2}\sqrt{(z_{10}^2+z_{00'}^2-z_{10'}^2)^2-4 z_{10}^2z_{00'}^2}\,.
\end{equation}
This vanishes when the points $z_1$, $z_0$ and $z_{0'}$ are collinear, however, then $z=\bar{z}$ and the numerator of \eqref{def BW} also vanishes. The function $B_2(u,v)$ is, in fact, analytic around every real configuration where points do not coincide.

Finally, we encounter an integral which produces a rather lengthy denominator:
\begin{align} \label{long integral}
    &\int_0^\infty\frac{\d\tau}{(\tau z_{10}+z_{10'})^2}\frac{z_{20'}^2-\tau z_{20}^2}{z_{20'}^2+\tau z_{20}^2} 
    \log \frac{(1+\tau)(\tau z_{10}^2+z_{10'}^2)}{\tau z_{00'}^2}
    \nonumber\\ &= \frac{z_{20}^2z_{20'}^2}{(z_{10}z_{20'}^2-z_{10'}z_{20}^2)^2}\Bigg\{
    F^{\rm long}_1 \equiv
    2\Li_2\left(1{-}\frac{z_{10}^2z_{20'}^2}{z_{10'}^2z_{20}^2}\right)
    -2\Li_2\left(1{-}\frac{z_{20}^2}{z_{20'}^2}\right)+\log \frac{z_{10}^2z_{20'}^4}{z_{10'}^2z_{20}^4}\log \frac{z_{00'}^2}{z_{10'}^2}
 \nonumber\\ &\hspace{55mm} + \left(\frac{z_{10}^2z_{20'}^2}{z_{10'}^2z_{20}^2}-\frac{z_{20}^2}{z_{20'}^2}\right)B_2\left(\frac{z_{10}^2}{z_{10'}^2},\frac{z_{00'}^2}{z_{10'}^2}\right)\Bigg\}\,.
\end{align}
The denominator can be written more explicitly as
\begin{equation}
    (z_{10}z_{20'}^2-z_{10'}z_{20}^2)^2=(z_{10'}^2z_{20}^2-z_{10}^2z_{20'}^2)(z_{20}^2-z_{20'}^2)+z_{20}^2z_{20'}^2z_{00'}^2\,.
\end{equation}
The function in \eqref{long integral} enjoys some nontrivial properties.
First, we note that it is possible for the denominator to vanish for physical kinematics; however, we find that the particular combination of transcendental functions in $F^{\rm long}_1$ then vanishes. 
Second, the function is odd under $z_0\leftrightarrow z_{0'}$,
as can be verified using dilogarithm identities.
In fact, both of these properties are manifest from the initial integral representation.  Finally, we note that $B_2$ and $F^{\rm long}_1$ constitute a basis of weight-two functions for all integrals with $(\tau z_{10}+z_{10'})^2$ and $(z_{20'}^2+\tau z_{20}^2)$ in the denominator. In effect, we chose (without loss of generality) to define $F^{\rm long}_1$ with a numerator that makes it antisymmetric.

All other integrals are either permutations of the above, namely
$B_2\left(\frac{z_{20'}^2}{z_{20}^2},\frac{z_{00'}^2}{z_{20}^2}\right)$ and $F^{\rm long}_2\equiv {\cal P}F^{\rm long}_1$, or produce denominators that are products of simple distances.

\paragraph{Results for double-real evolution kernels.} Examination of the evolution kernels from scalar, fermion, and gluons crossing the shock reveals that they share many terms.
The shared terms nicely follow the organization into the contributions from ${\cal N}=0,1,4$ supermultiplets, which can be used to span the three independent linear combinations of scalar, fermion and gluon amplitudes.
Factoring out a normalization similar to \eqref{eq:H1_102 pre},
we thus expand the double-real kernel as 
\begin{equation}\label{double result susy}
H^{(2)}_{[100'2]} = \left(\frac{z_{12}^2}{z_{10}^2z_{00'}^2z_{0'2}^2}\right)^{1-\varepsilon}
\tilde{H}^{(2)}_{[100'2]}\,,
\end{equation}
where
\begin{equation}\label{eq:DRHtilde}
    \tilde{H}^{(2)}_{[100'2]}\equiv\left[\begin{array}{l}
\phantom{+}(n_{\rm adj}^s{-}2n_{\rm adj}^f{+}2{-}2\varepsilon(1{-}\delta))
\tilde{H}^{(2){\cal N}=0}_{[100'2]}\\
+(n_{\rm adj}^f-4) \tilde{H}^{(2){\cal N}=1}_{[100'2]}
\\
+
\tilde{H}^{(2){\cal N}=4}_{[100'2]}
\end{array}\right]\,.
\end{equation}
In the DRED scheme with $\delta=1$ (see \eqref{eq:DREGvsDRED}), the first two lines happen to vanish when a full ${\cal N}=4$ supermultiplet runs in the loop ($n_{\rm adj}^s=6
$ and $n_{\rm adj}^f=4$);
the first line generally vanishes for any
${\cal N}=1$ matter content ($\Delta n_{\rm adj}^s=2\Delta n_{\rm adj}^f$, where $\Delta X$ quantifies the departure of $X$ from $\mathcal{N}=4$, i.e., $\Delta X=X-X|_{\mathcal{N}=4}$). 
We stress that the organization \eqref{double result susy} holds regardless of whether the theory of interest is supersymmetric; it is just a natural way to group the terms that arise from the calculation.

For the scalar contribution we then find:
\begin{align} \label{double-real scalar}
     \frac{z_{12}^2z_{00'}^2}{z_{10}^2z_{20'}^2}   \tilde{H}^{(2){\cal N}=0}_{[100'2]} &=
\left( 
\frac{v{+}1}{v{-}1}\log v
{-}2{-}8\varepsilon\right)
\left(1{-}\varepsilon\log uv
+2\varepsilon L_\mu
\right)
{+}\frac{8\varepsilon}{1{-}v}\Li_2(1{-}v)
\nonumber\\&\quad
{+}\varepsilon (1{+}{\cal P})\Bigg\{
z_{20}^2\frac{z_{20'}^2(z_{10}^2{+}z_{10'}^2{-}z_{00'}^2){-}2z_{10'}^2z_{20}^2}{(z_{10}z_{20'}^2{-}z_{10'}z_{20}^2)^2}
F^{\rm long}_1
\nonumber\\&\quad
{+} \frac{z_{20'}^2(z_{00'}^2(z_{10}^2{+}z_{10'}^2){-}(z_{10}^2{-}z_{10'}^2)^2){-}2z_{02}^2z_{10'}^2z_{00'}^2}
{z_{10'}^2z_{20'}^2z_{00'}^2}B_2\left(\frac{z_{10}^2}{z_{10'}^2},\frac{z_{00'}^2}{z_{10'}^2}\right)
\nonumber\\&\quad
{+}\frac{z_{00'}^2{+}z_{10'}^2{-}z_{10}^2}{z_{00'}^2}\left[2\Li_2\left(1{-}\frac{z_{10}^2}{z_{10'}^2}\right){+}\log\frac{z_{10}^2}{z_{10'}^2}\log\frac{z_{00'}^2}{z_{10'}^2}\right]
\nonumber\\&\quad
{+}\log\frac{z_{10}^2}{z_{10'}^2}\left[\log\frac{z_{10}^2z_{00'}^4}{z_{10'}^2z_{20}^4}{-}2\right]
\Bigg\}{+}{\cal O}(\varepsilon^2)\,,
\end{align}
where here and below we used the cross-ratios:
\begin{equation} \label{cross-ratios}
    u=\frac{z_{12}^2z_{00'}^2}{z_{10'}^2z_{20}^2},\qquad
    v=\frac{z_{10}^2z_{20'}^2}{z_{10'}^2z_{20}^2}\,.
\end{equation}
The amplitude for a ${\cal N}=1$ matter multiplet is comparatively simpler:
\begin{align} \label{double-real N=1}
\frac{z_{12}^2z_{00'}^2}{z_{10}^2z_{20'}^2}   \tilde{H}^{(2){\cal N}=1}_{[100'2]} &=
\frac{2u}{v{-}1}
\Bigg[
\log v
\left(1{-}\varepsilon\log uv+2\varepsilon L_\mu\right){-}4\varepsilon \Li_2(1{-}v)\Bigg]
\nonumber\\&\quad
{+} \varepsilon (1{+}{\cal P})\Bigg\{
z_{00'}^2{\cal N}_1
\left[\frac{F^{\rm long}_1}{(z_{10} z_{20'}^2{-}z_{10'}z_{20}^2)^2} {+} \frac{1}{z_{10'}^2z_{20}^2z_{20'}^2}B_2\Big(\frac{z_{10}^2}{z_{10'}^2},\frac{z_{00'}^2}{z_{10'}^2}\Big)\right]
\nonumber\\&\quad
{+}\frac{2z_{00'}^2(z_{20}^2{-}z_{12}^2{-}z_{10}^2)}{z_{10'}^2z_{20}^2}B_2\left(\frac{z_{10}^2}{z_{10'}^2},\frac{z_{00'}^2}{z_{10'}^2}\right) 
\Bigg\}{+}{\cal O}(\varepsilon^2)\,,
\end{align}
with ${\cal N}_1=z_{12}^2(z_{20'}^2-z_{20}^2)+z_{10}^2z_{20'}^2-z_{10'}^2z_{20}^2$.

\def\cchi{{\mathcal{X}}}
Somewhat to our surprise,
the ${\cal N}=4$ kernel (which contains the gluons) is less compact. Setting $\cchi=(z_{10}^2-z_{10'}^2-z_{20}^2+z_{20'}^2)/z_{12}^2$, we find:
\begin{align}\label{double-real N=4}
\tilde{H}^{(2){\cal N}=4}_{[100'2]} &=
    2\left(2\log u{+}\frac{u{+}v{-}1}{v{-}1}\log v\right)\left(1{-}\varepsilon\log uv+2\varepsilon L_\mu\right)
    \\&
    {+}\varepsilon\Bigg[
    \left( 2(2{-}\cchi)(1{+}v{-}u){-}8\frac{u{+}v{-}1}{v{-}1}\right){\rm Li}_2(1{-}v){+}2\zeta_2(\cchi{-}2)(1{+}u{-}v)
    \nonumber\\&
    \qquad {+}(1{+}(v{-}u)(2{-}\cchi))\log^2v{+}6\log u\log v{+}4\log u\log\frac{z_{00'}^4}{z_{10}^2z_{20'}^2}\Bigg]
    \nonumber\\&
    {+}\varepsilon(1{+}{\cal P})\Bigg\{
\frac{z_{20}^2}{z_{20'}^2}(z_{20}^2{-}z_{00'}^2{-}3z_{20'}^2)\left[ \frac{z_{10}^2z_{20'}^2 F^{\rm long}_1}{(z_{10} z_{20'}^2{-}z_{10'}z_{20}^2)^2}
{+}\frac{z_{10}^2}{z_{10'}^2z_{20}^2}B_2\left(\frac{z_{10}^2}{z_{10'}^2},\frac{z_{00'}^2}{z_{10'}^2}\right)\right]
\nonumber\\ &\quad
{+}\left(
\frac{z_{20'}^2}{z_{20}^2}(z_{00'}^2{-}z_{20'}^2{+}3z_{20}^2){+} z_{00'}^2\cchi\right)\left[\frac{z_{10}^2z_{20'}^2 F^{\rm long}_1}{(z_{10} z_{20'}^2{-}z_{10'}z_{20}^2)^2}
{-} \frac{z_{10}^2}{z_{10'}^2z_{20}^2}B_2\left(\frac{z_{10}^2}{z_{10'}^2},\frac{z_{00'}^2}{z_{10'}^2}\right)\right]
\nonumber\\ &\quad
{+} B_2\left(\frac{z_{10}^2}{z_{10'}^2},\frac{z_{00'}^2}{z_{10'}^2}\right)\left[\frac{z_{10}^2{+}z_{00'}^2{-}z_{10'}^2}{z_{10'}^2}\cchi{+}
\frac{z_{10}^2{+}z_{10'}^2{-}z_{00'}^2}{z_{10'}^2}
\left(\frac{z_{20'}^2{-}z_{00'}^2}{z_{20}^2}{+}\frac{z_{10}^2{-}z_{00'}^2}{z_{10'}^2}\right)\right]
\nonumber\\ &\quad
{+} \frac{z_{10}^2{-}z_{00'}^2{-}z_{10'}^2}{z_{10'}^2}\left[
2\log u \log\frac{z_{00'}^2}{z_{10}^2}{+}
\left(\log\frac{z_{10}^2}{z_{10'}^2} \log\frac{z_{00'}^2z_{10'}^2}{z_{20}^2z_{20'}^2}{+}(1\leftrightarrow2)\right){-}4\zeta_2\right]
\nonumber\\ &\quad
{+} \frac{z_{20}^2{-}z_{10}^2}{z_{12}^2}\left[\log\frac{z_{10}^2}{z_{20}^2}\log \frac{u^2 z_{20'}^2}{z_{20}^2}{+}\log\frac{z_{10}^2}{z_{00'}^2}\log v\right]{+}\log\frac{z_{10}^2}{z_{10'}^2}\log\frac{z_{20'}^2z_{00'}^4}{z_{20}^6}\Bigg\}{+}\mathcal{O}(\varepsilon^2)\,.\nonumber
\end{align}

The ${\cal N}=4$ theory is only conformal in $\D=4$ dimensions and we see that the ${\cal O}(\varepsilon)$ contributions indeed quite badly break conformal invariance (i.e., they cannot be expressed in terms of the $u$, $v$ cross-ratios in \eqref{cross-ratios}). 
The above expressions combine the $\D=4$ limit and ${\cal O}(\varepsilon)$ correction; in $\D=4$ they reduce to the known and conformal results in \cite{Balitsky:2009xg} (in the adjoint representation), namely:
\begin{equation}\label{double real 4d}\begin{aligned}
   \lim_{\varepsilon\to 0} \tilde{H}^{(2){\cal N}=0}_{[100'2]} &=
\frac{{v}}{u}\left(\frac{v+1}{v-1}\log v-2\right)\,,
\\
\lim_{\varepsilon\to 0} \tilde{H}^{(2){\cal N}=1}_{[100'2]} &=
\frac{2{v}\log v}{v-1}\,,
\\
\lim_{\varepsilon\to 0}
\tilde{H}^{(2){\cal N}=4}_{[100'2]} &=4\log u+2\frac{u+v-1}{v-1}\log v\,. 
\end{aligned}\end{equation}

\subsection{Comments on rapidity subtraction schemes and collinear divergences}\label{sssec:rapiditySubDR}

Here we briefly comment on two features of the obtained evolution kernel: its dependence on the choice of ordering variable (rapidity subtraction scheme) and its collinear limits.

\paragraph{Subtraction scheme.} How would the kernels change if we were to subtract the rapidity divergences using $k^+$ as our ordering variable? This would change the arguments in the step functions \eqref{double-real subtraction} to $\nu_{[ijk]}\mapsto \nu^{\,\rm +\,scheme}_{[ijk]}=k_j^+$ leading to
\begin{equation}
    \begin{split}
            H^{(2)+}_{[100'2]} &= \frac{{\cal N}_\varepsilon^2}{\cd^2}
\int_0^\infty \frac{\d\tau}{\tau}
\left[\begin{array}{ll}
{\cal I}^{(2)}_{[100'2]}(\tau)
&- {\cal I}^{(1)}_{[10'2]}{\cal I}^{(1)}_{[100']}\theta(1{>}\tau)
\\&- {\cal I}^{(1)}_{[102]}{\cal I}^{(1)}_{[00'2]}\theta(\tau{>}1)
\end{array}\right]\qquad\mbox{($k^+$ cut-off)}\\
&= \frac{{\cal N}_\varepsilon^2}{\cd^2}
\int_0^1 \frac{\d u}{[u(1-u)]_+} {\cal I}^{(2)}_{[100'2]}\left(\frac{u}{1-u}\right)
\hspace{2.6cm} \Big(\tau=\frac{u}{1-u}\Big)\,.
\label{plus example}
    \end{split}
\end{equation}
In the second line, the $+$-prescription has the usual meaning of subtracting $1/u$ times the $u\to 0$ limit of the integrand and $1/(1-u)$ times its $u\to 1$ limit.
The two lines are equal, thanks to the identity
\begin{equation}
    \int_0^1 \d u \left[ \frac{\theta(1/2>u)}{u(1-u)}- \frac{1}{u}\right]=0\,.
\end{equation}
Thus, using $k^+$ as our ordering variable, our procedure precisely reproduces the $+$-prescription used in the calculation of the two-loop evolution in \cite{Balitsky:2007feb,Balitsky:2009xg}.
On the other hand, because of the extra position dependence of $\nu_{[ijk]}$ in  \eqref{eq:orderingVarProp}, our actual $\tau$-integral in \eqref{double-real subtraction} differs from the +-prescription by an additional logarithm:
\begin{equation} \label{double-real scheme change}
  H^{(2)\Delta\rm scheme}_{[100'2]}=
    H^{(2)}_{[100'2]}-H^{(2)+}_{[100'2]} = (1+{\cal P})\left[\frac{{\cal N}_\varepsilon^2}{\cd^2}{\cal I}^{(1)}_{[10'2]}{\cal I}^{(1)}_{[100']}\log\sqrt{\frac{z_{12}^2z_{10}^2z_{00'}^2}{z_{10'}^4z_{0'2}^2}}\,\right]\,,
\end{equation}
which using \eqref{eq:I1_102} evaluates more explicitly as
\begin{align}\label{double-real scheme change 2}
 H^{(2)\Delta\rm scheme}_{[100'2]}  &=
\left(\frac{z_{12}^2}{z_{10}^2z_{00'}^2z_{0'2}^2}\right)^{1-\varepsilon}\,(1+{\cal P})
\Bigg\{2\log\frac{z_{12}^2z_{10}^2z_{00'}^2}{z_{10'}^4z_{0'2}^2}
\\&\hspace{-10mm}\times\Bigg[1
{+}\varepsilon\left(2\log\frac{\mubar^2z_{12}^2}{4\e^{-2\gamma_\text{E}}}
{+}\log\frac{z_{10'}^2}{z_{12}^2}
{-}\frac{z_{10'}^2{-}z_{20'}^2}{z_{12}^2}
\log\frac{z_{10'}^2}{z_{20'}^2}
{-}\frac{z_{10}^2{-}z_{00'}^2}{z_{10'}^2}
\log\frac{z_{10}^2}{z_{00'}^2}\right)\Bigg]{+}{\cal O}(\varepsilon)^2\Bigg\}\,.\nonumber
\end{align}
This contribution was automatically included in our calculation that started from \eqref{double-real subtraction}. Note that only gluon contributions are affected, because matter loops do not exhibit rapidity subdivergences since they do not contribute to $H^{(1)}_{[102]}$.

The ${\cal O}(\varepsilon^0)$ limit of \eqref{double-real scheme change 2} agrees precisely with the double-real part of the conversion to the ``composite conformal dipole'' in \cite[Sec.\, IV]{Balitsky:2009xg} (more precisely, with the $+2\log \frac{z_{12}^2z_{34}^2}{z_{14}^2z_{23}^2}$ term in (51) there). 
We conclude that our prescription for renormalizing rapidity divergences using $\nu_{[ijk]}$ in \eqref{eq:orderingVarProp} coincides precisely with the ``conformal dipole'' of \cite{Balitsky:2009xg}.
We stress that this prescription is applicable whether or not the theory is conformal, its essential point is to ensure that all sources of conformal symmetry-breaking are explicitly proportional to the $\beta$-function.

\paragraph{Collinear divergences.}

When taking the limit $\varepsilon\to0$ of the evolution equation, the transverse coordinate integration in \eqref{eq:evolEqU2} may produce divergences leading to $\varepsilon/\varepsilon$ contributions. To fully specify the $\varepsilon\to 0$ limit of the evolution, we must also record complete $\D$-dimensional expressions in singular limit(s). Thankfully, the amplitudes there dramatically simplify.

In the $z_{0'}\to z_0$ limit, only the $G[\frac{(\cdots)}{\tau z_{00'}^2}]$ terms in \eqref{double-real scalar}-\eqref{double-real N=4} are singular, and we can simplify them using the large-$x$ approximation
$G[x]\approx x^{\varepsilon}$ already mentioned below \eqref{amp ff position}.
The $\tau$-integral becomes elementary and we find: 
\begin{align} \label{double real Coll}
    \lim_{z_{0'}\to z_0} H^{(2)}_{[100'2]} &=
    {\cal N}_\varepsilon^2
{\cal A}_{[102]}^{(g,0)i} {\cal A}_{[102]}^{(g,0)j}
\frac{z_{00'}^i z_{00'}^j}{(z_{00'}^2)^{2-2\varepsilon}}
\frac{n_{\rm adj}^s-2n_{\rm adj}^f+2-2\varepsilon(1-\delta)}{6}
\nonumber\\ &\phantom{=}+H^{(1)}_{[102]} \frac{{\cal N}_\varepsilon\cd}{(z_{00'}^2)^{1-2\varepsilon}}
\left(\frac{n_{\rm adj}^f-4}{2} + \log\frac{z_{12}^2z_{00'}^2}{z_{10}^2z_{20}^2} \right)\,,
\end{align}
with the leading-order amplitude ${\cal A}^{(g,0)i}_{[102]}$ and the BK kernel $H^{(1)}_{[102]}$ recorded in \eqref{eq:LOamp} and \eqref{eq:H1_102}, respectively. All terms are only logarithmically divergent as $\varepsilon \to 0$; this equation is exact in $\varepsilon$ and is valid up to power-suppressed corrections in $z_{00'}$.
The last logarithm arises solely from switching to the conformal rapidity renormalization scheme in \eqref{double-real scheme change 2}.

Note that the second line is explicitly proportional to $H^{(1)}_{[102]}$, in accordance with what is expected from the collinear factorization. The same phenomenon occurs in the first line (which is equal to the scalar loop weighted by the total number of degrees of freedom) after symmetrical integration.

Another potentially dangerous limit is $z_0\to z_1$ with fixed (generic) $z_{0'}$. However, in that limit, we do not find any term that would be non-integrable at $\varepsilon\to 0$.
This is true whether or not \eqref{double-real scheme change 2} is included.

To make the evolution equation usable as $\varepsilon\to 0$, we introduce a subtraction term $H^{\rm coll}_{[100'2]}$ that captures the same divergence as \eqref{double real Coll}. We can then rewrite the second line of \eqref{eq:evolEqU2} in a form that is convergent as $\varepsilon\to 0$:
\begin{equation}\label{eq:subtractionDR}
    \int_{z_0,z_{0'}} \left( \bigl(\mathcal{U}_{10}\mathcal{U}_{00'}\mathcal{U}_{0'2}-\mathcal{U}_{12}\bigr)H_{[100'2]}
- \bigl(\mathcal{U}_{10}\mathcal{U}_{02}-\mathcal{U}_{12}\bigr) H_{[100'2]}^{\rm coll}
\right) = 
\mbox{finite}\,.
\end{equation}
The conditions on the subtraction are that it should not reintroduce other divergences, and furthermore, we would like its integral $\int_{z_{0'}}H_{[100'2]}^{\rm coll}$ to be relatively simple so that we can add it back analytically integrated to the virtual contribution below.  

We find that a good way to achieve this is to complete the expression \eqref{double real Coll} away from the limit $z_{0'}\to z_0$ in such a way that the results are given by conformal integrals,
meaning that the integrand is homogeneous of degree $-\Dperp=-2+2\varepsilon$ in distances $z_{i0'}^2$.
After considering several candidates, we settled on the following convenient basis of integrals:
\begin{subequations} \label{Icoll def}
\begin{align}
    {\cal I}^{{\rm coll},a,ij}_{[100'2]}&\equiv{
(z_{00'}^2)^{2\varepsilon}\frac{z_{10}^2z_{20}^2}{z_{10'}^2z_{20'}^2}
\bigg(\frac{z_{00'}^i}{z_{00'}^2}-\frac{z_{01}^i}{z_{01}^2}\bigg)
\bigg(\frac{z_{00'}^j}{z_{00'}^2}-\frac{z_{02}^j}{z_{02}^2}\bigg), } \\
 {\cal I}^{{\rm coll},b}_{[100'2]}&\equiv
 \frac{1}{(z_{00'}^2)^{1-2\varepsilon}}
\frac{z_{10}^2z_{20}^2}{z_{10}^2z_{20'}^2-z_{10'}^2z_{20}^2}\log\frac{z_{10}^2z_{20'}^2}{z_{10'}^2z_{20}^2}\,,\\
 {\cal I}^{{\rm coll},c}_{[100'2]}&\equiv
\frac{1}{(z_{00'}^2)^{1-2\varepsilon}} \frac{z_{20}^2}{z_{20'}^2}\log\frac{z_{12}^2z_{00'}^2}{z_{10'}^2z_{20}^2}\,.
\end{align}
\end{subequations}
Some comments are in order. The second integral was inspired by the $\varepsilon\to 0$ limit of the ${\cal N}=1$ kernel in \eqref{double real 4d}, which we simply multiplied by $(z_{00'}^2)^{2\varepsilon}$ to ensure the correct singular limit and homogeneity degree.
For the third integral, we chose a way to turn the logarithm in \eqref{double real Coll} into a conformal invariant cross-ratio, which happens to vanish when $z_{0'}\to z_2$; this enables multiplication by $z_{20}^2/z_{20'}^2$ to fix the degrees.
All have in common the following property: if we translate $z_0\to 0$ and perform a coordinate inversion $z_{0'}^i\mapsto x^i/x^2$ (under which $\int \d^{\Dperp}z_{0'}\mapsto \int \d^{\Dperp}x/(x^{2\Dperp})$), the integrand depends only on two distances: $(x-\tilde{z}_a)^2$ where $\tilde{z}_a=z_{0a}^i/z_{0a}^2$ for $a=1,2$. Thus, using translation invariance, the integral reduces to a simple function of the distance $\tilde{z}_1-\tilde{z}_2$.

In other words, conformal symmetry means that after integrating $z_{0'}$, what was potentially a function of three coordinates $z_{1}$, $z_2$ and $z_0$ is really just a function of a single distance.
In practice, this property also means that even though the integrals \eqref{Icoll def} may seem complicated, they can actually be computed analytically in any $\Dperp$.

We found that this can be done using elementary methods, such as exponentiating all denominators using Schwinger parameters (reintroducing the $\tau$-representation used in \eqref{short integral} for the second integral). Each integral then produces a straightforward product of $\Gamma$ functions (as opposed to hypergeometric functions which may have appeared for a non-conformal integral): 
\begin{equation} \label{subtraction integrals}
   \cd \int\limits_{z_{0'}}
\begin{bmatrix}
 {\cal I}^{{\rm coll},a,ij}_{[100'2]}\\
  {\cal I}^{{\rm coll},b}_{[100'2]}\\
   {\cal I}^{{\rm coll},c}_{[100'2]}
\end{bmatrix}
    = 
    \frac{\Gamma(1{-}\varepsilon)^3\Gamma(1{+}\varepsilon)}{\Gamma(1-2\varepsilon)}
    \left(\frac{z_{12}^2}{z_{10}^2z_{20}^2}\right)^{-\varepsilon}
 \times \begin{bmatrix}
     \frac{\delta^{ij}-2\varepsilon\tilde{z}_{12}^i\tilde{z}_{12}^j/\tilde{z}_{12}^2}{2\varepsilon(1-2\varepsilon)}
     \\
    \frac{1}{\varepsilon(1-2\varepsilon)}\\
    -\frac{1}{\varepsilon^2}
\end{bmatrix}\,,
\end{equation}
where $\tilde{z}_{12}^i = \frac{z_{01}^i}{z_{01}^2}-\frac{z_{02}^i}{z_{02}^2}$.
These results are exact in $\varepsilon$.

Now, using these functions, we can define a subtraction that
matches the $z_{0'}\to z_0$ limit of the integrand \eqref{double real Coll}:
\begin{align} \label{H2coll}
    H^{(2)\rm coll}_{[100'2]} &\equiv
{\cal N}_\varepsilon^2\frac{n_{\rm adj}^s-2n_{\rm adj}^f+2-2\varepsilon(1-\delta)}{6}
\,
{\cal A}_{[102]}^{(g,0)i} {\cal A}_{[102]}^{(g,0)j}
 \times {\cal I}^{{\rm coll},a,ij}_{[100'2]}
\nonumber\\
&\quad+ H^{(1)}_{[102]} {\cal N}_\varepsilon\cd
\left[
\frac{(n_{\rm adj}^f{-}4)}{2}
{\cal I}^{{\rm coll},b}_{[100'2]}
+{\cal I}^{{\rm coll},c}_{[100'2]}\right]\,.
\end{align}
As desired, this has the same singular limit $z_{0'}\to z_0$ as \eqref{double real Coll} to all orders in $\varepsilon$ and does not introduce new divergences as $z_{0'}\to z_1, z_2$.
Therefore, it is valid to use in \eqref{eq:subtractionDR}.
The integral over $z_{0'}$ is then given analytically by \eqref{subtraction integrals}.
Expanding the result to order $\varepsilon$ we obtain:
\begin{align}\label{eq:coll_div}
    \cd \int\limits_{z_{0'}}H^{(2)\rm coll}_{[100'2]}&=
H^{(1)}_{[102]}
\left(\frac{\mubar^2z_{10}^2z_{20}^2}{4 z_{12}^2\e^{-2\gamma_\text{E}}}\right)^{\varepsilon}
\left[-\frac{2}{\varepsilon^2}{-}\frac{b_0}{\varepsilon}
{+}{2\nf{-}\frac{25{-}\delta}{3}{-}\frac{\pi^2}{6}}{+}c_1^{\rm coll}\varepsilon
\right]{+}\mathcal{O}(\varepsilon^2)\,,
\end{align}
with $b_0=\frac{11}{3}-\frac{1}{6}n^s_{\rm adj}-\frac{2}{3}n^f_{\rm adj}$ in the class of theories under consideration and
\begin{equation}
c_1^{\rm coll}=
4(\nf-4)-b_0\frac{\pi^2}{12}+{\frac{10}{3}\zeta_3 }\,.    
\end{equation}
We note that while the coefficient of the single pole is the same as the $\beta$-function, physical distinctions between collinear and running coupling effects have been emphasized in \cite{Balitsky:2007feb}.

Finally, although obtaining \eqref{eq:coll_div} required the full finite-$\varepsilon$ behavior of \eqref{H2coll} as $z_{0'}\to z_0$, its application in \eqref{eq:subtractionDR} only requires the $\varepsilon$-expanded integrand, which we record 
here for convenience (in the normalization \eqref{double result susy}):
\begin{align}\label{H2coll expanded}
    \tilde{H}^{(2)\rm coll}_{[100'2]}&=
    (1{+}\varepsilon L_\mu^{\rm coll})
    \Big[
    (2{-}2\nf{+}\ns{-}2\varepsilon(1{-}\delta))\frac{(1{+}u{-}v)(1{-}u{-}v)}{6u}{+}4\log u{+}(4{-}\nf)\frac{2v \log v}{1-v}
    \Big]\nonumber
    \\&
    \phantom{+}+\varepsilon\frac{z_{20}^2}{z_{00'}^2z_{12}^2z_{20'}^2}\log\frac{z_{20}^2}{z_{10}^2}\Big[
    \frac{2z_{00'}^2z_{20'}^2(z_{10}^2{+}z_{12}^2{-}z_{20}^2)(2(1{-}v)\log u{+}(4{-}\nf)v\log v)}{z_{20}^2(1-v)}\nonumber
    \\&
    \phantom{+}-\frac{(2{-}2\nf{+}\ns)}{6}\Big(
    ((1-u-v)z_{00'}^2+(u^2+(2v-u)(1-v))z_{10'}^2)z_{20'}^2
    \\&
    \phantom{+}+v z_{10'}^2((1+u-v)z_{00'}^2-2(1-v)z_{20}^2)
    \Big)
    \Big]+\mathcal{O}(\varepsilon^2)\,,\nonumber
\end{align}
where $L_\mu^{\rm coll}\equiv \log\Big[\frac{\overline{\mu}^{4}z_{00'}^2z_{10}^2z_{12}^2}{16 \e^{-4\gamma_{\text{E}}}z_{20'}^2}\Big]$ and $u$ and $v$ are the cross-ratios from \eqref{cross-ratios}.

\begin{figure}
    \centering
 \begin{subfigure}[t]{.18\linewidth}
    \centering\adjustbox{valign=c}{\includegraphics[scale=0.4]{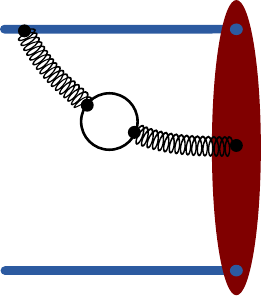}}
    \caption*{$(\Pi)$}
  \end{subfigure}
  \begin{subfigure}[t]{.18\linewidth}
    \centering \adjustbox{valign=c}{\includegraphics[scale=0.4]{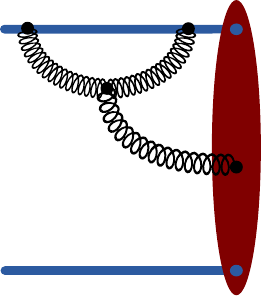}}
   \caption*{$(\Upsilon)$}
  \end{subfigure}
  \begin{subfigure}[t]{.38\linewidth}
    \centering$\adjustbox{valign=c}{\includegraphics[scale=0.4]{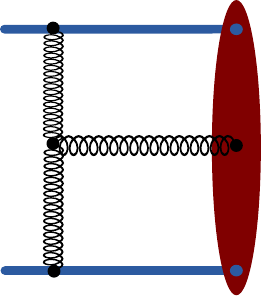}}\quad
            +\adjustbox{valign=c}{\includegraphics[scale=0.4]{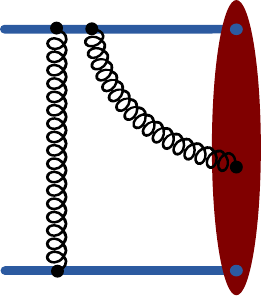}}$
    \caption*{$(\T)$}
  \end{subfigure}
 \begin{subfigure}[t]{.18\linewidth}
    \centering\adjustbox{valign=c}{\includegraphics[scale=0.4]{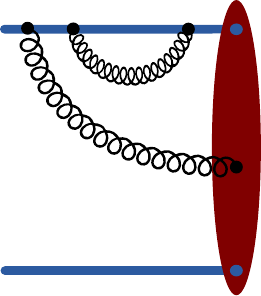}}
    \caption*{$0$}
  \end{subfigure}
  \caption{
       The set of planar Feynman diagrams contributing to the NLO real-virtual amplitude on the left of the shock, separated into self-energy ($\Pi$), single-Wilson line ($\Upsilon$) and two Wilson-line diagrams ($\T$).  The rightmost diagram (0) vanishes  in dimensional regularization and we have omitted other graphs that similarly vanish. As before, solid black lines accounts for all possible pair of intermediate states: $(g,g), (f,\bar{f}), (s,s)$.
       }
        \label{fig:diagRV}
\end{figure}

{\subsection{Real-virtual evolution kernel}\label{sssec:rapiditySubRV}}
The real-virtual contributions to the two-loop BK evolution contain the one-loop corrections to the one-gluon amplitude shown in Fig.~\ref{fig:diagRV}.
The calculation parallels
the steps just described for the double real diagrams and is detailed in App.\,\ref{app:detailsHRV}. Here, we highlight the main ideas and results.

Starting from the same Feynman rules, we first integrate out $k_i^-$ to obtain a Fourier representation for the one-gluon amplitude ${\cal A}^{(g,1)i}_{[102]}(\tau)$,
which depends on the transverse position $z_0$ where the gluon crosses the shock,
two transverse momenta $k_1,k_2$, and 
the energy ratio $\tau=k_0^+/k_{0'}^+$ between the real and virtual partons.
We then integrate over transverse momenta $k_1$, $k_2$ using a combination of standard Schwinger parameter techniques and the differential equation method. We finally multiply with the tree-level amplitude ${\cal A}^{(g,0)i}_{[102]}$ from the other side of the shock to obtain the integrand ${\cal I}_{[102]}^{(2)}$ as defined earlier in \eqref{I from A}, and integrate over $\tau$.

As for double-real contributions, the $\tau$ integral in the double-real contributions exhibits rapidity subdivergences as $\tau\to0$ and $\tau\to\infty$, arising from the integration regions where the integrand factorizes into successive iterations of the one-loop evolution.
The actual evolution kernel $H_{[102]}^{(2)}$ is the sum of several pieces. In order: the real-virtual integrand \eqref{eq:L2RV}, the coupling constant renormalization counter-term from \eqref{bare coupling}, plus the integrated double-real collinear subtraction from \eqref{eq:subtractionDR}:\footnote{Note that, unlike in \eqref{eq:L2RV}, the counterterm subtraction is now made explicit and the double‑real collinear term has been included.}
\begin{equation}\label{eq:H102 def}
H_{[102]}^{(2)}=
\frac{{\cal N}_\varepsilon^2}{\cd}
    \int_0^{\infty} 
    \frac{\d\tau}{\tau}
\left[{\cal I}^{(2)}_{[102]}(\tau)
- {\cal S}_{[102]}^{(2)}(\tau)\right]
-\frac{b_0}{\varepsilon}H^{(1)}_{[102]}
+\cd\int\limits_{z_{0'}} H^{(2)\rm coll}_{[100'2]}\,,
\end{equation}
with, as before, ${\cal N}_\varepsilon=2\pi\mu^{2\varepsilon}$, $\cd$ defined in \eqref{def cd} and where the rapidity subtraction is given by
\begin{equation}\label{H1 sub}
\hspace{-0.3cm}
    {\cal S}_{[102]}^{(2)}(\tau)=
\int\limits_{z_{0'}} {\cal I}_{[102]}
\left(\mathcal{I}_{[10'2]}^{(1)}\theta\Big(1>\tfrac{\nu_{[102]}}{\nu_{[10'2]}}\Big)+
\mathcal{I}_{[00'2]}^{(1)}\theta\Big(\tfrac{\nu_{[102]}}{\nu_{[00'2]}}>1\Big)+
\mathcal{I}_{[10'0]}^{(1)}\theta\Big(\tfrac{\nu_{[102]}}{\nu_{[10'0]}}>1\Big)
\right).
\end{equation}
The scale ratio $\frac{\nu_{[i0j]}}{\nu_{[k0'\ell]}}\propto \tau$ depends on the transverse coordinates in a way that enforces conformal symmetry (see \eqref{eq:conformalScheme}). It is important to note that the subtraction features an auxiliary point $z_{0'}$ that is not present in the original integrand ${\cal I}_{[102]}^{(2)}(\tau)$. Nevertheless, the square bracket of \eqref{eq:H102 def} vanishes as $\tau\to0$ and $\tau\to\infty$ thanks to the identity \eqref{real-virtual one-loop} and factorization property \eqref{virtual factorization}. 
In practice, this property also provides a useful cross-check of the overall normalization of each individual diagram.

Since ${\cal A}^{(g,0)i}_{[102]}$ is $\tau$ independent, a useful feature
of one-loop virtual corrections is that the energy‐ratio integration over $\tau$ and the Fourier transforms in momentum space can both be performed at the amplitude level— i.e., \emph{before} multiplying the amplitudes from either side of the shock.
In App.~\ref{app:detailsHRV}, we initially adopt the so-called $+$-scheme (also used in \cite{Balitsky:2007feb}), in which $\tau \;\equiv\;\frac{\nu_{[i0j]}^+}{\nu_{[k0'\ell]}^+}$
can be integrated without introducing the auxiliary point $z_{0'}$.

The kernel \eqref{eq:H102 def} can thus be organized in the following way:
\begin{equation}\label{eq:H102 def 2}
 H_{[102]}^{(2)} =
 \frac{{\cal N}_\varepsilon}{\cd}
 2
 \left({\cal A}_{[102]}^{(g,1)i+}\big|_{\Pi+\Upsilon+{\rm c.t.}}+{\cal A}_{[102]}^{(g,1)i+}\big|_{\T}\right)
 {\cal A}_{[102]}^{(g,0)i}
+H_{[102]}^{(2)\Delta {\rm scheme}}
+\cd\int\limits_{z_{0'}} H^{(2)\rm coll}_{[100'2]}\,.
\end{equation}
The main results of App.~\ref{app:detailsHRV}
are the first two amplitudes in \eqref{eq:H102 def 2}:
\begin{subequations}
\begin{align} \label{virtual 1}
    {\cal A}_{[102]}^{(g,1)i+}&\Big|_{\Pi+\Upsilon+{\rm c.t.}} = 
 \frac{z_{10}^i\cd}{(z_{10}^2)^{1-\varepsilon}}
 \left[
 \e^{-\varepsilon\gamma_\text{E}}\Gamma(1{-}\varepsilon)
\left(\frac{\mubar^2z_{10}^2}{4\e^{-2\gamma_\text{E}}}\right)^\varepsilon
C_{\Pi+\Upsilon} - \frac{b_0}{2\varepsilon}
\right]-(z_1{\leftrightarrow}z_2)\,,
\\
{\cal A}_{[102]}^{(g,1)i+}&\Big|_{\T}\!\!=\!
\frac{-z_{10}^i \cd}{(z_{10}^2)^{1-\varepsilon}}\!\left[\log\frac{z_{12}^2}{z_{10}^2}\log\frac{z_{12}^2}{z_{20}^2}\!\left[\!1{+}\frac{\varepsilon}{2}\log\frac{\mubar^4z_{12}^2z_{20}^2}{16\e^{-4\gamma_{\text{E}}}}\right]\!
{-}\varepsilon {\cal L}_{1,0,1}\right]\! {-}(z_1{\leftrightarrow}z_2){+}{\cal O}(\varepsilon^2)\,.
\end{align}
\end{subequations}
The first amplitude is obtained exactly in $\varepsilon$ and expressed in terms of the following constant
\begin{equation}
 C_{\Pi+\Upsilon}=
\frac{h(\varepsilon)+3h(-2\varepsilon)-3h(-\varepsilon)}{\varepsilon}
-\frac{\nf-4}{\varepsilon(1-2\varepsilon)}
 -\frac{\ns-2\nf+2-2\varepsilon(1{-}\delta)}{2\varepsilon(1-2\varepsilon)(3-2\varepsilon)}\,,
\end{equation}
with $h(x)=\Gamma'(x+1)/\Gamma(x+1)+\gamma_\text{E}$.
The amplitude for the $\T$ diagram, on the other hand, depends more nontrivially on the transverse coordinates and could only be obtained as a series in $\varepsilon$. For example, it includes more sophisticated functions, such as the single-valued multiple polylogarithm
\cite{Dixon:2012yy}
\begin{equation}\label{L101}
\mathcal{L}_{1,0,1}(\zeta,\bar\zeta)=
H_{1,0,1}+\overline{H}_{1,0,1}+H_{1,0}\overline{H}_{1}+H_{1}\overline{H}_{1,0}\,,
\end{equation}
where $\bar{H}_\bullet\equiv H_\bullet(\bar\zeta)$,
\begin{equation}\nonumber
H_{1,0,1}\!=\!2\Li_3\bigl(1{{-}}\zeta\bigr)-\log\bigl(1{{-}}\zeta\bigr)\,\bigl(\Li_2\bigl(1{{-}}\zeta\bigr)+\zeta_2\bigr)-2\zeta_3,\quad
H_{1,0}\!=\!\Li_2\bigl(1{{-}}\zeta\bigr)-\zeta_2,
\quad
H_{1}\!=\!-\log\bigl(1{{-}}\zeta\bigr),
\end{equation}
and the (complex) variables $\zeta$ and $\overline{\zeta}$ are related to the transverse distances according to
\begin{equation}\label{eq:zetadef}
   z_{10}^2 = \zeta \overline{\zeta} z_{12}^2\,,
   \quad  z_{20}^2 = (1{-}\zeta) (1{-}\overline{\zeta}) z_{12}^2
   \,\,\Rightarrow\,\,
\zeta,\overline{\zeta}=
\frac{z_{12}^2{+}z_{10}^2{-}z_{20}^2+ \lambda^{\pm}}{2z_{12}^2}\,,
\end{equation}
with the square roots given by $\lambda^{\pm}\equiv \pm i\sqrt{4z_{12}^2z_{10}^2-(z_{12}^2{+}z_{10}^2{-}z_{20}^2)^2}$.
\paragraph{Real-virtual Hamiltonian in the conformal scheme.}
The integral of $H^{(2){\rm coll}}$ was given in \eqref{eq:coll_div}, and so the last remaining ingredient in \eqref{eq:H102 def 2} is the conversion from the $+$-rapidity subtraction scheme to the conformal one. The difference, similarly to \eqref{double-real scheme change}, originates from changing the ordering variables in the step functions in \eqref{H1 sub}:
\begin{equation}
    \begin{split}
        \label{Delta scheme virtual}
  H_{[102]}^{(2)\Delta {\rm scheme}}  &=\frac{{\cal N}_\varepsilon^2}{\cd} \int_0^\infty \frac{\d\tau}{\tau}
\left[{\cal S}_{[102]}^{(2)+}(\tau)
-{\cal S}_{[102]}^{(2)}(\tau)\right]
\\ &= \frac{{\cal N}_\varepsilon^2}{\cd}\int\limits_{z_{0'}}\int_0^\infty \frac{\d\tau}{\tau}\,
\mathcal{I}_{[102]}^{(1)}\, 
    \left[\begin{array}{rll}
&\mathcal{I}_{[10'2]}^{(1)}&\Big[\theta\Big(1>\frac{\nu_{[102]}}{\nu_{[10'2]}}\Big){-}\theta(1>\tau)\Big]\\
+&\mathcal{I}_{[00'2]}^{(1)}&\Big[\theta\Big(\frac{\nu_{[102]}}{\nu_{[00'2]}}>1\Big){-}\theta(\tau>1)\Big]
\\+&\mathcal{I}_{[10'0]}^{(1)}&\Big[\theta\Big(\frac{\nu_{[102]}}{\nu_{[10'0]}}>1\Big){-}\theta(\tau>1)\Big]
\end{array}\right].
    \end{split}
\end{equation}
Since the leading-order integrands ${\cal I}^{(1)}$ are independent of energies (see \eqref{eq:I1_102}) and $\frac{\nu_{[i0j]}}{\nu_{[k0'\ell]}}\propto \tau$,
the $\tau$ integral gives simple logarithms of transverse positions. The $z_{0'}$ integral can then be computed using the standard bubble integral (and its derivatives with respect to the exponents):\footnote{Although the complete integral \eqref{Delta scheme virtual} is well defined in dimensional regularization, when expanded to use the standard form \eqref{bubble integral} one finds ill-defined integrals $\propto \d^{2-2\varepsilon}z_{0'}/z_{0'a}^{2-2\varepsilon}$.
They can be dealt with by adding a small quantity $\lambda$ to the exponents and taking $\lambda\to 0$ in the end.}
\begin{equation}\label{bubble integral}
 \cd   
 \int\limits_{z_{0'}}\frac{1}{(z_{i0'}^2)^\alpha (z_{0'k}^2)^\beta}=\Gamma(1{-}\varepsilon)\frac{\Gamma(1{-}\varepsilon{-}\alpha)\Gamma(1{-}\varepsilon{-}\beta)\Gamma(\alpha{+}\beta{-}(1{-}\varepsilon))}{\Gamma(\alpha)\Gamma(\beta)\Gamma(2{-}\alpha{-}\beta{-}2\varepsilon)}z_{ik}^{2(1{-}\alpha{-}\beta{-}\varepsilon)}\,.
\end{equation}
Introducing the abbreviations
\begin{equation}
    L_\mu=\log \frac{\mubar^2z_{12}^2}{4\e^{-2\gamma_\text{E}}}\,,\quad
    L_1=\log\frac{z_{12}^4}{z_{01}^2z_{02}^2}\,,\quad
    L_2=\log\frac{z_{12}^2}{z_{10}^2}\log\frac{z_{12}^2}{z_{20}^2}\,,
\end{equation}
for recurring logarithms, we find in this way:
\begin{equation}\label{H1 scheme change}
H_{[102]}^{(2)\Delta {\rm scheme}}
=H_{[102]}^{(1)}\left(\frac{\mubar^2z_{10}^2z_{20}^2}{4\e^{-2\gamma_\text{E}}z_{12}^2}\right)^\varepsilon
\left[\frac{2}{\varepsilon^2}+\frac{\pi^2}{6}+
2(1+\varepsilon L_1)L_2+\frac{14}{3}\varepsilon\zeta_3
+{\cal O}(\varepsilon^2)
\right].
\end{equation}
Finally, assembling \eqref{virtual 1}, \eqref{H1 scheme change} and \eqref{eq:coll_div} gives us the $\varepsilon$-expansion of the single-real kernel:
\begin{equation}
    H_{[102]}^{(2)} = \left( \frac{z_{12}^2}{z_{10}^2z_{02}^2}\right)^{1-\varepsilon} 
    \left[\tilde{H}_{[102]}^{(2,0)}+
    \varepsilon\tilde{H}_{[102]}^{(2,1)}+\ldots
    \right],
\end{equation}
where the right-hand side is given by the explicit functions of transverse coordinates:
\begin{subequations}\label{rv result}
\begin{align}\label{rv result 1}
    \tilde{H}_{[102]}^{(2,0)}&=
    b_0 \tilde{H}_{[102]}^{(1,1)} +
    \left(c^{(2,0)}\equiv 2\frac{59-2\nf-8\ns-3\delta}{9}-\frac{4\pi^2}{3}\right)\,,
\\ \label{rv result 2}
        \tilde{H}_{[102]}^{(2,1)}&=
8 \frac{94{-}\nf{-}13\ns{-}12\delta}{27}-24\zeta_3+c^{(2,0)}\tilde{H}_{[102]}^{(1,1)}+b_0(2\tilde{H}_{[102]}^{(1,2)}{+}L_\mu^2-\zeta_2)+(L_1{-}2b_0)L_2
\nonumber\\
&{+} \left(2\zeta_2{+}\frac{4}{9}(\ns{-}2\nf{+}2)\right)\!(\tilde{H}_{[102]}^{(1,1)}{+}2L_1{-}2L_\mu)
+ \left(\frac12\tilde{H}_{[102]}^{(1,1)}-L_\mu\right)\left(b_0(2L_\mu{-}L_1)+L_2\right)\nonumber\\
&+2\frac{z_{12}^2+z_{20}^2-z_{10}^2}{z_{12}^2}{\cal L}_{1,0,1}(\zeta,\overline{\zeta})
+2\frac{z_{12}^2+z_{10}^2-z_{20}^2}{z_{12}^2}{\cal L}_{1,0,1}(1{-}\zeta,1{-}\overline{\zeta})\,.
\end{align}
\end{subequations}

It is noteworthy that, at $\varepsilon=0$, the only source of conformal symmetry-breaking in \eqref{rv result 1}, namely $\tilde{H}_{[102]}^{(1,1)}$, is proportional to the one-loop $\beta$-function coefficient $b_0$.\footnote{Since there exists no conformal invariant cross-ratios of three coordinates $z_{1}$, $z_2$ and $z_0$, any nonconstant contribution to $\tilde{H}_{[102]}$ violates conformal invariance.}
Indeed, this ensures that the evolution is conformal in the conformal dimension \eqref{eq:WF_point_def}: $\varepsilon=-ab_0+\mathcal{O}(a^2)$, as was discussed recently in \cite{Balitsky:2024xvi}.
In Sec.~\ref{sec:renormalon}, we will extend this observation to three-loop order, at least for terms quadratic in $\ns$, $\nf$.

\subsection{Summary of two-loop evolution} 
To recapitulate, to two-loop accuracy we can write the BK rapidity evolution equation in a generic planar gauge theory as:\footnote{Equivalently, the collinear subtraction term could be written with $z_{0'}$ and $z_0$ exchanged, e.g.
$(\mathcal{U}_{10'}\mathcal{U}_{0'2}-\mathcal{U}_{12})
H_{[10'02]}^{\rm coll}$, or symmetrized. This would not change the value of $\tilde{H}_{[102]}$.}
\begin{align}
\label{eq:summary_res}
        \nu\partial_\nu\,\mathcal{U}_{12}
    &= C_\varepsilon\!\int\limits_{z_0}\,
 \left(\frac{z_{12}^2}{z_{10}^2z_{02}^2}\right)^{1-\varepsilon}
    \big(\Uu_{10}\Uu_{02}-\Uu_{12}\big)\,\tilde{H}_{[102]}
    \\&
    +\cd^2 \int\limits_{z_0,z_{0'}}
    \left(\frac{z_{12}^2}{z_{10}^2z_{00'}^2z_{0'2}^2}\right)^{1-\varepsilon}
    \left( (\mathcal{U}_{10}\mathcal{U}_{00'}\mathcal{U}_{0'2}-\mathcal{U}_{12})\tilde{H}_{[100'2]}
-(\mathcal{U}_{10}\mathcal{U}_{02}-\mathcal{U}_{12})
\tilde{H}_{[100'2]}^{\rm coll}
\right)\,.\notag
\end{align}
Each kernel can be expanded in the gauge coupling $a=\frac{\alpha_s N}{4\pi}$ and $\varepsilon$ as
\begin{equation}
\tilde{H}_{[102]}=\sum_{\ell=1}^\infty \sum_{k\geq 0} a^\ell\varepsilon^k\tilde{H}^{(\ell,k)}_{[102]}\,,\qquad \tilde{H}_{[100'2]}=\sum_{\ell=2}^\infty \sum_{k\geq 0} a^\ell\varepsilon^k\tilde{H}^{(\ell,k)}_{[100'2]}\,.
\end{equation}
The one-loop single-emission kernel
$\tilde{H}_{[102]}^{(1,k)}$ is given in \eqref{eq:H1_102exp} for $k=0,1,2$ and the two-loop kernel is given in \eqref{rv result} for $k=0,1$.
For double-emission kernels, the analogous formulas including $k=0,1$ are in \eqref{double result susy}-\eqref{double-real N=4}. Finally, the collinear subtraction is described in \eqref{H2coll expanded}.
The $\D=4$ part of the result ($k=0$) has been known for some time \cite{Balitsky:2006wa,Balitsky:2007feb,Balitsky:2009xg} and the new result here are the $\varepsilon$ corrections ($k=1$ terms).

To convert the above to $n_S$ complex scalars and $n_F$ fundamental Dirac fermions instead of adjoint matter, one simply has to replace $\ns\mapsto n_S/N$, $\nf\mapsto n_F/N$, and set $\Uu_{00'}\mapsto 1$ in any term that these multiply.

\paragraph{Comparison with the literature.}

The reader familiar with the $\D=4$ evolution kernels obtained in \cite{Balitsky:2007feb,Balitsky:2009xg} 
may notice that the constant in our single-virtual contribution is different.
For convenience, let us recall the result of these references for the evolution of the so-called conformal dipole, which matches to our conformal scheme (see also Eq.~(2.9) of \cite{Balitsky:2024xvi}), converted to our theory:
\begin{align} \label{BC result}
 \nu\partial_\nu \Uu_{12}^{\rm there}&=a\int \frac{\d^2z_0}{\pi}\frac{z_{12}^2}{z_{10}^2z_{02}^2}(\Uu_{10}\Uu_{02}-\Uu_{12}) \left[2+a \left(b_0 \tilde{H}_{[102]}^{(1,1)}+2\Gamma_{\rm c}\right)\right] 
 \\ & +a^2\int\frac{\d^2z_0\,\d^2z_{0'}}{\pi^2}\frac{z_{12}^2}{z_{10}^2z_{00'}^2z_{0'2}^2}
\left(\Uu_{10}\Uu_{00'}\Uu_{0'2}{-}\tfrac12 \Uu_{10}\Uu_{02}
{-}\tfrac12 \Uu_{10'}\Uu_{0'2}\right)\tilde{H}^{(2,0)}_{[100'2]}{+}{\cal O}(a^3,\varepsilon)\,.\nonumber
\end{align}
with $\Gamma_{\rm c}=\frac{67-3\delta-10\nf-4\ns}{9}-\frac{\pi^2}{3}$.
As noted already, the $\D=4$ limit of our double-real kernel (see $\tilde{H}^{(2,0)}_{[100'2]}$ in \eqref{double result susy} and \eqref{double real 4d}) agrees precisely with these references. However, one can see that the constant $\Gamma_{\rm c}$ differs from that in \eqref{rv result 1}.

This apparent mismatch can be attributed to the choice of collinear subtraction in \eqref{eq:summary_res}, where Refs.~\cite{Balitsky:2007feb,Balitsky:2009xg} use the full kernels: $H^{(2)\rm coll,there}_{[100'2]}=H^{(2)}_{[100'2]}$
as shown in \eqref{BC result}.
This changes the single-emission term by a total integral:
\begin{equation}\label{eq:HdiffLit}
    \begin{split}
            \tilde{H}^{(2)\rm there}_{[102]}
-\tilde{H}^{(2)\rm here}_{[102]}&= \int \frac{\d^2z_{0'}}{\pi} \frac{z_{02}^2}{z_{00'}^2z_{0'2}^2}
\left(\tilde{H}_{[100'2]}^{(2,0)}-\tilde{H}_{[100'2]}^{(2,0)\rm coll}\right)
\\ &
\stackrel{?}{=} \textcolor{black}{\frac{8}{9}(n_{\rm adj}^s{-}2n_{\rm adj}^f{+}2)+\frac{2\pi^2}{3}}\,, 
    \end{split}
\end{equation}
where the second line is the difference in the constants.
The integrand in \eqref{eq:HdiffLit} can be read directly from \eqref{eq:DRHtilde}-\eqref{double real 4d}; in terms of the cross-ratios \eqref{cross-ratios}:
\begin{equation}
\tilde{H}_{[100'2]}^{(2,0)}-\tilde{H}_{[100'2]}^{(2,0)\rm coll}=\frac{n_{\rm adj}^s{-}2n_{\rm adj}^f{+}2}{u}
\left[\frac{v+1}{v-1}v\log v-2v-\frac{(1-v)^2-u^2}{6}\right]
+2\frac{u+v-1}{v-1}\log v\,.
\nonumber\end{equation}
The integral \eqref{eq:HdiffLit} is absolutely convergent and produces exactly the constant on the second line, confirming the equivalence of the two ways of writing the evolution equation. Let us briefly comment on how the integral can be straightforwardly verified numerically. Since it reduces to a numerical constant, we can gauge‑fix the three points $(z_1,z_2,z_0)=(0,\infty,1)$ by
conformal invariance and set $z_{0'}=r\,\e^{i\theta},$ where $ r\in(0,\infty)$ and $\theta\in(0,2\pi)$. The resulting integrals over $r$ and $\theta$ are manifestly finite and can be carried out numerically to high precision (using, e.g., \textsc{Mathematica}'s \texttt{NIntegrate[]}), producing the second line of \eqref{eq:HdiffLit}.

{\section{The spacelike-timelike correspondence}\label{sec:correspondence}}

The spacelike-timelike correspondence (see also \cite{Hofman:2008ar,Hatta:2008st,Caron-Huot:2015bja,Mueller:2018llt,Caron-Huot:2022eqs,Balitsky:2024xvi}) is a general equivalence, in any conformal field theory, between measurements performed at null infinity and measurements performed on a generic null plane $x^+=0$.  In the present context, it relates the integral kernel governing the evolution/resummation of large non-global logarithms (NGLs) in jet physics with the rapidity evolution of high‑energy forward scattering that is the focus of this paper.

As we will review, although this correspondence is not exact in QCD due to the running of the coupling, its validity in the critical dimension \eqref{eq:WF_point_def} still implies nontrivial relations to exploit in perturbation theory.
 
\subsection{Review: non-global logarithms at leading order}\label{ssec:NGL}

Let us briefly review the setup which gives rise to so-called ``non-global logarithms.''  For complementary discussions, see, for example, \cite{Marchesini:2015ica,Caron-Huot:2015bja,Balitsky:2024xvi}. 

\begin{figure}
    \centering
\tikzset{every picture/.style={line width=0.75pt}}
\begin{tikzpicture}[x=0.75pt,y=0.75pt,yscale=-1,xscale=1]
%Shape: Triangle 
\draw  [draw opacity=0][fill=mgGreen  ,fill opacity=0.22 ] (263.13,149.54) -- (414.66,238.86) -- (111.6,238.86) -- cycle ;
%Straight Lines 
\draw [color=sidsblue  ,draw opacity=1 ][decorate, decoration={coil, segment length=3pt, amplitude=2pt}]   (160,150) -- (134.57,185.07) ;
%Straight Lines 
\draw [color=sidsblue  ,draw opacity=1 ][decorate, decoration={coil, segment length=3pt, amplitude=2pt}]   (142.86,210.54) -- (134.57,185.07) ;
%Straight Lines 
\draw [color=sidsblue  ,draw opacity=1 ][decorate, decoration={coil, segment length=3pt, amplitude=2pt}]   (191,223.5) -- (142.86,210.54) ;
%Straight Lines 
\draw [color=sidsblue  ,draw opacity=1 ][decorate, decoration={coil, segment length=3pt, amplitude=2pt}]   (348,209) -- (326.57,150.07) ;
%Straight Lines 
\draw [color=sidsblue  ,draw opacity=1 ][decorate, decoration={coil, segment length=3pt, amplitude=2pt}]   (409.57,189.07) -- (358.57,149.07) ;
%Straight Lines 
\draw [color=sidsblue  ,draw opacity=1 ][decorate, decoration={coil, segment length=3pt, amplitude=2pt}]   (370.14,228.86) -- (409.57,189.07) ;
%Straight Lines 
\draw [color=sidsblue  ,draw opacity=1 ][decorate, decoration={coil, segment length=3pt, amplitude=2pt}]   (220.79,188.04) -- (225.57,150.07) ;
%Straight Lines 
\draw [color=sidsblue  ,draw opacity=1 ][decorate, decoration={coil, segment length=3pt, amplitude=2pt}]   (160.14,125.43) -- (194.57,149.07) ;
%Straight Lines 
\draw [color=sidsblue  ,draw opacity=1 ][decorate, decoration={coil, segment length=3pt, amplitude=2pt}]   (301.14,148.86) -- (338.14,128.86) ;
%Straight Lines 
\draw [color=sidsblue  ,draw opacity=1 ][decorate, decoration={coil, segment length=3pt, amplitude=2pt}]   (338.14,128.86) -- (354.14,104.86) ;
%Straight Lines 
\draw [color=sidsblue  ,draw opacity=1 ][decorate, decoration={coil, segment length=3pt, amplitude=2pt}]   (160.14,125.43) -- (155,104.5) ;
\draw [color=sidsMaroon  ,draw opacity=1 ]   (263.13,149.54) -- (98,149.54) ;
\draw [shift={(175.76,149.54)}, rotate = 360] [color=sidsMaroon  ,draw opacity=1 ][line width=0.75]    (8.74,-2.63) .. controls (5.56,-1.12) and (2.65,-0.24) .. (0,0) .. controls (2.65,0.24) and (5.56,1.12) .. (8.74,2.63)   ;
\draw [color=sidsMaroon  ,draw opacity=1 ]   (428.26,149.54) -- (263.13,149.54) ;
\draw [shift={(351.5,149.54)}, rotate = 180] [color=sidsMaroon  ,draw opacity=1 ][line width=0.75]    (8.74,-2.63) .. controls (5.56,-1.12) and (2.65,-0.24) .. (0,0) .. controls (2.65,0.24) and (5.56,1.12) .. (8.74,2.63)   ;
%Shape: Ellipse 
\draw  [color=mgGreen  ,draw opacity=1 ][fill={rgb, 255:red, 255; green, 255; blue, 255 }  ,fill opacity=1 ][dash pattern={on 4.5pt off 4.5pt}] (414.66,238.86) .. controls (414.66,243.45) and (346.82,247.18) .. (263.13,247.18) .. controls (179.44,247.18) and (111.6,243.45) .. (111.6,238.86) .. controls (111.6,234.26) and (179.44,230.54) .. (263.13,230.54) .. controls (346.82,230.54) and (414.66,234.26) .. (414.66,238.86) -- cycle ;
%Straight Lines 
\draw [color=sidsblue  ,draw opacity=1 ][decorate, decoration={coil, segment length=3pt, amplitude=2pt}]   (437,231.5) -- (427.57,202.07) ;
%Straight Lines 
\draw [color=sidsblue  ,draw opacity=1 ][decorate, decoration={coil, segment length=3pt, amplitude=2pt}]   (134.57,185.07) -- (101,212) ;
%Straight Lines 
\draw [color=sidsblue  ,draw opacity=1 ][decorate, decoration={coil, segment length=3pt, amplitude=2pt}]   (142,227.5) -- (142.86,210.54) ;
%Straight Lines 
\draw [color=sidsblue  ,draw opacity=1 ][decorate, decoration={coil, segment length=3pt, amplitude=2pt}]   (138,115.5) -- (160.14,125.43) ;
%Straight Lines 
\draw [color=sidsblue  ,draw opacity=1 ][decorate, decoration={coil, segment length=3pt, amplitude=2pt}]   (338.14,128.86) -- (375,111.5) ;
%Straight Lines 
\draw [color=sidsblue  ,draw opacity=1 ][decorate, decoration={coil, segment length=3pt, amplitude=2pt}]   (427.57,202.07) -- (409.57,189.07) ;
%Straight Lines 
\draw [color=sidsblue  ,draw opacity=1 ][decorate, decoration={coil, segment length=3pt, amplitude=2pt}]   (475,219.5) -- (427.57,202.07) ;
%Straight Lines 
\draw [color=sidsblue  ,draw opacity=1 ][decorate, decoration={coil, segment length=3pt, amplitude=2pt}]   (138.71,197.8) -- (105.14,224.73) ;
%Straight Lines 
\draw [color=sidsblue  ,draw opacity=1 ][decorate, decoration={coil, segment length=3pt, amplitude=2pt}]   (220.79,188.04) -- (240,223) ;
%Straight Lines 
\draw [color=sidsblue  ,draw opacity=1 ][decorate, decoration={coil, segment length=3pt, amplitude=2pt}]   (348,209) -- (326,225) ;
%Straight Lines 
\draw [color=sidsblue  ,draw opacity=1 ][decorate, decoration={coil, segment length=3pt, amplitude=2pt}]   (216,226) -- (220.79,188.04) ;
%Straight Lines 
\draw [color=sidsblue  ,draw opacity=1 ][decorate, decoration={coil, segment length=3pt, amplitude=2pt}]   (358,227) -- (348,209) ;
%Straight Lines 
\draw [color=sidsblue  ,draw opacity=1 ][decorate, decoration={coil, segment length=3pt, amplitude=2pt}]   (136,149) -- (117,111) ;
%Shape: Right Angle
\draw   (450,93.5) -- (430,93.5) -- (430,73.5) ;
%Shape: circle
\draw  [draw opacity=0][fill={rgb, 255:red, 74; green, 74; blue, 74 }  ,fill opacity=1 ] (260.38,149.54) .. controls (260.38,148.02) and (261.61,146.79) .. (263.13,146.79) .. controls (264.65,146.79) and (265.88,148.02) .. (265.88,149.54) .. controls (265.88,151.06) and (264.65,152.29) .. (263.13,152.29) .. controls (261.61,152.29) and (260.38,151.06) .. (260.38,149.54) -- cycle ;
% Text Node
\draw (84,141.4) node [anchor=north west][inner sep=0.75pt]  [color=sidsMaroon  ,opacity=1 ]  {$q$};
% Text Node
\draw (432,141.4) node [anchor=north west][inner sep=0.75pt]  [color=sidsMaroon  ,opacity=1 ]  {$\overline{q}$};
% Text Node
\draw (264,175.4)
  node[anchor=center,align=center,inner sep=0.75pt] {
    $\mathcal{C}_{\text{out}}$\\
    $(E<\mu)$
  };
% Text Node
\draw (249,115.4) node [anchor=north west][inner sep=0.75pt]    {$\mathcal{C}_{\text{in}}$};
% Text Node
\draw (433,74.4) node [anchor=north west][inner sep=0.75pt]    {$S^{2}$};
% Text Node
\draw (129,200.94) node [anchor=north west][inner sep=0.75pt]  [color=Orange  ,opacity=1 ]  {\huge $\times $};
% Text Node
\draw (210,165.94) node [anchor=north west][inner sep=0.75pt]  [color=Orange  ,opacity=1 ]  {\huge $\times $};
% Text Node
\draw (328,173.94) node [anchor=north west][inner sep=0.75pt]  [color=Orange  ,opacity=1 ]  {\huge $\times $};
% Text Node
\draw (375,200.4) node [anchor=north west][inner sep=0.75pt]  [color=Orange  ,opacity=1 ]  {\huge $\times$};
\end{tikzpicture}
    \caption{Back-to-back dijet production (e.g., in the center‑of‑mass frame of $e^+e^-\to\gamma^*\to q\bar q$) provides the canonical setup for the emergence of non‑global logarithms, which are resummed by the BMS equation discussed in the text. One considers the probability to \emph{not} radiate quanta of energy larger than a veto scale $\mu$ in the exclusion region $\mathcal{C}_{\mathrm{out}}$ of the detector sphere $S^2$. Its complement, $\mathcal{C}_{\mathrm{in}} = S^2\setminus\mathcal{C}_{\mathrm{out}}$, denotes the allowed region. Forbidden gluons (subsequent emissions included) are marked with a ``$\times$.''}
    \label{fig:NGL}
\end{figure}
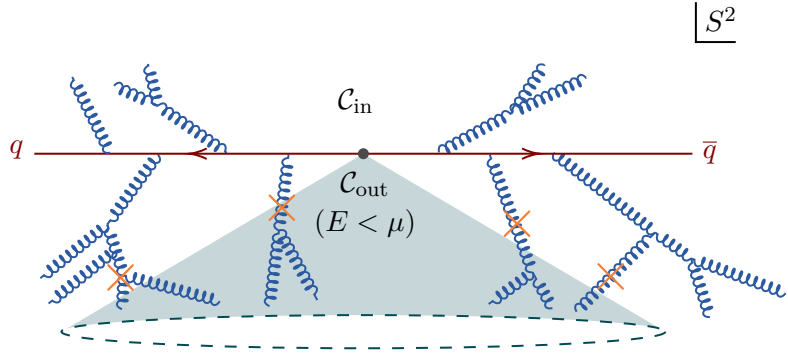

NGLs control the probability that a scattering process does \emph{not} radiate outside a fixed allowed angular region $\mathcal{C}_{\text{in}}$, up to some veto energy $\mu$. (That is, only quanta with energy less than $\mu$ are allowed in the excluded region $\mathcal{C}_{\text{out}}=S^2\setminus \mathcal{C}_{\text{in}}$.)
For example, in a $e^+e^-\to \gamma^* \to$ dijet process, $\mathcal{C}_{\text{out}}$ could be defined as any region not intersecting the jets; see Fig.~\ref{fig:NGL} \cite{Hatta:2017fwr}.

The presence of a large hierarchy $\mu\ll Q$, where $Q$ is the overall center-of-mass energy, suppresses the cross section by large logarithms $\log(Q/\mu)$.
In the planar (large-$N$) limit, these can be resummed by the
\emph{Banfi--Marchesini--Smye} (BMS) evolution equation, given at leading order in the strong-coupling as \cite{Banfi:2002hw}:
\begin{equation}
  -\mu\partial_\mu \,
  \sigma_{ik}  \;=\;
  a
  \int \frac{\d\Omega_j}{4\pi}\,
 H^{(1)}_{[ijk],\rm NGL}
  \Big[\sigma_{ij}\,\sigma_{jk}\,\theta(j\in {\mathcal{C}_{\mathrm{in}}})-\sigma_{ik}\Big]\,.
  \label{eq:BMS}
\end{equation}
In this equation, $\sigma_{ab}$ is interpreted as the probability that a parent dipole formed by partons $a$ and $b$ (propagating along the null directions $n_a$ and $n_b$) does \emph{not} radiate more than energy $\mu$ outside $\mathcal{C}_{\text{in}}$.
The initiating dijet (e.g., the $\gamma^*\to q\bar q$ dipole) corresponds to $a=i$ and $b=k$, while $j$ represents a daughter parton.

The BMS equation describes the physics underlying the fact that, as one lowers the veto energy $\mu$, more and more quanta radiated in the allowed region (in which $\theta(j\in {\mathcal{C}_{\mathrm{in}}})=1$) can act as a potential source of radiation. The equation is nonlinear because we are interested in the probability that \emph{none} of these sources radiate into the excluded region. 
Because emissions are integrated only over $\mathcal{C}_{\mathrm{in}}$ and not throughout the whole sphere $S^2$, the resulting logarithms are called ``non-global.''

The kernel, namely
\begin{equation}
 H^{(1)}_{[ijk],\rm NGL}
  = \frac{1 - n_i\!\cdot\!n_k}
         {(1 - n_i\!\cdot\!n_j)\,(1 - n_j\!\cdot\!n_k)}=
        \frac{2\alpha_{ik}}{\alpha_{ij}\alpha_{jk}}\quad\mbox{with}\quad 
\alpha_{ij}=\frac{1 - \cos\theta_{ij}}{2}\,,
\end{equation}
has a simple physical interpretation: it encodes the differential probability of \emph{timelike} wide‐angle soft emissions from a parent dipole $\sigma_{ik}$ in the direction $n_j$. The way that the $\sigma$'s arrange in the square brackets of \eqref{eq:BMS} is also easy to understand physically: according to the KLN theorem \cite{Kinoshita:1962ur,Lee:1964is}, infrared singularities cancel in a fully inclusive cross section; real and virtual contributions are precisely balanced. Thus, for $\mathcal{C}_{\text{in}}=S^2$, the choice $\sigma_{ij}=1$ must remain a fixed point of the evolution.

As explained in \cite{Caron-Huot:2016tzz}, to extend \eqref{eq:BMS} beyond the leading order and planar limits it is useful to consider a ``weighted cross section,'' where the wavefunction of each colored final-state parton is multiplied by a color rotation $V(n_i)^{a}_b$ inserted between the amplitude and the complex conjugate amplitude. The NGL problem is recovered formally by setting $V(n_i)\mapsto 0$ for $n_i\in \mathcal{C}_{\text{out}}$, but since soft radiation is sensitive to color, the more general weighted cross sections offer a natural bookkeeping device to track the color connections between the radiated partons. In the planar limit, these color factors become products of dipoles $\Vv_{ij}\simeq \sigma_{ij}$.\footnote{For clarity, we use different letters to denote the dipoles $\Uu_{ij}$ and $\Vv_{ij}$ entering the forward scattering amplitude and weighted cross sections.}

The leading-order evolution equation for the dressed cross section in the planar limit is thus given by \eqref{eq:BMS} with $\sigma_{ij}\mapsto \Uu_{ij}$ and the step function set to unity \cite{Caron-Huot:2015bja}. It becomes extremely reminiscent of the BK rapidity evolution encountered earlier in \eqref{eq:BK0}, which we recall here for convenience: 
\begin{subequations}
    \begin{align}
\label{eq:BMS dipole}
  -\mu\partial_\mu\,\Vv_{12} &=
a \int \frac{\d^2\Omega_0}{4\pi}\,
  H_{[102],\rm NGL}^{(1)}
  \Big[\Vv_{10}\Vv_{02}-\Vv_{12}\Big]\,,
\\
  \nu\partial_\nu\, \Uu_{12}&=
 a
  \int \frac{\d^2 z_0}{\pi}\,
  H_{[102],\rm BK}^{(1)}
  \Big[\Uu_{10}\Uu_{02}-\Uu_{12}\Big]\,,
  \label{eq:BK for NGL}
\end{align}
\end{subequations}
where, as before, $H_{[ijk],\rm BK}^{(1)}=\frac{2 z_{ik}^2}{z_{ij}^2\,z_{jk}^2}$.\footnote{In this section we include a subscript in $H_{\rm BK}$ to help distinguish the BK and BMS equations; any $H$ without a subscript should be understood to refer to the BK case, as in the rest of this paper.}
Similarly to the case of the BMS equation, one can interpret the dipole amplitude ${\cal U}_{ij}$ as the probability that ``something does not happen,’’ namely the probability that a highly boosted dipole of transverse size $z_{ab}$ traverses the target unscathed \cite{Marchesini:2015ica,Mueller:2018llt}.
The equation \eqref{eq:BK for NGL}, as reviewed in Sec.~\ref{sec:heuristic}, thus captures the physics underlying the fact that any virtual (or \emph{spacelike}) parton in the projectile's wavefunction can potentially be disrupted by the target.

The similarity between \eqref{eq:BMS dipole} and \eqref{eq:BK for NGL}, made evident by their juxtaposition, is not coincidental; in fact, it precisely exemplifies the \emph{spacelike-timelike correspondence}, which we discuss next.

\subsection{The correspondence}\label{ssec:correspondence}

The relationship between rapidity evolution in high-energy forward scattering, and the energy evolution of weighted cross section, in \eqref{eq:BMS dipole} and \eqref{eq:BK for NGL}, originates in the (approximate) conformal symmetry of perturbative QCD.  Here, we formally explain this statement and stress its predictions for the actual non-conformal theory of perturbative QCD.

The key idea is that, in a conformal theory, null infinity (where the timelike measurements of the weighted cross sections are performed) is equivalent to any other null plane. This is realized by the change of coordinates between $(x^+,x^-,z^i)$ and $(y^+,y^-,y_\perp^i)$ as \cite{Hofman:2008ar,Hatta:2008st}
\begin{equation} \label{x from y}
    x^+ = -\frac{\ell^2}{y^+}\,, \quad x^- = y^--\frac{y_\perp^2}{y^+}\,,\quad z^i = \ell\frac{y^i}{y^+}\,,
\end{equation}
where we denote $x^\pm= x^0\pm x^1$, $y^\pm=y^0\pm y^1$
and Latin indices are transverse coordinates, and $\ell$ is some fixed (but arbitrary) length scale introduced to make the units match, but which will not play a role below. This is a ``conformal transformation'' because it takes the Minkowski metric to a multiple of itself:
\begin{equation}
    -\d y^-\d y^+ + \d y_\perp^i \d y_\perp^i = \frac{-\d x^- \d x^+ +\d z^i \d z^i}{(x^+/\ell)^2}\,.
\end{equation}
Let us first pretend that the theory of interest is conformally invariant, in which case the rescaling by $(x^+/\ell)^2$ does not change the physics, and work out how the weighted cross section in $y$-space maps to a diffraction problem in $x$-space.
According to \eqref{x from y}, a parton moving along the null direction $(n^0,\vec{n})=(1,\hat{n})$, with $\hat{n}=(n^1,n^i)$ a unit vector,
will reach null infinity in $y^\mu$ space
at a position that maps to a specific location in the $x^\mu$ transverse plane:
\begin{equation}\label{eq:proj}
y^\mu\propto n^\mu \quad\Rightarrow\quad
x^\pm\to 0 \quad \mbox{ and }\quad z^i \to \ell \frac{n^i}{1+n^1} \quad 
\mbox{ as } \quad y^+\to\infty\,.
\end{equation}
Geometrically, this means that the sphere at infinity is stereographically mapped to the $x$ transverse plane; see Fig.~\ref{fig:tsc}. 
\definecolor{asparagus}{rgb}{0.53, 0.66, 0.42}
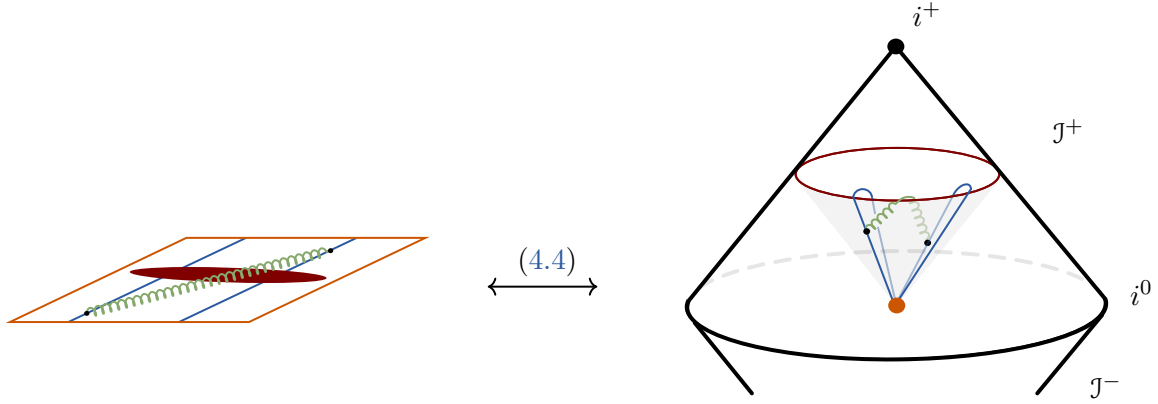
\begin{figure}
    \centering
\adjustbox{valign=c,scale=1}{
\tikzset{every picture/.style={line width=0.75pt}} 
\begin{tikzpicture}[x=0.75pt,y=0.75pt,yscale=-1.3,xscale=1.4]
\begin{scope}[xshift=-3cm,scale=1]
\begin{scope}[xshift=0.14cm,yshift=-0.52cm]
%bbb
\draw [color=sidsblue  ,draw opacity=0.4 ]   (532,162) -- (533,167.67) ;
\draw [color=sidsblue  ,draw opacity=0.4 ]   (534,172.67) -- (540.73,203.28) ;
%Straight Lines
\draw[color=asparagus  ,draw opacity=0.4 ][decorate, decoration={coil, segment length=3pt, amplitude=1.5pt}]  (545.85,160.06) -- (552,179.33);
\end{scope}
\begin{scope}[xshift=0.03cm,yshift=-0.27cm]
%Shape: Circle
\draw  [draw opacity=0][fill=black  ,fill opacity=1 ] (557.23,169.33) .. controls (557.23,168.65) and (556.68,168.1) .. (556,168.1) .. controls (555.32,168.1) and (554.77,168.65) .. (554.77,169.33) .. controls (554.77,170.02) and (555.32,170.57) .. (556,170.57) .. controls (556.68,170.57) and (557.23,170.02) .. (557.23,169.33) -- cycle ;
\end{scope}
%Shape: Arc
\draw  [draw opacity=0] (618.7,177.68) .. controls (619.93,179.15) and (620.59,180.69) .. (620.61,182.28) .. controls (620.73,193.77) and (587.21,203.45) .. (545.74,203.89) .. controls (504.27,204.33) and (470.56,195.37) .. (470.44,183.88) .. controls (470.43,182.92) and (470.65,181.97) .. (471.09,181.04) -- (545.52,183.08) -- cycle ; \draw   (618.7,177.68) .. controls (619.93,179.15) and (620.59,180.69) .. (620.61,182.28) .. controls (620.73,193.77) and (587.21,203.45) .. (545.74,203.89) .. controls (504.27,204.33) and (470.56,195.37) .. (470.44,183.88) .. controls (470.43,182.92) and (470.65,181.97) .. (471.09,181.04) ;  
%Shape: Arc
\draw  [draw opacity=0][line width=1.5]  (620.91,180.88) .. controls (621.06,181.41) and (621.14,181.94) .. (621.14,182.48) .. controls (621.27,193.97) and (587.75,203.64) .. (546.28,204.09) .. controls (504.81,204.53) and (471.09,195.57) .. (470.97,184.08) .. controls (470.96,183.11) and (471.18,182.17) .. (471.63,181.24) -- (546.06,183.28) -- cycle ; \draw  [color=black  ,draw opacity=1 ][line width=1.5][line cap=round]  (620.91,180.88) .. controls (621.06,181.41) and (621.14,181.94) .. (621.14,182.48) .. controls (621.27,193.97) and (587.75,203.64) .. (546.28,204.09) .. controls (504.81,204.53) and (471.09,195.57) .. (470.97,184.08) .. controls (470.96,183.11) and (471.18,182.17) .. (471.63,181.24) ;  
%Shape: Arc
\draw  [draw opacity=0][dash pattern={on 5.63pt off 4.5pt}][line width=1.5]  (475.23,176.43) .. controls (485.91,168.48) and (513.22,162.61) .. (545.3,162.27) .. controls (581.76,161.88) and (612.24,168.76) .. (619.14,178.24) -- (545.52,183.08) -- cycle ; \draw  [color=black  ,draw opacity=0.1 ][dash pattern={on 5.63pt off 4.5pt}][line cap=round][line width=1.5]  (475.23,176.43) .. controls (485.91,168.48) and (513.22,162.61) .. (545.3,162.27) .. controls (581.76,161.88) and (612.24,168.76) .. (619.14,178.24) ;  
%Shape: Triangle
\draw  [draw opacity=0][fill=black  ,fill opacity=0.05 ] (545.95,183.41) -- (508,132.08) -- (585,132.08) -- cycle ;
%Shape: Chord rep
\draw  [color=sidsMaroon  ,draw opacity=1 ][fill={rgb, 255:red, 255; green, 255; blue, 255 }  ,fill opacity=1 ][line width=0.75]  (509.82,132.6) .. controls (509.82,126.98) and (526.16,122.43) .. (546.31,122.43) .. controls (566.47,122.43) and (582.81,126.98) .. (582.81,132.6) .. controls (582.81,138.21) and (566.47,142.76) .. (546.31,142.76) .. controls (526.16,142.76) and (509.82,138.21) .. (509.82,132.6) -- cycle ;
%Shape: Arc
\draw  [draw opacity=0] (572.67,138.33) .. controls (572.64,136.51) and (571.15,135.95) .. (569.3,137.08) .. controls (568.05,137.86) and (566.95,139.22) .. (566.37,140.6) -- (569.3,140.45) -- cycle ; \draw  [color=sidsblue  ,draw opacity=1 ][line cap=round] (572.67,138.33) .. controls (572.64,136.51) and (571.15,135.95) .. (569.3,137.08) .. controls (568.05,137.86) and (566.95,139.22) .. (566.37,140.6) ;  
%Straight Lines
\draw [color=sidsblue  ,draw opacity=0.4 ][line cap=round]   (545.52,183.08) -- (566.37,140.6) ;
%Straight Lines
\draw [line width=1.5][line cap=round]    (494.15,217.2) -- (473.67,190) ;
%Straight Lines
\draw [line width=1.5][line cap=round]    (596.44,217.2) -- (619.67,187) ;
%Shape: Ellipse
\draw  [color=sidsMaroon  ,draw opacity=1 ][line width=0.75]  (509.82,132.6) .. controls (509.82,126.98) and (526.16,122.43) .. (546.31,122.43) .. controls (566.47,122.43) and (582.81,126.98) .. (582.81,132.6) .. controls (582.81,138.21) and (566.47,142.76) .. (546.31,142.76) .. controls (526.16,142.76) and (509.82,138.21) .. (509.82,132.6) -- cycle ;
%Straight Lines
\draw [color=sidsblue  ,draw opacity=1 ][line cap=round]   (572.67,138.33) -- (545.37,183.41) ;
%Straight Lines
\draw [color=sidsblue  ,draw opacity=1 ][line cap=round]   (530.49,141.38) -- (545.37,183.41) ;
%Shape: Ellipse
\draw  [draw opacity=0][fill=sidsOrange  ,fill opacity=1 ] (543.01,183.28) .. controls (543.01,181.59) and (544.37,180.22) .. (546.06,180.22) .. controls (547.74,180.22) and (549.11,181.59) .. (549.11,183.28) .. controls (549.11,184.96) and (547.74,186.33) .. (546.06,186.33) .. controls (544.37,186.33) and (543.01,184.96) .. (543.01,183.28) -- cycle ;
%Shape: Ellipse
\draw  [draw opacity=0][fill=black  ,fill opacity=1 ] (542.7,83.28) .. controls (542.7,81.6) and (544.06,80.23) .. (545.75,80.23) .. controls (547.43,80.23) and (548.8,81.6) .. (548.8,83.28) .. controls (548.8,84.97) and (547.43,86.33) .. (545.75,86.33) .. controls (544.06,86.33) and (542.7,84.97) .. (542.7,83.28) -- cycle ;
%Straight Lines
\draw [line width=1.5]    (471.63,181.24) -- (545.75,83.28) ;
%Straight Lines
\draw [line width=1.5]    (620.91,180.88) -- (545.75,83.28) ;
% Text Node
\draw (614.38,208.75) node [anchor=north west][inner sep=0.75pt]    {$\mathscr I^-$};
% Text Node
\draw (600.9,110.18) node [anchor=north west][inner sep=0.75pt]    {$\mathscr I^+$};
% Text Node
\draw (550.66,65.03) node [anchor=north west][inner sep=0.75pt]    {$i^{+}$};
% Text Node
\draw (628.66,172.4) node [anchor=north west][inner sep=0.75pt]    {$i^{0}$};
%NEW
\begin{scope}[xshift=0.14cm,yshift=-0.52cm]
%Straight Lines
\draw [color=sidsblue  ,draw opacity=1](531.89,161.3) -- (532,161.9) ;  
%Shape: Arc
\draw  [draw opacity=0] (531.89,161.35) .. controls (531.84,159.53) and (530.36,158.08) .. (528.53,158.08) .. controls (526.69,158.08) and (525.2,159.55) .. (525.17,161.38) -- (528.53,161.44) -- cycle ; \draw  [color=sidsblue  ,draw opacity=1 ] (531.89,161.35) .. controls (531.84,159.53) and (530.36,158.08) .. (528.53,158.08) .. controls (526.69,158.08) and (525.2,159.55) .. (525.17,161.38) ;  
\end{scope}
\begin{scope}[xshift=0.15cm,yshift=-0.52cm]
%Straight Lines
\draw [color=asparagus  ,draw opacity=1][decorate, decoration={coil, segment length=3pt, amplitude=1.5pt}]   (529.6,174.4) -- (540.99,162.76) ;
%Shape: Arc 
\draw  [draw opacity=0] (546.74,161.58) .. controls (546.58,161.48) and (546.4,161.4) .. (546.2,161.32) .. controls (544.42,160.65) and (542.1,161.29) .. (540.99,162.76) -- (544.22,164.04) -- cycle ; \draw  [color=asparagus ,draw opacity=1 ][line cap=round] (546.74,161.58) .. controls (546.58,161.48) and (546.4,161.4) .. (546.2,161.32) .. controls (544.42,160.65) and (542.1,161.29) .. (540.99,162.76) ;  
%Shape: Circle
\draw  [draw opacity=0][fill=black  ,fill opacity=1 ] (530.84,174.4) .. controls (530.84,173.72) and (530.29,173.16) .. (529.6,173.16) .. controls (528.92,173.16) and (528.37,173.72) .. (528.37,174.4) .. controls (528.37,175.08) and (528.92,175.63) .. (529.6,175.63) .. controls (530.29,175.63) and (530.84,175.08) .. (530.84,174.4) -- cycle ;
\end{scope}
\begin{scope}[xshift=2.7cm]
%Straight Lines 
\draw[<->] 
(297,176) -- (336.33,176)
node[midway,above][align=center]{
~~\eqref{x from y}
};
\end{scope}
\end{scope}
%left part
\begin{scope}[xshift=2cm,yshift=1.35cm,scale=0.65]
%Straight Lines 
\draw [color=sidsblue  ,draw opacity=1 ]   (91.42,213) -- (189.42,163) ;
%Straight Lines
\draw [color=sidsblue  ,draw opacity=1 ]   (152.42,213) -- (250.42,163) ;
%Shape: Ellipse
\draw  [draw opacity=0][fill=sidsMaroon  ,fill opacity=1 ] (129.39,185.06) .. controls (117.98,182.44) and (131.18,180.31) .. (158.89,180.31) .. controls (186.6,180.31) and (218.31,182.44) .. (229.72,185.06) .. controls (241.14,187.68) and (227.93,189.81) .. (200.22,189.81) .. controls (172.52,189.81) and (140.8,187.68) .. (129.39,185.06) -- cycle ;
%Straight Lines
\draw [color=asparagus  ,draw opacity=1 ][decorate, decoration={coil, segment length=3pt, amplitude=1.7pt}]   (101.59,207.84) -- (236.04,170.61) ;
%Shape: Circle
\draw  [draw opacity=0][fill=black  ,fill opacity=1 ] (103.06,207.84) .. controls (103.06,207.03) and (102.4,206.37) .. (101.59,206.37) .. controls (100.78,206.37) and (100.13,207.03) .. (100.13,207.84) .. controls (100.13,208.65) and (100.78,209.3) .. (101.59,209.3) .. controls (102.4,209.3) and (103.06,208.65) .. (103.06,207.84) -- cycle ;
%Shape: Circle
\draw  [draw opacity=0][fill=black  ,fill opacity=1 ] (237.5,170.61) .. controls (237.5,169.8) and (236.85,169.14) .. (236.04,169.14) .. controls (235.23,169.14) and (234.57,169.8) .. (234.57,170.61) .. controls (234.57,171.42) and (235.23,172.08) .. (236.04,172.08) .. controls (236.85,172.08) and (237.5,171.42) .. (237.5,170.61) -- cycle ;
\begin{scope}[xshift=0.45cm]
\draw  [color=sidsOrange  ,draw opacity=1 ] (139.42,163) -- (271.26,163) -- (173.97,213) -- (42.13,213) -- cycle ;
\end{scope}
\end{scope}
\end{tikzpicture}
}
    \caption{The coordinate change \eqref{x from y} maps the shock (red blob) in the left panel to the detector (red circle) on the celestial sphere at null infinity $\mathscr I^+$ in the right panel, thereby interchanging $H_{\rm BK}(\varepsilon_*)$ with $H_{\rm NGL}$. Both panels depict the same sample diagram with a radiated gluon.  Note that the two sides of the shock get ``folded'' onto the same cone, on which they represent respectively the amplitude and complex conjugate amplitude; we depict the latter with shaded lines.}
    \label{fig:tsc}
\end{figure}
Under this map, the angular measure and separations are easily seen to transform as
\begin{equation}
  \frac{\d^\Dperp\Omega_0}{(4\pi)^{\Dperp/2}} =\frac{\,\d^{\Dperp} z_0}{\pi^{\Dperp/2}}\frac{1}{(\ell+ |z_0|^2/\ell)^\Dperp}\,,\quad
\alpha_{ij} = \frac{z_{ij}^2}{(\ell+|z_i|^2/\ell)(\ell+|z_j|^2/\ell)}\,.
\end{equation}
All $\ell$-dependent factors nicely cancel for integrands that have the correct homogeneity degree
($\sim \d^\Dperp\Omega_0/\alpha_{i0}^{\Dperp}$)
to be Lorentz-invariant, such as the BMS kernel \eqref{eq:BMS}, which then gets precisely mapped to the BK kernel:
\begin{equation}
    \frac{\d^2\Omega_0}{4\pi} H^{(1)\rm NGL}_{[102]}
= \frac{\d^2\Omega_0}{4\pi} \frac{2\alpha_{12}}{\alpha_{10}\alpha_{02}}
\quad=\quad
\frac{\d^2z_0}{\pi} \frac{2z_{12}^2}{z_{10}^2z_{02}^2}
= \frac{\d^2z_0}{\pi} H^{(1,0)}_{[102]}\,.
\end{equation}
The conformal transformation \eqref{x from y}, together with the observation that perturbative QCD is conformal up to leading order (where the running of the coupling can be neglected), thus explains the precise match of \eqref{eq:BMS dipole} and \eqref{eq:BK for NGL}.
The definition of ``weighted cross sections'' in \cite{Caron-Huot:2015bja}, essentially
as a shockwave at infinity, was designed to map to a conventional shockwave at $x^+=0$ under this transformation.

\paragraph{What about higher loops in QCD?}
A crucial observation, first emphasized in \cite{Vladimirov:2016dll} to our knowledge, is that
rapidity divergences exist in any dimension, and therefore the corresponding rapidity anomalous dimensions are a nontrivial function of $\varepsilon$.
On the other hand, the non-global logarithm evolution equation, like other anomalous dimensions, does \emph{not} depend on the dimensional regularization parameter $\varepsilon$ in a minimal subtraction scheme.
Thus, the evolution equations admit, respectively, double and single expansions:
\begin{equation}
\nu\partial_\nu\,\Uu_{12} =\sum_{\ell=1}^\infty\sum_{k=0}^\infty a ^{\ell}\varepsilon^k H_{\rm BK}^{(\ell,k)}
\quad \mbox{ and }\quad
-\mu\partial_\mu\,\Vv_{12}=
\sum_{\ell=1}^\infty a ^{\ell} H^{\overline{\text{MS}}\,(\ell)}_{\rm NGL}\,.\label{eq:NGLexp2}
\end{equation}
The power of this observation lies in the fact that the two equations must coincide in the conformal dimension (Wilson--Fisher point; see the discussion around \eqref{eq:WF_point_def}):
\begin{equation}\label{eq:conjecture}
    H_{\rm BK}(\varepsilon=\varepsilon_{*}=
    -\bar{\beta}(\alpha_s))\simeq H_{\rm NGL}^{\overline{\text{MS}}}\,,
\end{equation}
which is represented pictorially in Fig.\,\ref{fig:tsc}.
Here, the symbol $\simeq$ is used to mean equal under
the stereographic projection \eqref{eq:proj}, $\Uu\leftrightarrow\Vv$, \emph{and} up to possible (but physically irrelevant) terms that depend on the choice of a subtraction scheme.

Expanding \eqref{eq:conjecture} to the second order in the gauge coupling $a$ gives:
\begin{equation}\label{eq:stc-2loop} H_{\rm BK}^{(2,0)}-
b_0
H_{\text{BK}}^{(1,1)}\simeq H_{\text{NGL}}^{\overline{\text{MS}}\,(2)}\,.
\end{equation}
It was initially observed by one of the present authors
that $H_{\rm BK}^{(2,0)}-H_{\text{NGL}}^{\overline{\text{MS}}\,(2)}$ is relatively simple and proportional to $b_0$ \cite[Sec.~4.2]{Caron-Huot:2015bja}; the precise relation \eqref{eq:stc-2loop} was recently verified in \cite{Balitsky:2024xvi}.

\subsubsection{Prediction: the non-conformal part of three-loop BK evolution}

At the next order, the same relation \eqref{eq:conjecture} implies that:
\begin{equation}\label{eq:conjecture_exp}
H_{\text{BK}}^{(3,0)}-
             b_0
             H_{\text{BK}}^{(2,1)}-
             b_1 H_{\text{BK}}^{(1,1)}+
             b_0^2 H_{\text{BK}}^{(1,2)}\simeq H_{\text{NGL}}^{\overline{\text{MS}}\,(3)}\,.
\end{equation}
A similar formula was used in \cite{Vladimirov:2016dll,Vladimirov:2017ksc} to extract the three‑loop rapidity anomalous dimension for null parallel \emph{semi-infinite} Wilson lines (as opposed to the Wilson lines \eqref{eq:WL0} extending from past to future infinity used in our context).

The relation \eqref{eq:conjecture_exp} constitutes the main physical motivation for the calculation of $H_{\text{BK}}^{(2,1)}$ (and of  $H_{\text{BK}}^{(1,1)}$, $H_{\text{BK}}^{(1,2)}$) performed in this paper, and whose result was recorded above in \eqref{BC result}. Thanks to this relation, knowledge of $H_{\text{BK}}^{(3,0)}$ in QCD becomes equivalent to knowledge of $H_{\text{NGL}}^{\overline{\text{MS}}\,(3)}$ in the same theory. To be fully explicit:
\begin{equation}
\label{eq:resultthree-loop}
\begin{split}
H_{\mathrm{BK}}^{(3,0)}\simeq H_{\text{NGL}}^{\overline{\text{MS}}\,(3)}&+\int \frac{\d^2z_0}{\pi}\frac{z_{12}^2}{z_{10}^2z_{02}^2}(\Uu_{10}\Uu_{02}-\Uu_{12})\Big[b_0 \Tilde{H}_{[102]}^{(2,1)}+b_1\Tilde{H}_{[102]}^{(1,1)}-b_0^2\Tilde{H}_{[102]}^{(1,2)}\Big]
    \\&    \phantom{a}+b_0\int\frac{\d^2z_0\,\d^2z_{0'}}{\pi^2}\frac{z_{12}^2}{z_{10}^2z_{00'}^2z_{0'2}^2}
    \left[
\begin{array}{rl}
&(\Uu_{10}\Uu_{00'}\Uu_{0'2}-\Uu_{12})\Tilde{H}_{[100'2]}^{(2,1)}
\\&
-(\Uu_{10}\Uu_{02}-\Uu_{12})\Tilde{H}_{[100'2]}^{(2,1)\mathrm{coll}}
\end{array}
\right]\,.
\end{split}
\end{equation}
The one-loop single-emission kernels $\tilde{H}_{[102]}^{(1,k)}$ (for $k=1,2$) can be found in \eqref{eq:H1_102exp}, and the two-loop kernel $\tilde{H}_{[102]}^{(2,1)}$ is given in \eqref{rv result 2}. The corresponding double-emission kernel $\tilde{H}_{[100'2]}^{(2,1)}$ is obtained from the coefficient of $\varepsilon$ in \eqref{double result susy}-\eqref{double-real N=4}, and the collinear subtraction term $\tilde{H}_{[100'2]}^{(2,1)\mathrm{coll}}$ can similarly be read off from \eqref{H2coll expanded}.
All of these ingredients are provided in Wolfram Language syntax in the ancillary file accompanying the arXiv submission of this paper.

\paragraph{What is known about  $H_{\text{NGL}}^{\overline{\text{MS}}\,(3)}$?}
A general fact is that it can only involve conformally invariant functions. This is because the conformal symmetry of the BK transverse plane maps to the Lorentz symmetry of the NGL calculation, which is preserved by dimensional regularization. Hence, \eqref{eq:resultthree-loop} predicts all non-conformal terms in the three-loop BK evolution.

Another relevant fact is that calculations of the infrared divergences leading to $H_{\text{NGL}}$ tend to be technically simpler because they are regulated by standard dimensional regularization, and the factorization of amplitudes in soft limits is well understood.
For example, the calculation of the two-loop BMS equation generalized to full color in \cite{Caron-Huot:2015bja} is arguably technically simpler than that of the corresponding rapidity evolution equation  \cite{Balitsky:2013fea,Kovner:2013ona,Kovner:2014lca}, even though the final results differ only by simple terms proportional to $b_0$ \cite{Caron-Huot:2015bja} (which, with hindsight, we now expect to be explained by \eqref{eq:stc-2loop}\footnote{While we find that \eqref{eq:stc-2loop} fully explains the first line of (4.6) in \cite{Caron-Huot:2015bja}, the second line there produces an extra phase that can be nonzero for sufficiently complicated color structures and whose relation to \eqref{eq:stc-2loop} remains to be explained.}).

The three-loop version of $H_{\text{NGL}}^{\overline{\text{MS}}\,(3)}$ in the planar limit of $\mathcal{N}=4$ supersymmetric Yang-Mills (SYM) theory was derived in \cite{Caron-Huot:2016tzz}.
This linearized eigenvalue derived from this result was nontrivially matched with the three-loop BFKL trajectory found using completely different integrability-based methods \cite{Gromov:2015vua}.
Subtracting this known quantity, we can thus express the missing term in \eqref{eq:resultthree-loop} as
\begin{equation} \label{QCD minus N4}
H_{\text{NGL}}^{\overline{\text{MS}}\,(3)}\Big|_{\rm planar\, QCD}
= H_{\text{NGL}}^{\overline{\text{MS}}\,(3)}\Big|_{\rm planar\, {\cal N}=4}+ \left(H_{\text{NGL}}^{\overline{\text{MS}}\,(3)}\Big|_{\rm planar\, QCD}-H_{\text{NGL}}^{\overline{\text{MS}}\,(3)}\Big|_{\rm planar\, {\cal N}=4}\right)\,,
\end{equation}
where the first term is given in \cite[Eq. (4.34c)]{Caron-Huot:2016tzz}.
The two theories, planar ${\cal N}=4$ SYM and QCD, share the same gluon interactions, and the difference is their matter content.  Thus, the parentheses only receive contributions from matter loops (possibly involving the Yukawa scalar-fermion-fermion vertex of ${\cal N}=4$ SYM). Hence, this formula is interesting because the most complicated and numerous diagrams, which contain only gluons, all reside in the first term.

In future work, our aim is to calculate the parentheses in \eqref{QCD minus N4} and thus assemble the complete three-loop nonlinear BK evolution equation in planar QCD. Meanwhile, we provide a nontrivial check of \eqref{eq:resultthree-loop} by explicit calculation in the large-flavor limit, where only diagrams with the maximal number of self-energy insertions contribute.

{\subsection{Check for three-loop bubble chain in large-\texorpdfstring{$n_f$}{dum2} theory}}
\label{sec:renormalon}

To test the spacelike-timelike correspondence just outlined, we consider the bubble chain (called ``renormalon chain'' in \cite{Balitsky:2007feb}, see also \cite{Balitsky:2006wa,Gardi:2006rp,Kovchegov:2006vj}), which controls the limit of large number flavors $\nf, \ns$. At three loops, we thus focus on diagrams that are quadratic in $\nf$, $\ns$.  We will treat both $\ns$ and $\nf$ as large to obtain a more stringent test, but we will simply refer to this as the ``large-$n_f$ limit.'' As a proof of principle, we concentrate on the double-real contributions.

The relevant three-loop diagrams contain either a self-energy or counter-term, giving the amplitude:
\begin{align}\label{eq:renAmp}
       \Aa_{[100'2]}^{(aa,1)}(\tau)& = \, \Aa_{[100'2]}^{(aa,1)}(\tau)_\Pi - \frac{1}{\varepsilon}b_0\big|_n
       \Aa_{[100'2]}^{(aa,0)}(\tau) \\
    &= \left[\adjustbox{valign=c}{\includegraphics[scale=0.4]{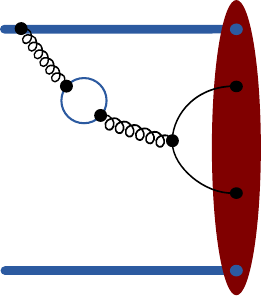}}\quad+\adjustbox{valign=c}{\includegraphics[scale=0.4]{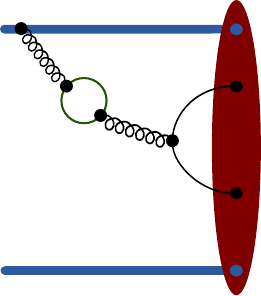}}\quad+\text{perms}\right] -\frac{b_0\big|_n}{\varepsilon }\left[\adjustbox{valign=c}{\includegraphics[scale=0.4]{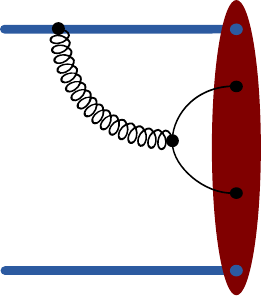}}\quad+\text{perms}\right]\notag
\end{align}
where, at large $n_f$, the particles that cross the shock are either a pair of scalars or fermions ($aa=ss$ or $ff$). The respective integrands $\Aa_{[100'2]}^{(aa,0)}(\tau)$ are given in \eqref{eq:dr_sf_amp}, and the self-energy correction inserts an extra factor of $(\mubar^2/(k_1+k_2)^2)^\varepsilon \Pi$:
\begin{subequations}
\label{large flavor bubble chain amplitudes}
    \begin{align}
\hspace{-0.7cm}
\label{large flavor bubble chain amplitude scalar}
 \Aa_{[100'2]}^{(ss,1)}(\tau)_\Pi
 &= \Pi\big|_n\int\limits_{k_1,k_2}
\left(\e^{ik_1{\cdot}z_{01}+ik_2{\cdot}z_{0'1}}
-(z_1{\mapsto}z_2)\right)
\frac{2\tau}{(1{+}\tau)}\frac{(k_2^2-k_1^2)\mubar^{2\varepsilon}\,}{(k_1^2{+}\tau k_2^2)(k_1{+}k_2)^{2+2\varepsilon}}\,, \\
\label{large flavor bubble chain amplitude fer}
 \Aa_{[100'2]}^{(ff,1)}(\tau)_\Pi
&= \Pi\big|_n \int\limits_{k_1,k_2}
\left(\e^{ik_1{\cdot}z_{01}+ik_2{\cdot}z_{0'1}}
-(z_1{\mapsto}z_2)\right)
\frac{2\sqrt{\tau}\, \kslash_1\kslash_2\,\mubar^{2\varepsilon}}{(k_1^2+\tau k_2^2)(k_1+k_2)^{2+2\varepsilon}}\,.
\end{align}
\end{subequations}
In the above the large-$n_f$ contributions to the self-energy \eqref{Pi function} and $\beta$-function are:
\begin{equation} \label{Pi largen}
    \Pi\big|_n =  \e^{\varepsilon\gamma_\text{E}}\frac{\Gamma(1+\varepsilon)\Gamma^2(1-\varepsilon)}{\Gamma(1-2\varepsilon)}
\left[\frac{-4\nf(1-\varepsilon)-\ns}{2\varepsilon(1-2\varepsilon)(3-2\varepsilon)}\right],
\qquad
    b_0\big|_n = -\frac{2}{3}\nf -\frac{1}{6}\ns.
\end{equation}

To perform the Fourier transform to impact parameter space, we can study the corresponding integral family using twisted cohomology \cite{Mastrolia:2018uzb,Frellesvig:2019kgj,Frellesvig:2019uqt,Frellesvig:2020qot,Chestnov:2022xsy,Fontana:2023amt,Brunello:2023rpq,Brunello:2024tqf}, using the same method as in Sec.~\ref{ssec:splitting amps}.
We find that the basic Fourier transform that appears in \eqref{large flavor bubble chain amplitudes} is given by
\begin{equation}
\frac{1}{\cd^2} \int\limits_{k_1,k_2}
\e^{ik_1{\cdot}z_{01}+ik_2{\cdot}z_{0'1}}
\frac{1}{(k_1^2{+}\tau k_2^2)(k_1{+}k_2)^{2+2\varepsilon}}
= \frac{\tau^{2\varepsilon}(z_{00'}^2)^{3\varepsilon}}{16(1{+}\tau)^{1+4\varepsilon}\varepsilon^2}
\left(F_2^\varepsilon[x]+\frac16 F_1^\varepsilon[x]\right)\,,
\end{equation}
with $x=\frac{(\tau z_{10}+ z_{10'})^2}{z_{00'}^2 \tau}$
and where the pure functions $F_i^\varepsilon$, generalizing \eqref{eq:LOFT} to this case,
are
\begin{equation}
    \begin{bmatrix} 
    F_1^\varepsilon \\ F_2^\varepsilon
    \end{bmatrix}
     =
     \frac{\Gamma(1-3\varepsilon)\Gamma(1-2\varepsilon)}{4^{1+\varepsilon}\Gamma(1-\varepsilon)^3}
     \begin{bmatrix}
     6  \ \ {-}2\\
     {-}1 \ \ 1
     \end{bmatrix}\cdot
     \begin{bmatrix} 
(1+x)\,{}_2F_1(1{-}3\varepsilon,1{-}2\varepsilon;1{-}\varepsilon;-x)\, \\
{}_2F_1({-}3\varepsilon,{-}2\varepsilon;1{-}\varepsilon;-x)  \end{bmatrix}.
\end{equation}
These satisfy the following system of canonical differential equations:
\begin {align}\label{F12ren diff eq}
 \frac {\d} {\d x}\begin {bmatrix}
F_ {1}^\varepsilon \\ F_ {2}^\varepsilon
 \end {bmatrix} = \varepsilon\begin {bmatrix}\frac {6} {1 + 
         x} & \frac{12}{1+x} \\\frac {-1} { 
       1+x} &\frac {1-x} {x(1+x)}\end {bmatrix}\cdot\begin {bmatrix}
F_ {1}^\varepsilon \\ F_{2}^\varepsilon
 \end {bmatrix}\,,
\end {align}
from which the $\varepsilon$ expansions can be readily derived, e,g.,
\begin{equation}\begin{aligned}
    F_1^\varepsilon & = (4\e^{-2\gamma_{\rm E}})^{-\varepsilon}
\bigl[1+6\varepsilon\log(1+x)+\varepsilon^2(5\zeta_2+12\log^2(1+x)+\mathcal{O}(\varepsilon^3)\bigr]\,,\nonumber \\
    F_2^\varepsilon & =
    (4\e^{-2\gamma_{\rm E}})^{-\varepsilon}\bigl[
    -\varepsilon\log(1+x)+\varepsilon^2(\text{Li}_2(-x)-2\log^2(1+x)+\mathcal{O}(\varepsilon^3)\bigr]\,.
\end{aligned}\end{equation}
Below, we only used the $\mathcal{O}(\varepsilon)$ terms of this expansion. 

Evaluating the numerator of \eqref{large flavor bubble chain amplitudes} as derivatives (see the discussion around \eqref{eq:forDer}), we then obtain the impact parameter space amplitudes: 
\begin{subequations}
\begin{align}
\label{ large flavor bubble chain scalar}
\Aa_{[100'2]}^{(ss,1)}(\tau)_\Pi  &= 
\frac{\cd^2\tau^{1+2\varepsilon}\mubar^{2\varepsilon}\Pi\big|_n}{(1+\tau)^{1+4\varepsilon}(z_{00'}^2)^{1-3\varepsilon}}
\bigg\{ \frac12\frac{z_{10}^2-z_{10'}^2}{\tau z_{10}^2+z_{10'}^2}F_1^\varepsilon\left[\frac{(\tau z_{10}+ z_{10'})^2}{z_{00'}^2 \tau}\right] +
\biggl( -2 + \frac{(\tau+1) z_{10}^2}{\tau z_{10}^2+  z_{10'}^2} \nonumber \\ 
&
+ \frac{(\tau +1)^2 z_{10}^2-z_{00'}^2}{(\tau +1) \left(\tau z_{10}^2+z_{10'}^2\right)-\tau 
   z_{00'}^2}\biggr)F_2^\varepsilon\left[\frac{(\tau z_{10}+ z_{10'})^2}{z_{00'}^2 \tau}\right]
   \bigg\}- (z_1 \leftrightarrow z_2)\,,
   \\
\label{large flavor bubble chain amplitude integrated}
\Aa_{[100'2]}^{(ff,1)}(\tau)_\Pi &= 
\frac{\cd^2\tau^{\frac12+2\varepsilon}\mubar^{2\varepsilon}\Pi\big|_n}{(1+\tau)^{1+4\varepsilon}(z_{00'}^2)^{1-3\varepsilon}}\biggl\{
  \frac12 \left( 1-\frac{(\tau +1) \zslash_{1 0}\zslash_{10'}}{\tau z_{10}^2+z_{10'}^2}\right)F_1^\varepsilon\left[ \frac{(\tau z_{10}+ z_{10'})^2}{z_{00'}^2 \tau}\right] \nonumber \\ 
 & \quad+\biggl(
   \frac{(\tau +1)\zslash_{1 0}\zslash_{10'}\left(\tau  z_{00'}^2-2 (\tau +1) \left(\tau z_{10}^2+  z_{10'}^2\right)\right)}{\left(\tau z_{10}^2+ 
   z_{10'}^2\right) \left((\tau +1) \left(\tau z_{10}^2+z_{10'}^2\right)-\tau  z_{00'}^2\right)} \\
   &
  \qquad + \frac{2 (\tau +1) \left(\tau z_{10}^2+  z_{10'}^2\right)-3 \tau  z_{00'}^2}{(\tau +1) \left(\tau z_{10}^2+ 
   z_{10'}^2\right)-\tau  z_{00'}^2}
  \biggr)F_2^\varepsilon\left[\frac{(\tau z_{10}+ z_{10'})^2}{z_{00'}^2 \tau}\right] \biggr\}- (z_1\leftrightarrow z_2)\,.\nonumber
\end{align}
\end{subequations}

As in Sec.~\ref{ex:BKmain}, we obtain the contribution to the evolution kernel by multiplying the amplitudes on the two sides of the shock to get an integrand ${\cal I}$, and integrating over relative energies: 
\begin{subequations}
    \begin{align} \label{I3 large n}
     \mathcal{I}^{(3)}_{[100'2]}(\tau)\bigg|_n &= 
     2\ns \mathcal{A}_{[100'2]}^{(ss,1)}(\tau)\mathcal{A}_{[100'2]}^{(ss,0)}(\tau)+
     2\nf \text{tr}_f[\mathcal{A}_{[100'2]}^{(ff,1)}(\tau)\mathcal{A}_{[100'2]}^{(ff,0)}(\tau)^\dagger]\,,
     \\
    H_{[100'2]}^{(3)}\bigg|_n &= \frac{(2\pi\mu^{2\varepsilon})^2}{\cd^2}
    \int_0^\infty \frac{\d\tau}{\tau} \mathcal{I}^{(3)}_{[100'2]}(\tau)\bigg|_n
    \equiv \left(\frac{z_{12}^2}{z_{10}^2z_{00'}^2z_{0'2}^2}\right)^{1-\varepsilon}
    \tilde{H}_{[100'2]}^{(3)}\bigg|_n  \,.
\end{align}
\end{subequations}
There are no rapidity subdivergences as $\tau\to 0,\infty$, and the integrand is finite as $\varepsilon\to 0$, so we can simply take the limit before integrating.
The integrals are quite formidable, but thankfully we find that these are almost completely equivalent to the ${\cal O}(\varepsilon)$ contribution to the two-loop kernels \eqref{double-real scalar}-\eqref{double-real N=1}!
In fact, we find that we can recognize and subtract what was computed earlier and are left with a simple (conformal) remainder:
\begin{align}
\label{large flavor bubble chain result}
\tilde{H}^{(3,0)}_{[100'2]}\bigg|_n & = \,  b_0\big|_n\times \left(
(\ns-2\nf)\tilde{H}^{(2,1){\cal N}=0}_{[100'2]}+
\nf \tilde{H}^{(2,1){\cal N}=1}_{[100'2]}\right) + \tilde{H}^{(3,0)}_{[100'2],\text{conf}}\,,
\end{align}
where:
\begin{align} \label{H3 conf result}
\tilde{H}^{(3,0)}_{[100'2],\text{conf}}
&=
\left(
\frac{4\nf+\ns}{6}\log \frac{z_{12}^2 z_{00'}^2}{z_{10}^2z_{0'2}^2}-\frac{20 \nf+8 \ns}{9}\right)\nonumber\\
    &\hspace{-18mm}{\times}\Bigg[
    (\ns{-}2\nf)
    \frac{z_{10}^2z_{20'}^2}{z_{12}^2z_{00'}^2}
    \left(\frac{z_{10}^2 z_{20'}^2+z_{10'}^2 z_{20}^2}{ z_{10}^2 z_{20'}^2- z_{10'}^2
   z_{20}^2}\log \frac{z_{10}^2 z_{20'}^2}{z_{10'}^2 z_{20}^2} {-}2\right){+}2\nf\frac{z_{10}^2 z_{20'}^2}{ z_{10}^2 z_{20'}^2- z_{10'}^2
   z_{20}^2}\log \frac{z_{10}^2 z_{20'}^2}{z_{10'}^2 z_{20}^2} \Bigg]
   \nonumber \\ &= \left(
b_0\big|_n\times \log\frac{v}{u}-\frac{20 \nf+8 \ns}{9}\right)
\times \tilde{H}^{(2,0)}_{[100'2]}\bigg|_{n}\,.
\end{align}
To write the second line, we have recognized that the square bracket is precisely equal to the matter contribution to the two-loop evolution in $\D=4$ (see \eqref{double real 4d}).

We can already test the spacelike-timelike correspondence 
between the BK and non-global logarithm Hamiltonians with what we have derived so far.
Indeed, taking the large-$n_f$ terms in \eqref{eq:conjecture_exp}, the correspondence predicts that 
\begin{align}
    \tilde{H}^{(3,0)}_{[100'2]}\bigg\vert_n - b_0\big\vert_n \times\tilde{H}^{(2,1)]}_{[100'2]}\bigg\vert_n =  \, \tilde{H}^{{\rm NGL}(3,0)}_{[100'2]}\bigg\vert_n\,.
\end{align}
Compared with the result of the direct calculation in \eqref{large flavor bubble chain result}, we see that the correspondence is valid provided that
\begin{equation}
 \tilde{H}^{(3,0)}_{[100'2],\text{conf}}\bigg\vert_n \stackrel{?}{=}
   \tilde{H}^{(3,0)}_{[100'2],\rm NGL}\bigg\vert_n\,. \label{correspondence check}
\end{equation}
Although we have not yet computed the right-hand side, it is highly nontrivial that the left-hand side, given in \eqref{H3 conf result}, should be conformal invariant (i.e., can be written as a function of the cross-ratios $u$ and $v$ in \eqref{cross-ratios}), which it indeed is!  In the following, we will further compute the right-hand side and precisely confirm this equality.

It should be noted that the two-loop $\mathcal{O}(\varepsilon)$ terms
in \eqref{H3 conf result} (namely $\tilde{H}^{(2,1){\cal N}=0}_{[100'2]}$
and $\tilde{H}^{(2,1){\cal N}=1}_{[100'2]}$ given in \eqref{double-real scalar} and \eqref{double-real N=1}), which together take about a page, predict the most massive chunk of \eqref{large flavor bubble chain result}; the remainder $\tilde{H}^{(3,0)}_{[100'2],\text{conf}}$ is a comparatively compact one-line expression.

\subsection{Explicit match with three-loop large-\texorpdfstring{$n_f$}{dum3} non-global logarithms}

Here, we independently compute $\tilde{H}^{(3,0)}_{[100'2],\rm NGL}$ appearing in \eqref{correspondence check} in the large-$n_f$ limit, using the methods of \cite{Caron-Huot:2015bja,Caron-Huot:2016tzz} designed for the NGL renormalization group equation (see \cite[Eq. (1.3)]{Caron-Huot:2016tzz}).
We focus on the double-real terms in the large-$n_f$ limit.
These come from the amplitude to emit two scalars or two fermions off parent $q\bar{q}$ pair going along momenta $p_1$ and $p_2$: 
\begin{equation}
    {\mathcal M}^{(aa)}(p_0,p_{0'})= 
 \adjustbox{valign=c}{\includegraphics[scale=0.5]{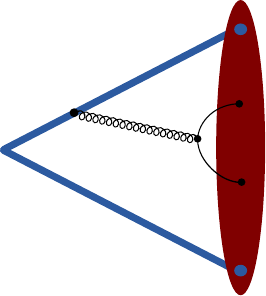}}
 \quad+\quad
    \adjustbox{valign=c}{\includegraphics[scale=0.5]
    {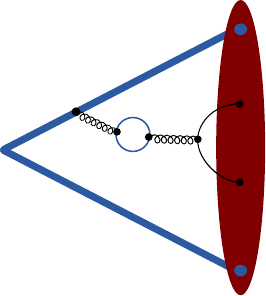}}
\quad+\mbox{perms}+\mbox{higher orders}.
\end{equation}
Explicitly,
\begin{subequations}\label{amps resummed}
    \begin{align} 
    {\mathcal M}^{(ss)}(p_0,p_{0'}) &=
    \frac{g_0^2}{1-a_0\Pi\big|_n (2p_0{\cdot}p_{0'}/\mubar^2)^{-\varepsilon}} \left(\frac{p_1^\mu}{p_1{\cdot} p_0}-\frac{p_2^\mu}{p_2{\cdot} p_0} \right) \frac{(p_0-p_{0'})_\mu}{2p_0{\cdot}p_{0'}}\,, \\
     {\mathcal M}^{(ff)}(p_0,p_{0'}) &=
    \frac{g_0^2}{1-a_0\Pi\big|_n (2p_0{\cdot}p_{0'}/\mubar^2)^{-\varepsilon}} \left(\frac{p_1^\mu}{p_1{\cdot} p_0}-\frac{p_2^\mu}{p_2{\cdot} p_0} \right) \frac{\bar{u}(p_{0'})\gamma_\mu u(p_0)}{2p_0{\cdot}p_{0'}}\,,
\end{align}
\end{subequations}
where all momenta are on-shell, $p_i^2=0$, and the bare coupling $g_0$ is as in \eqref{bare coupling}.
These scalar and fermion amplitudes have previously been used in the calculation of two-loop NGLs (see \cite[Eq. (3.5)]{Caron-Huot:2015bja}); the only difference here is that we have resummed the large-$n_f$ self-energy corrections to the gluon propagator, with $\Pi|_n$ given in \eqref{Pi largen}.

As reviewed in Sec.~\ref{ssec:NGL}, the evolution equation for NGLs can be obtained by looking at the infrared divergences of the cross section with the final states weighted by color operators $\Vv_{ab}$.
The contribution to the weighted cross section from the preceding amplitudes squared is: 
\begin{equation} \label{sigma NGL}
\hspace{-0.3cm}
    \sigma \supset \frac{N^2}{4}\int \frac{\d^{\D-1}p_0}{(2\pi)^{\D-1}2p_0^0} \frac{\d^{\D-1}p_{0'}}{(2\pi)^{\D-1}2p_{0'}^0}
\Vv_{10}\Vv_{00'}\Vv_{0'2}
\biggl[\ns \bigl|{\mathcal M}^{(ss)}\bigr|^2
+\nf {\rm tr}_f\bigl[{\mathcal M}^{(ff)}{\mathcal M}^{(ff)\dagger}\bigr]\biggr]\,.
\end{equation}

Following \cite{Caron-Huot:2016tzz}, we slice the phase space using the geometric means of energies measured in a Lorentz-covariant way (abbreviating $s_{ij}\equiv -2p_i{\cdot}p_j$):
\begin{equation}
    Q_{[102]}=\left(\frac{s_{10}s_{02}}{s_{12}}\right)^{1/2},
    \quad
    Q_{[100'2]}=\left(\frac{s_{10}s_{00'}s_{0'2}}{s_{12}}\right)^{1/4},\quad\ldots
\end{equation}
These are analogous to the conformally covariant longitudinal energies $\nu$ in \eqref{eq:conformalScheme}. Here, this is designed to maintain Lorentz symmetry of the NGL
evolution equation, rather than the conformal symmetry of the rapidity evolution equation, but note that these two symmetries are interchanged by the coordinate inversion \eqref{x from y}.
Changing variables to $Q\equiv Q_{[100'2]}$, energy fraction $\tau=p_0^0/p_{0'}^0$, and angles, \eqref{sigma NGL} becomes:
\begin{equation} \label{sigma NGL 2}
    \sigma \supset \frac{N^2/4}{(4\pi)^{4-4\varepsilon}}\int \frac{\d Q}{Q^{1+4\varepsilon}}
\frac{\d\tau}{\tau} \frac{\d\Omega_0}{4\pi} \frac{\d\Omega_{0'}}{4\pi}
\Vv_{10}\Vv_{00'}\Vv_{0'2}
\left(\frac{\alpha_{12}}{\alpha_{10}\alpha_{00'}\alpha_{0'2}}\right)^{1-\varepsilon}
\frac{s_{10}s_{00'}s_{0'2}}{s_{12}}\biggl[\ldots\biggr]\,.
\end{equation}
As explained in \cite{Caron-Huot:2016tzz}, in the absence of subdivergences, the overall energy integral $\d Q/Q^{1+4\varepsilon}$ produces the infrared divergences, and what it multiplies, evaluated with $\varepsilon\to 0$ and $\mubar=Q$, gives the corresponding contribution to the anomalous dimension $-H_{\rm NGL}$ in a Lorentz-invariant minimal subtraction scheme. The limit can be taken in the amplitudes \eqref{amps resummed} directly and gives a renormalized resummed propagator:\footnote{
The $i\pi$ term is a remnant of the branch cut in 
$\log\frac{\mubar^2}{2p_0{\cdot}p_{0'}} \equiv \log\frac{\mubar^2}{-s_{00'}-i0}$ with $s_{00'}>0$.
}
\begin{equation}
\begin{split}
  & \lim_{\varepsilon\to 0} \frac{g_0^2}{1-a_0\Pi\big|_n (2p_0{\cdot}p_{0'}/Q_{[100'2]}^2)^{-\varepsilon}}
\equiv \frac{g^2}{1-a \Pi^{(1)}_{\rm ren}}\,,
\\
\label{Pi1}
\mbox{with}\qquad &\Pi^{(1)}_{\rm ren}=
b_0|_n\left(\frac12\log\frac{\alpha_{10}\alpha_{20'}}{\alpha_{12}\alpha_{00'}}+i\pi\right) - \frac{4\ns+10\nf}{9}
\qquad\mbox{(large-$n_f$)\,.}
\end{split}\end{equation}
Taking the squared amplitude and fermion trace in \eqref{sigma NGL} we thus get
\begin{align}
    H_{\rm NGL}&\supset \int \frac{\d\Omega_{0}}{4\pi} \frac{\d\Omega_{0'}}{4\pi} \frac{\alpha_{12}}{\alpha_{10}\alpha_{00'}\alpha_{0'2}}
    \Vv_{10}\Vv_{00'}\Vv_{0'2}\tilde{H}_{[100'2],\rm NGL}\,,
\end{align}
where
\begin{align} \label{K2 double real}
\tilde{H}_{[100'2],\rm NGL}&=
\frac{a^2}{\bigl|1-\Pi^{(1)}_{\rm ren}\bigr|^2}
\int_0^\infty \d\tau
\left[\begin{array}{l}\displaystyle (\ns-2\nf)\frac{v}{u} \frac{\tau(\alpha_{10}\alpha_{20'}-\alpha_{10'}\alpha_{20})^2}{(\tau \alpha_{10}+\alpha_{10'})^2(\tau \alpha_{20}+\alpha_{20'})^2}
\\\displaystyle
+\nf \frac{2\alpha_{10}\alpha_{20'}}{(\tau \alpha_{10}+\alpha_{10'})(\tau \alpha_{20}+\alpha_{20'})}\end{array}\right]
\nonumber\\ &=\frac{a^2}{\bigl|1-\Pi^{(1)}_{\rm ren}\bigr|^2} \left[(\ns-2\nf)\frac{v}{u}\left(\frac{(v+1)\log v}{v-1}-2\right)+\nf \frac{2v\log v}{v-1}\right]\,,
\end{align}
with the cross-ratios $u$ and $v$ defined identically to \eqref{cross-ratios} with $z^2_{ij}\mapsto \alpha_{ij}$. The square bracket is precisely the two-loop matter contribution to the NGL anomalous dimension given previously in \cite[Eq. (3.35)]{Caron-Huot:2015bja}, specialized to the planar theory with adjoint matter; it also coincides with the square bracket in \eqref{H3 conf result}. What is new here is the inclusion of large-$n_f$ self-energies through $\Pi^{(1)}_{\rm ren}$ in \eqref{Pi1}.

The result \eqref{K2 double real} is to be compared with the three-loop large-$n_f$ contribution $\tilde{H}^{(3)}_{[100'2],\rm conf}$ in \eqref{H3 conf result}, using the dictionary in \eqref{eq:proj}.
By expanding
\begin{equation}
    \frac{1}{\bigl|1-\Pi^{(1)}_{\rm ren}\bigr|^2}
\approx 1+a\left(b_0|_n\log\frac{v}{u}- \frac{8\ns+20\nf}{9}
\right)+\ldots,
\end{equation}
we observe perfect agreement, that is:
\begin{equation}
    \tilde{H}^{(3)}_{[100'2],\rm conf} =
    \tilde{H}^{(3)}_{[100'2],\rm NGL}\,.
\end{equation}
This nontrivially confirms the identity \eqref{correspondence check} in the large-$n_f$ theory: not only is the left-hand-side conformal, but it can be calculated independently by studying the  infrared divergences of ``non-global'' weighted cross sections! The conformal scheme for rapidity divergences used in this paper maps precisely to the Lorentz scheme for infrared divergences introduced in \cite{Caron-Huot:2016tzz}.

The large-$n_f$ theory also contains real-virtual corrections, where only a gluon crosses the shock. We leave their evaluation to future work.

{\section{Conclusion}\label{sec:conclusion}}

We have, for the first time, derived the non-conformal component of the three-loop (NNLO) evolution equation, resumming large logarithms in the forward scattering of a color-singlet off a strongly interacting target (e.g., deep inelastic scattering off a nucleus).
We worked in the `t Hooft--Veneziano planar limit and considered a generic gauge theory with fermion and scalar matter fields.  From the QCD perspective, this ``planar QCD'' retains quarks but approximates all color connections as planar products of color dipoles---the same approximation which gives rise to the Balitsky--Kovchegov equation \cite{Balitsky:1995ub,Kovchegov:1999yj}.

The ``non-conformal'' component we computed is, in fact, the $\mathcal{O}(\varepsilon)$ piece of the \emph{two-loop} BK equation in $\D=4-2\varepsilon$ spacetime dimensions.  It appears as a contribution to the three-loop evolution in the context of the spacelike-timelike correspondence reviewed in Sec.\,\ref{ssec:correspondence} (see \eqref{eq:conjecture} and \eqref{eq:resultthree-loop}), which is an exact identity in the conformal dimension $\varepsilon_*=-\bar\beta(a)$ with useful implications away from it.  In short, the full three-loop BK equation in (planar) QCD in $\D=4$ is equal to the $\mathcal{O}(\varepsilon)$ lower-loop pieces calculated in this paper, plus a manifestly conformal three-loop piece that can be calculated using a different setup involving ``non-global logarithms.'' We nontrivially confirmed this prediction for $\mathcal{O}(n_f^2)$ terms. In the $\varepsilon=0$ limit, our results agree with the existing literature.

We also introduced new conformally covariant rapidity variables to regulate the null Wilson lines (see Sec.\,\ref{ssec:renDipole}), proportional to but not \emph{equal} to longitudinal energies. This ensures that all breaking of conformal symmetry in our results is associated with the physical running of the coupling rather than being artifacts of the regularization scheme.

% add a one-sentence comment re:Mueller's proposal to make the spacelike-timelike correspondence manifest?

Compared with the landmark calculation of the two-loop BK equation in \cite{Balitsky:2007feb,Balitsky:2009xg}, we started from the same momentum space and energy integrals, which we independently re-obtained from the Feynman rules. We then applied a variety of integration techniques, ranging from Schwinger parameters to the methods of intersection theory, integration-by-parts and differential equations. In particular, the latter approaches were generalized beyond their usual scope to accommodate the special functions (such as exponentials and logarithms) appearing in the considered integrands. This allowed for the exact computation of Fourier transforms, followed by integration over energies as a series in $\varepsilon$. If desired, the same techniques could be used to generate higher orders in the $\varepsilon$-expansion.

Several natural future directions remain. We hope in a future paper to complete the calculation of the conformal contribution and thereby complete the three-loop BK equation in planar QCD. Since the three-loop equation is already known in $\mathcal{N}=4$ supersymmetric Yang-Mills theory \cite{Caron-Huot:2016tzz}, we note that only diagrams involving matter loops remain to be calculated. The linearization of this result will characterize the so-called ``BFKL Pomeron'' in impact parameter space, and its eigenvalue will yield the (hard) Pomeron trajectory (the eigenfunctions being
already known \cite{Chirilli:2013kca}).
Comparison with collinear (DGLAP) evolution \cite{Moch:2004pa,Blumlein:2021enk} in the overlapping regime (see e.g., Sec.\,5.2 of \cite{Caron-Huot:2016tzz} or Sec.\,3.3 of \cite{Balitsky:2024xvi})
will provide important cross-checks.

Going beyond the planar limit could be technically feasible as far as integral computations are concerned, but bookkeeping all color interconnections could become challenging.
Our planar results apply to the so-called ``dilute-dense'' scaling where the projectile carries a fixed $\mathcal{O}(N^0)$ number of Wilson lines, whereas each dipole amplitude $\frac{1}{N}\text{Tr}\bigl[U_1U_2^\dagger\bigr]$ is treated as $\mathcal{O}(1)$ as $N\to\infty$. This accounts for all effects enhanced by the complexity of the target, but it ignores, e.g., $4\to 2$ Reggeon transitions that close the Pomeron loop \cite{Caron-Huot:2013fea}.
Better understanding these nonlinear effects at finite coupling, perhaps in models such as AdS/CFT, where the linear evolution is understood in terms of bulk graviton exchanges \cite{Brower:2006ea}, could be interesting.

One of our main motivations for obtaining the three-loop BK equation is to provide a means to test the convergence of both the perturbative series and the BK/BFKL resummation, as discussed in the introduction.  Numerical analysis of the full NNLO BK kernel, once available, could provide much insight: whether the kernel produces stable evolution under realistic initial conditions, or whether there exist regions where it becomes negative (signaling a breakdown of fixed-order perturbation theory) are important questions of phenomenological interest. If such instabilities or negativities arise, it would be crucial to understand if they can be cured—at least partially—by including collinear resummations.  Addressing these questions will clarify the practical utility of rapidity evolution equations and their interplay with collinear equations.

\acknowledgments 
%We are grateful to Ian Balitsky and Giovanni A. Chirilli for comments on the drafts.
We thank the organizers of the 2024 Amplitudes conference at the Institute for Advanced Study, where this work began, and the organizers and participants of the 2025 LoopFest and Saturated Glue (SURGE) Collaboration Workshop, where the content of this work was presented and benefited from useful discussions.
M.G.’s work is supported by the National Science and Engineering Council of Canada (NSERC) and the Canada Research Chair program, reference CRC-2022-00421.
G.B. research is supported by the European Research Council, under grant
ERC–AdG–885414 Ampl2Einstein, by the Università Italo-Francese, under grant Vinci, by the Italian MIUR under contract 20223ANFHR (PRIN2022), and partially by the Amplitudes INFN scientific initiative.
G.C.'s research is supported by the United Kingdom Research and Innovation grant UKRI FLF MR/Y003829/1.
S.S. research is partially supported by the Amplitudes INFN scientific initiative.

\appendix
\section{Lightcone gauge Feynman rules and a simple example}\label{App:lcFR}

The purpose of this appendix is to provide, for the readers' convenience, the set of rules used in the main text to construct the impact parameter space amplitudes leading to the BK Hamiltonians in the main text. A detailed discussion can be found in \cite{Kovchegov:2012mbw} (see also \cite{BRODSKY1998299}).

In any gauge theory with gauge algebra $\mathfrak{su}(N)$, one may couple $n_\text{adj}^s$ real scalars $\{\phi_p^a\}_{p=1}^{n_\text{adj}^s}$ in the adjoint of $\mathfrak{su}(N)$, and $n_\text{adj}^f$ (Weyl) fermions $\{\psi_q^a\}_{q=1}^{n_\text{adj}^f}$ in the adjoint of $\mathfrak{su}(N)$ via the Lagrangian density
\begin{equation}
\label{eq:generic-lagrangian}
\mathcal{L}_{\rm GT}
=
-\frac14\,F_{\mu\nu}^a F^{a\,\mu\nu}
+
\sum_{p=1}^{n_\text{adj}^s}\frac12\,(D_\mu\phi_p^a)(D^\mu\phi_p^a)
+
\sum_{q=1}^{n_\text{adj}^f}i\,\bar\psi_q^a\,\bar\sigma^\mu\,D_\mu\psi_q^a
+\mathcal{L}_{\rm int}(\phi,\psi)\,,
\end{equation}
where $D_\mu \Phi^a=\partial_\mu\Phi^a + g f^{abc} A_\mu^b \Phi^c$ 
and where $\bar\sigma^\mu = \bigl(\mathbb{1}_{2\times2},\,-\sigma\bigr)$, with $\sigma$ denoting the standard Pauli matrices.

Note that all fields are effectively massless since we work at high energies. Above, $\mathcal{L}_{\rm int}$
denotes any allowed Yukawa couplings or scalar quartic terms. The latter are only relevant starting at three loops, since massless tadpoles vanish in DREG.  Moreover, we work in the lightcone gauge, defined by $n\cdot A = 0$, where $n^\mu$ is a fixed null ($n^2 = 0$) but nonzero four‐vector. Hence, the standard Faddeev--Popov determinant is independent of $A_\mu$, and the associated ghost term $ \mathcal{L}_{\text{ghost}} \;=\; \bar{c}\,(n \cdot \partial)\,c
$ is left implicit, since it does not couple to the gauge field.

To assemble the lightcone-gauge amplitudes quoted in the main text, one may use the set of elementary Feynman rules given in \cite[Ch.~1.3.1]{Kovchegov:2012mbw}. It is worth mentioning that almost all these rules coincide with the standard (covariant) ones quoted in QFT textbooks (e.g., \cite{Peskin:1995ev}). In fact, the only gauge‐dependent Feynman rule is that of the gluon propagator, so it is the sole elementary rule we will quote explicitly:
\begin{equation}
            \adjustbox{valign=c}{\includegraphics[scale=0.6]{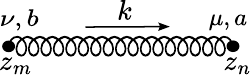}}=\frac{-i\delta^{ab}}{k^2- i\,\epsilon}D_{\mu\nu}(k) \e^{i k_\perp\cdot z_{nm}} \quad \text{with} \quad D_{\mu\nu}(k)=\eta_{\mu\nu}-\frac{n_\mu k_\nu+n_\nu k_\mu}{n\cdot k}\,.\label{eq:gluProp}
\end{equation}

In the main text, we follow Balitsky and Chirilli \cite{Balitsky:2006wa,Balitsky:2007feb,Balitsky:2009xg} by choosing $n^\mu$ so that momenta decompose as follows:
\begin{equation}
    k^\mu=k^- \overline{n}^\mu+k^+ n^\mu+k_\perp^\mu \quad \text{where}\quad n\cdot n=0 \implies \quad k_\perp\cdot\{n,\overline{n}\}=0 \quad \text{and} \quad n\cdot \overline{n}=-1\,.
\end{equation}
As an example, in this basis and after relabeling $k^\pm \to \frac{2}{s}\{\beta,\alpha\}$, the gluon propagator numerator $D_{\mu\nu}(k)$ in \eqref{eq:gluProp} is seen to coincide exactly with \cite[Eq. (10)]{Balitsky:2007feb}. Hence, after the $k^\pm \to \frac{2}{s}\{\beta,\alpha\}$ identification is made, one can directly use the shock‐background version of the above propagators as done implicitly in the main text. That is, the fermion, scalar, and gluon propagators—given explicitly by the integrands in \cite[Eq. (12)]{Balitsky:2006wa}, \cite[Eq. (22)]{Balitsky:2009xg}, and \cite[Eq. (9)]{Balitsky:2007feb}, respectively. In particular, from these rules in the $N\to \infty$ limit the phase factors $\Uu_{ij}$ in \eqref{eq:evolEqU2} eventually arise.

\paragraph{A simple example: building block for the one-loop amplitudes.}
We now demonstrate how the usual (covariant) Feynman rules can be used to construct the impact-parameter space amplitudes in the main text. For simplicity, we focus on deriving the expression from which the one-loop amplitudes in \eqref{amplitude single real} and \eqref{amplitude single virtual} are obtained.

We begin with the amplitude for an on‑shell massless fermion (quark) to emit a gluon of fixed longitudinal momentum $k^+$. After applying the Feynman rules, the amplitude (as a function of $k^+$ and the transverse separation $z$) is expressed as 
\begin{equation}\label{eq:subamp1}
\begin{split}
    \mathcal{M}(z,k^+)
    \;&=\adjustbox{valign=c}{\includegraphics[scale=0.7]{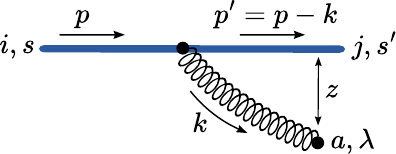}}
\\&
= \int \frac{\d k^-}{2\pi}\, \,\,\dbar^{\Dperp}k\,\frac{\overline{u}_{s'}(p')\,(g f^{aij})\,\slashed{\varepsilon}(k)\,u_s(p)}{[k^2-i0][(p-k)^2-i0]}\,\,\e^{i k_\perp\cdot z}\,\,.\;
\end{split}
\end{equation}
In this expression $a,i,j$ denote color indices, $s,s'$ spin indices, and $\lambda$ the gluon helicity index. Note that we have \emph{not yet} imposed the gluon to be much softer than the quark ($k^+\ll p^+$), and we have retained the standard Feynman $i0$ prescriptions. We omit the $k^+$ integral here because $k^+$ is conserved across the shock and so it can only be integrated over \emph{after} multiplying \eqref{eq:subamp1} with a corresponding amplitude from the other side of the shock.

Before performing the $k^-$ integral, it is useful to first go in the frame where $p^\mu=(p^+,p^-=0,p_\perp=0)$ and introduce the parameter $\xi$ such that
\begin{equation}
    (p')^+=(1-\xi)\,p^+ \quad \text{and} \quad k^+=\xi\, p^+\quad \text{where the soft limit corresponds to $\xi\to 0$}\,.
\end{equation}
Carrying out standard numerator algebra, one finds that the numerator is independent of $k^-$
\begin{equation}
    \overline{u}_{s'}(p')\,(g f^{aij})\,\slashed{\varepsilon}(k)\,u_s(p)=\frac{2 g f^{aij}}{\xi \sqrt{1-\xi}}(\delta_{\lambda,2s}+(1-\xi)\delta_{\lambda,-2s})(k_\perp-\xi p_\perp)\cdot \varepsilon_\perp^\lambda(k)\,.
\end{equation}
Consequently, the $k^-$ integral in \eqref{eq:subamp1} is systematically performed by picking the residue at $k^-=\frac{k_\perp^2-i0}{k^+}$, which is consistent with having the two poles in the integrand's denominator in different half-planes when $0<k^+<p^+$. 
As a result, the second propagator in \eqref{eq:subamp1} becomes the so-called ``energy denominator'' (see \cite[Eq. (1.55)]{Kovchegov:2012mbw} for more details) and
\begin{equation*}
    \mathcal{M}(z,k^+)=g f^{aij} \int\,\dbar^{\Dperp}k\,\,\frac{[2i(\delta_{\lambda,2s}{+}(1{-}\xi)\delta_{\lambda,-2s})(k_\perp{-}\xi p_\perp){\cdot} \varepsilon_\perp^\lambda(k)]\theta(k^+)\theta(p^+{-}k^+)}{\xi \sqrt{1-\xi}\,k^+\,[p^+-k^+][\frac{k_\perp^2}{k^+}+\frac{(p_\perp-k_\perp)^2}{p^+-k^+}]}\,\,\e^{i k_\perp\cdot z}\,.
\end{equation*}
In the soft limit, this amplitude effectively becomes independent of $p^+$ and reads:
\begin{equation}\label{eq:Msoft}
\mathcal{M}(z, k^+)\approx g f^{aij}\frac{\theta(k^+)}{ k^+}\int\dbar^{\Dperp}k\,\,
\tilde{\mathcal{A}}^i(k)\cdot\varepsilon_\perp(k)\,\,\e^{i k_\perp\cdot z}
\quad
\text{with}
\quad \tilde{\mathcal{A}}^i(k)\equiv\frac{2i k_\perp^i}{k_\perp^2}\,.
\end{equation}
The integrand $\tilde{\mathcal{A}}^i$ is precisely that entering \eqref{amplitude single real} and \eqref{amplitude single virtual}. In the main text, we omit the color and polarization factors appearing in \eqref{eq:Msoft}, since upon squaring the amplitude they reduce to numerical constants absorbed into the normalization conventions outlined in Sec.\,\ref{sec:setup}.

{\section{Details of real-virtual calculation}\label{app:detailsHRV}}

In this appendix, we detail our evaluation of the two-loop real-virtual diagrams shown in Fig.~\ref{fig:diagRV}, in a way that closely parallels the double-real calculation in the main text.
We first describe momentum space integrands, Fourier transform to obtain position space amplitude. We then include the details of the differential equation method used to evaluate the integrals obtained.

\subsection{Momentum space integrand}

An integrand (defined generally in \eqref{eq:bareDipole}) for the diagrams in Fig.~\ref{fig:diagRV} can be obtained by applying the lightcone gauge Feynman rules just reviewed
and integrating over $\ell_i^-$ momenta. This leaves integrals over two transverse momenta and the energy ratio $\tau=\ell_{0}^+/\ell_{0'}^+$.
We have verified that the result of this step matches precisely with equations (68), (70), (77), and (95) in \cite{Balitsky:2007feb}, to which we therefore refer for more details on this part of the derivation.

Organizing the diagrams into self-energies ($\Pi$), single-source ($\Upsilon$) and double-sources ($\T$) following Fig.~\ref{fig:diagRV}, we have:
\begin{equation} \label{virtual integrand}
    {\cal A}_{[102]}^{(g,1)i}(\{z\};\tau)=
\left[{\cal A}_{[102]}^{(g,1)i}(z_{01};\tau)_\Pi+
{\cal A}_{[102]}^{(g,1)i}(z_{01};\tau)_\Upsilon
    +{\cal A}_{[102]}^{(g,1)i}(z_{01},z_{02};\tau)_{\T}\right]-(z_1\leftrightarrow z_2),
\end{equation}
where 
\begin{subequations}
\begin{align}
\frac{{\cal A}_{[102]}^{(g,1)i}(z_{01};\tau)_\Pi}{-8i\,\tau}&=
\!\!\!\int\limits_{k_1,k_2}\e^{i(k_1+k_2){\cdot}z_{01}}
\left[
\begin{array}{rl}
&\displaystyle\frac{\tau k_1^i (k_2^2{-}k_1^2)\big(\ns{-}2\nf{+}2{-}2\varepsilon(1{-}\delta)\big)}{(1+\tau)^2(k_1+k_2)^4(\tau k_1^2+ k_2^2)}
\\&
+\displaystyle\frac{ k_1^i\big(\nf-4\big)}{(1{+}\tau)(k_1{+}k_2)^2(\tau k_1^2{+}k_2^2)}
\\&
+\displaystyle\frac{k_1^i(\tau^2+5\tau+2)}{\tau (1{+}\tau)(k_1{+}k_2)^2(\tau k_1^2{+}k_2^2)}
\end{array}
\right]
\,,\label{virtual integrand PI}
\\
\frac{{\cal A}_{[102]}^{(g,1)i}(z_{01};\tau)_\Upsilon}{-4i \,\tau}&=
\!\!\!\int\limits_{k_1,k_2}\e^{i(k_1+k_2){\cdot}z_{01}}
\left[
\begin{array}{rl}
&\!\!\!2\displaystyle\frac{k_1^i ((1{+}\tau)k_1^2{+}k_1{\cdot}k_2){-}
k_2^i (\tau k_1{\cdot}k_2{+}(1{+}\tau)k_2^2)}{(1{+}\tau)(k_1{+}k_2)^2k_2^2(\tau k_1^2{+} k_2^2)}
\\&
\!\!\!
+\displaystyle\frac{2k_2^i(k_1{\cdot}k_2(1{-}\tau){-}\tau k_2^2){-}(1{+}\tau)k_1^i(k_1^2{+}k_2^2)}
{(1+\tau)(k_1+k_2)^2k_2^2(\tau k_1^2+(k_1+k_2)^2)}
\end{array}
\right],
\label{virtual integrand UP}
    \\
    {\cal A}_{[102]}^{(g,1)i}(z_{01},z_{02};\tau)_{\T}&=
-4i\,\tau\int\limits_{k_1,k_2}
    \e^{ik_1{\cdot}z_{01}+ik_2{\cdot}z_{02}}
\left[
\begin{array}{rl}
&\displaystyle\frac{\frac{k_2^i}{k_2^2}+\frac{(k_1^2{+}k_2^2)(k_1{+}k_2)^i}{k_2^2(k_1{+}k_2)^2}+\frac{2(k_1{+}k_2)^i}{\tau k_1^2}}{\tau k_1^2{+}(k_1{+}k_2)^2}
\\&+
\displaystyle\frac{k_2^i}{(1+\tau)k_1^2k_2^2}
\end{array}
\right]
\,. \label{virtual integrand T}
\end{align}
\end{subequations}
In the above, we have dropped scaleless integrals, which vanish in dimensional regularization (examples of which are given in Fig.~\ref{fig:diagRV}$(0)$).
The permutation of the $\T$ contribution in \eqref{virtual integrand} is not a typo: in \eqref{virtual integrand T} we included only part of the first diagram in Fig.~\ref{fig:diagRV}$(\T)$;
the permutation adds its second half.

\begin{figure}
    \centering
    \includegraphics[scale=0.5]{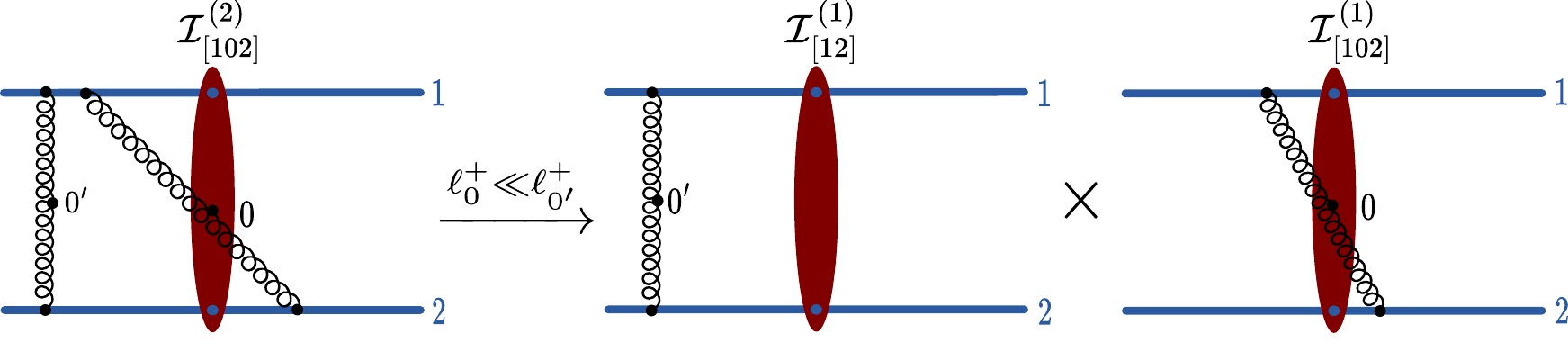}
    \caption{Similarly to the factorization of double-real integrands shown in Fig.~\ref{fig:facto1}, real-virtual integrands also factorize as the singular limits $\tau\to 0$ and $\tau\to\infty$ are approached.}
    \label{fig:facto2}
\end{figure}

Above, we have also applied judicious permutations to various terms in order to manifest the rapidity factorization limits:
\begin{subequations}
\label{virtual factorization}
\begin{align}
 \lim_{\tau\to 0}{\cal A}_{[102]}^{(g,1)i}(\{z\};\tau) &=
 \int\limits_{k_1}
\frac{2ik_1^i}{k_1^2} \left(\e^{ik_1{\cdot}z_{01}}-\e^{ik_1{\cdot}z_{02}}\right)
\int\limits_{k_2} \frac{4}{k_2^2} \e^{ik_2{\cdot}z_{12}}={\cal A}_{[102]}^{(g,0)i}{\cal A}_{12}^{(\varnothing,1)}\,,
\label{virtual factorizationA}
\\
\lim_{\tau\to \infty}{\cal A}_{[102]}^{(g,1)i}(\{z\};\tau) &=  \int\limits_{k_1}
\frac{2ik_1^i}{k_1^2} \left(\e^{ik_1{\cdot}z_{01}}-\e^{ik_1{\cdot}z_{02}}\right)
\int\limits_{k_2} \frac{4}{k_2^2} \left(\e^{ik_2{\cdot}z_{01}}+\e^{ik_2{\cdot}z_{02}}\right)
\nonumber\\ &=
{\cal A}_{[102]}^{(g,0)i} \left({\cal A}_{10}^{(\varnothing,1)}+{\cal A}_{02}^{(\varnothing,1)}\right)\,.
\label{virtual factorizationB}
\end{align}
\end{subequations}
Like for the double-real contributions, these factorization limits are essential to our extraction of a two-loop kernel. These hold modulo scaleless integrals and shifts in $k_i$. Note that \eqref{virtual factorizationA} is represented pictorially in Fig.~\ref{fig:facto2}. (In particular, from Fig.~\ref{fig:facto2}, one sees why no triple‑dipole term precedes $\mathcal{I}_{[102]}^{(1)}\,\mathcal{I}_{[10'2]}^{(1)}$ in the subtraction term \eqref{eq:Sdef}.  Were such a term present, it would subtract double‑real diagrams without three‑gluon vertices: in those cases the formation‑time ratios on either side of the target scale inversely to each other, so the $\tau$-dependent factors would cancel upon taking the product of amplitudes.  In contrast, for the subset of diagrams shown in Fig.~\ref{fig:facto1} no cancellation of such inverse ratio occurs, and therefore an explicit subtraction is required.)

As we will see below, the integrals for $\Pi$ and $\Upsilon$ give simple $z$-dependent factors $z_{01}^i/(z_{02}^2)^{1-\varepsilon}$ times constants that can be calculated exactly in $\varepsilon$. Although finite, the $\T$ integral is more complicated and will only be computed order-by-order in $\varepsilon$.

\subsection{Position-space amplitudes}

For the real-virtual corrections, we can integrate \eqref{virtual integrand} over both $k_1$, $k_2$ and $\tau$ at the amplitude level before multiplying by the right amplitude.  There is some freedom in the orders in which we do it, and the $\tau$ integral also depends on the choice of a scheme that we use to subtract the rapidity subdivergences. In this appendix, we perform the integrals in the ``+'' scheme described in Sec.~\ref{ssec:renDipole} (i.e., a hard momentum cutoff), and in Sec.~\ref{sssec:rapiditySubRV} we explain the conversion to the conformal scheme.

Adding the counter-term from \eqref{bare coupling}, let us thus define the integrated virtual amplitude as:
\begin{equation} \label{eq:1FT_res1}
    {\cal A}_{[102]}^{(g,1)i+}(\{z\}) = {\cal N}_\varepsilon \int_0^\infty \left[\frac{\d\tau}{\tau}\right]_+{\cal A}_{[102]}^{(g,1)i}(\{z\};\tau) 
    -\frac{b_0}{2\varepsilon}{\cal A}_{[102]}^{(g,0)i}\,,
\end{equation}
with ${\cal N}_\varepsilon=2\pi\mu^{2\varepsilon}$ as in \eqref{def cd}.
The self-energy contribution to the integrand ${\cal A}_{[102]}^{(g,1)i}(\{z\};\tau)$ in \eqref{virtual integrand} is free of rapidity subdivergences, but for the other contributions, the $+$-prescription means that divergences at $\tau\to 0$ and $\tau\to\infty$ (see \eqref{plus example}) are subtracted using the identities:\footnote{
In terms of the energy fraction $u=\frac{\tau}{\tau+1}$, 
these amount to the more familiar $\int_0^1 \frac{\d u}{u_+}=0=\int_0^1 \frac{\d u}{(1-u)_+}$.}
\begin{equation}
    \int_0^\infty \left[\frac{\d\tau}{\tau}\right]_+ \frac{1}{\tau+1} \equiv 0
=\int_0^\infty \left[\frac{\d\tau}{\tau}\right]_+ \frac{\tau}{\tau+1}\,.
\end{equation}
Applying these to perform the integrals over $k_2$ (with fixed $k_1+k_2$), the self-energy and one-parton contributions ($\Pi$ and $\Upsilon$) can be integrated using standard techniques, e.g., Schwinger parameters. Postponing the intermediate steps to the next subsection, we find:
\begin{subequations} \label{virtual 1app}
\begin{align}
\hspace{-0.3cm}
    {\cal A}_{[102]}^{(g,1)i+}(\{z\})\Big|_{\Pi+\Upsilon+{\rm c.t.}}&=
    \int\limits_{k_1} \frac{2ik_1^i}{k_1^2}
(\e^{ik_1\cdot z_{01}}{-}\e^{ik_1\cdot z_{02}})\left[
 \left(\frac{\mubar^2}{k_1^2}\right)^\varepsilon
\left( \Pi(\varepsilon)+ \Upsilon(\varepsilon)\right) - \frac{b_0}{2\varepsilon}\right]\label{virtual 1A}
 \\ &
 =\frac{z_{10}^i\cd}{(z_{10}^2)^{1-\varepsilon}}
 \left[\frac{
 \Gamma(1{-}2\varepsilon)(\mubar^2z_{10}^2/4)^\varepsilon}
 {\Gamma(1{-}\varepsilon)\Gamma(1{+}\varepsilon)}
\left( \Pi(\varepsilon){+} \Upsilon(\varepsilon)\right)
 {-}\frac{b_0}{2\varepsilon}\right]-(z_1{\leftrightarrow}z_2)\,,\label{virtual 1B}
\end{align}
\end{subequations}
where as before, $b_0=\frac{11}{3}-\frac{2}{3}n_{\rm adj}^f-\frac{1}{6}
n_{\rm adj}^s$, and
\begin{subequations}\label{eq:PiAndUp}
\begin{align}
    \Pi(\varepsilon)&=
 \e^{\varepsilon\gamma_{\text{E}}} \frac{\Gamma(1+\varepsilon)\Gamma^2(1-\varepsilon)}{\Gamma(1-2\varepsilon)}
 \left[\frac{4}{\varepsilon^2}-\frac{\nf-4}{\varepsilon(1-2\varepsilon)}
 -\frac{\ns-2\nf+2-2\varepsilon(1{-}\delta)}{2\varepsilon(1-2\varepsilon)(3-2\varepsilon)} \right],
\label{Pi function}
 \\
\Upsilon(\varepsilon)&=
\e^{\varepsilon\gamma_{\text{E}}} \frac{\Gamma(1+\varepsilon)\Gamma^2(1-\varepsilon)}{\Gamma(1-2\varepsilon)}
\left[-\frac{4}{\varepsilon^2}+\frac{h(\varepsilon)+3h(-2\varepsilon)-3h(-\varepsilon)}{\varepsilon}\right].
\label{Up function}
\end{align}
\end{subequations}
Both lines in \eqref{eq:PiAndUp} are exact results in $\varepsilon$. Here, $h(n)$ denotes the $n^\text{th}$ harmonic number, which is analytically continued in terms of the digamma function: $h(n)=\psi(n+1)-\psi(1)$ and $\Pi(\varepsilon)$ is the 
gluon self-energy in lightcone gauge. Finally, the $k_1$ integral in \eqref{virtual 1app} was performed using the standard Fourier transform of a power law,
\begin{equation}\label{eq:powerLawFT}
    \int\limits_{k} \frac{\e^{i k\cdot z}\, k^i}{(k^2)^n}=
    i\frac{\Gamma(2-\varepsilon-n)}{2^{2n-1}\pi^{1-\varepsilon}\Gamma(n)}\frac{z^i}{(z^2)^{2-n-\varepsilon}}\,.
\end{equation}
The gluon self-energy in lightcone gauge has a double pole $1/\varepsilon^2$, which can be attributed to soft and collinear divergences; it nicely cancels against a double pole in $\Upsilon$. The square bracket in \eqref{Pi function} further contains a single pole with coefficient $+b_0/\varepsilon$.
Note that the counter-term $\frac{-b_0}{2\varepsilon}$ cancels only \emph{half} it: the remaining half will cancel below against real collinear divergences via a variant of the KLN theorem.\footnote{In our calculation we treat ${\cal U}\sim 1$. An alternative bookkeeping method is to treat perturbatively the interactions with the target, leading to ${\cal U}\sim g^2$ and a doubled $b_0$ counterterm. Then one obtains the same final answer upon accounting from additional ``loop inside the shock'' diagrams \cite{Balitsky:2006wa,Balitsky:2007feb}.}

For the $\T$ contribution, we use a different strategy. We first integrate \eqref{virtual integrand T} over $\tau$ in the $+$-prescription:
\begin{equation}
\label{RV amplitude left}
    \frac{{\cal A}_{[102]}^{(g,1)i+}\Big|_{\T}}{{-}8\pi i \mu^{2\varepsilon}}=
\int\limits_{k_1,k_2}\!\! 
\e^{i(k_{1}\cdot z_{01}+k_{2}\cdot z_{12})}
\left[\frac{(k_{1}-k_{2})^i}{k_{2}^{2}(k_{1}{-}k_{2})^{2}}{+}\frac{k_{1}^i}{k_{1}^{2}(k_{1}{-}k_{2})^{2}}{-}\frac{k_{1}^i}{k_{1}^{2}k_{2}^{2}}\right]
\log\frac{k_{1}^{2}}{k_{2}^{2}} -(z_1{\leftrightarrow}z_2).
\end{equation}
This is then handled by adapting modern Feynman‑calculus methods to Fourier transforms (see also \cite{Brunello:2023fef}), including integration-by-parts \cite{Chetyrkin:1981qh,Laporta:2000dsw} and (canonical) differential equations \cite{Remiddi:1997ny,Henn:2013pwa}, as detailed in the next subsection. We quote the final result used in the main text:
\begin{equation}\label{virtual amplitude}
\hspace{-0.37cm}
{\cal A}_{[102]}^{(g,1)i+}\Big|_{\T}\!\!=\!
\frac{-z_{10}^i \cd}{(z_{10}^2)^{1-\varepsilon}}\!\left[\log\frac{z_{12}^2}{z_{10}^2}\log\frac{z_{12}^2}{z_{20}^2}\!\left[\!1{+}\frac{\varepsilon}{2}\log\frac{\mubar^4z_{12}^2z_{20}^2}{16\e^{-4\gamma_{\text{E}}}}\right]\!
{-}\varepsilon {\cal L}_{1,0,1}\right]\! {-}(z_1{\leftrightarrow}z_2){+}{\cal O}(\varepsilon^2)\,,
\end{equation}
where ${\mathcal{L}}_{1,0,1}$ is the single-valued multiple polylogarithm defined in \eqref{L101}.
If desired, higher-order corrections in $\varepsilon$ can be obtained systematically using the differential equation method now described.

\subsection{Some details on the Schwinger parameter method}

In this subsection, we provide additional details on how \eqref{virtual 1app} is obtained from \eqref{eq:1FT_res1}.

We start with the \eqref{virtual integrand PI} contribution to \eqref{eq:1FT_res1}. Integrating over $\tau$ in the $+$-prescription and applying suitable momentum shifts to convert the $k_2$ Fourier transform into a ``standard'' loop integral yields
\begin{equation}\label{eq:todoA1}
    \frac{-i\pi}{2^{1-2\varepsilon}}\int\limits_{k_1,k_2}\e^{ik_1{\cdot}z_{01}}\frac{k_1^i}{k_1^4}\Big[1{+}\frac{k_1\cdot k_2}{k_1^2}\Big]
\left[\begin{array}{rl}
          &8k_1^2\Big(\frac{1}{k_1\cdot k_2}{-}\frac{4}{(k_1{-}k_2)^2}{+}\frac{2}{(k_1{+}k_2)^2}\Big)
          \\&
          {+}\frac{4(\ns-2\nf+2-2\varepsilon(1-\delta))k_1^2}{k_1\cdot k_2}
          \\&
          +\frac{(\nf-4)(k_1-k_2)^2}{k_1\cdot k_2}
    \end{array}
    \right]
    \log\frac{(k_1{+}k_2)^2}{(k_1{-}k_2)^2}\,.
\end{equation}
Using $X^y=1 + y \log X + \ldots$ to thread the logarithm for a power-law, the $k_2$ integral reduces to a single Schwinger-parameter integral for multiple choices of exponents $a, b$ and $c$
\begin{equation}\label{eq:todoA2}
    \int\limits_{k_2}\frac{1}{[k_1\cdot k_2]^{a}[(k_1{-}k_2)^2]^b[(k_1{+}k_2)^2]^c}=\int\frac{\d \Vec{\alpha}}{\text{GL}(1)}\frac{(\alpha_2{+}\alpha_3)^{1-\varepsilon}\e^{\frac{k_1^2(\alpha_1-4\alpha_2)(\alpha_1+4\alpha_3)}{4(\alpha_2+\alpha_3)}}}{(4\pi)^{1-\varepsilon}\Gamma(a)\Gamma(b)\Gamma(c)\alpha_1^{1-a}\alpha_2^{1-b}\alpha_3^{1-c}}\,,
\end{equation}
where $b$ and $c$ can depend on $y$ and $a\in \mathbb{Z}$. 
Here we quotient the measure $\d \Vec{\alpha}=\d \alpha_1\,\d \alpha_2\,\d \alpha_3$ by the GL(1) action, which encodes the projective/reparametrization invariance of the Schwinger parameters. In practice, it is dealt with by rescaling $\alpha_i \;\to\; \lambda\,\alpha_i$ for all $i$ and send
\begin{equation}
    \int\frac{\d \Vec{\alpha}}{\text{GL}(1)}\mapsto \alpha_3 \int_0^\infty \lambda^{2}\,\d \alpha_1\,\d \alpha_2\,\d\lambda\,.
\end{equation}
For the choices of $a, b$ and $c$ relevant to \eqref{eq:todoA1}, \eqref{eq:todoA2} can be carried out analytically (e.g., using \textsc{Mathematica}'s \texttt{Integrate[]}). Since the explicit form is not particularly illuminating, we do not quote it explicitly here. Inserting the result of \eqref{eq:todoA2} where appropriate in \eqref{eq:todoA1} produces the term $\sim\Pi$ in \eqref{virtual 1A}.

We applied analogous manipulations to the contribution of \eqref{virtual integrand UP} in \eqref{eq:1FT_res1}. After performing suitable momentum shifts and integrating over $\tau$, we obtain an expression similar to \eqref{eq:todoA1}, albeit more cumbersome, for which the $k_2$ loop integral can still be written solely in terms of \eqref{eq:todoA2}. However, in contrast to \eqref{virtual integrand PI}, needed instances of $a$, $b$, and $c$ yield hypergeometric functions in the final Schwinger-parameter integral, which \textsc{Mathematica}’s \texttt{Integrate[]} fails to evaluate directly. Our solution is therefore to expand the integrand in \eqref{eq:todoA2} in $\varepsilon$ and integrate the result order-by-order. The $\Upsilon$ term in \eqref{virtual 1A} then arises after isolating the $1/\varepsilon^2$ pole in \eqref{Up function} and using \textsc{Mathematica}’s \texttt{GeneratingFunction[]} to reconstruct the finite (harmonic-number dependent) part as an exact function of $\varepsilon$.

\subsection{Details on the differential equation method}\label{section:RV_details}

In this section, we present details on the calculation of the double emission Fourier transform in the real-virtual amplitude \eqref{RV amplitude left}. We use the method of canonical differential equations \cite{Henn:2013pwa} and work, as always, in $\Dperp=2-2\varepsilon$ dimensions. 

\paragraph{Form factor decomposition.} We start from the integral in \eqref{RV amplitude left} modulo permutation:
\begin{equation}\label{eq:vectorInt}
    \mathscr{F}^{i}(z_{01},z_{12}) = \int\limits_{k_{1},k_{2}}\e^{i(k_{1}\cdot z_{01}+k_{2}\cdot z_{12})}\left(\frac{(k_{1}-k_{2})^{i}}{k_{2}^{2}(k_{1}-k_{2})^{2}}+\frac{k_{1}^{i}}{k_{1}^{2}(k_{1}-k_{2})^{2}}-\frac{k_{1}^{i}}{k_{1}^{2}k_{2}^{2}}\right)\log\frac{k_{1}^{2}}{k_{2}^{2}}\,,
\end{equation}
which can be decomposed onto a set of ``form factors'' 
\begin{equation}
    \mathscr{F}^{i}(z_{10},z_{12}) = F_{1}(z_{10},z_{12})z_{10}^{i}+F_{2}(z_{10},z_{12})z_{12}^{i}\,,
\end{equation}
where the scalar integrals $F_{j}(z_{10},z_{12})$ are given by
\begin{equation}\label{eq:form_factors_def}
    \begin{bmatrix}
        F_{1}(z_{10},z_{12})\\
        F_{2}(z_{10},z_{12})
    \end{bmatrix} = \begin{bmatrix}
        z_{10}^{2}&z_{10}\cdot z_{12}\\
        z_{10}\cdot z_{12}&z_{12}^{2}
    \end{bmatrix}^{-1}\cdot \begin{bmatrix}
        \mathscr{F}\cdot z_{10}\\
        \mathscr{F}\cdot z_{12}
    \end{bmatrix}\,.
\end{equation}
The resulting scalar integrals comprising $F_{1,2}$ belong to a family of integrals given by
\begin{equation}\label{eq:family}
    \mathscr{L}_{n_{1}{\ldots} n_{7}}^{\delta} = \int\limits_{k_{1},k_{2}}\left(\prod_{i=1}^{7}\textsf{D}_{i}^{-n_{i}}\right)\e^{\textsf{D}_{4}+\textsf{D}_{7}}\log^{\delta}\frac{\textsf{D}_{1}}{\textsf{D}_{3}}\,,
\end{equation}
with the propagators defined as
\begin{equation}
    \begin{split}
    \textsf{D}_{1} &= k_{1}^{2}, \quad \textsf{D}_{2} = (k_{1}-k_{2})^{2}, \quad \textsf{D}_{3} = k_{2}^{2}\,,\\
    \textsf{D}_{4} = ik_{1}\cdot z_{01}, &\quad \textsf{D}_{5} = ik_{1}\cdot z_{12}, \quad \textsf{D}_{6} = ik_{2}\cdot z_{01}, \quad \textsf{D}_{7} = ik_{2}\cdot z_{12}\,.
    \end{split}
\end{equation}
We observe that $\textsf{D}_{1},{\ldots}, \textsf{D}_{3}$ represent the propagators appearing in \eqref{eq:vectorInt}, whereas $\textsf{D}_{4},{\ldots},\textsf{D}_{7}$ are auxiliary denominators (irreducible scalar products) chosen to complete the basis of scalar products. As such, we restrict these propagators to appear in the numerator only, namely $n_{4},\cdots, n_{7} \leq 0$. To capture the full range of scalar integrals in $F_{1,2}$, we further allow our basis elements to contain a logarithm with power $\delta \in \{0,1\}.$

\paragraph{Master integrals.} In a similar manner to ordinary Feynman integrals, we can build linear relations between elements of $\mathscr{L}$ using integration-by-parts identities (IBPs) \cite{Laporta:2000dsw}, generated by the relation
\begin{equation}\label{eq:core_ibp_identity}
    0 = \int\limits_{k_{1},k_{2}}\frac{\partial}{\partial k_{a}^{i}}\left(q_{j}^{i}\prod_{k=1}^{7}\textsf{D}_{k}^{-n_{k}}\e^{\textsf{D}_{4}+\textsf{D}_{7}}\log^{\delta}\frac{\textsf{D}_{1}}{\textsf{D}_{3}}\right)\,,
\end{equation}
where $a=1,2$ and $q_{j}\in\{k_{1},k_{2},z_{10},z_{12}\}$, which holds in dimensional regularization (see, e.g., \cite{Weinzierl:2020gda}). Using the Laporta algorithm \cite{Laporta:2000dsw} via the automated package \texttt{LiteRed} \cite{Lee:2012cn} to generate the identities and \texttt{FiniteFlow} \cite{Peraro:2019svx} to solve them, we find that all elements of $\mathscr{L}$ admit a decomposition in terms of eight master integrals, for which a valid basis choice is given by a valid basis.
\begin{equation}\label{eq:initial_basis}
\begin{aligned}
    \mathcal{T}_1 &= \mathscr{L}^{0}_{0110000}\,, \quad \mathcal{T}_2 = \mathscr{L}^{0}_{1010000}\,, \quad
    \mathcal{T}_3 = \mathscr{L}^{0}_{1100000}\,, \quad
    \mathcal{T}_4 = \mathscr{L}^{0}_{1110000}\,, \quad
    \\
    \mathcal{T}_5 &= \mathscr{L}^{1}_{0110000}\,, \quad
    \mathcal{T}_6 = \mathscr{L}^{1}_{1010000}\,, \quad
    \mathcal{T}_7 = \mathscr{L}^{1}_{1100000}\,, \quad
    \mathcal{T}_8 = \mathscr{L}^{1}_{1110000}\,.
\end{aligned}
\end{equation}
When reducing the form factors $F_{1,2}$ on this basis, we find that the master integral $\mathcal{T}_8$ never appears in the reduction, which can be seen directly by the fact that no term in \eqref{RV amplitude left} contains at the same time all of the three denominators. Thus, for the purposes of the evaluation of $F_{1,2}$, the computation of the first seven master integrals is totally sufficient.

\paragraph{Differential equations system.}
In order to evaluate the integrals $\mathcal{T}_1, {\ldots}, \mathcal{T}_7$ in \eqref{eq:initial_basis}, we construct the to-be-solved differential equation system obeyed by the master integrals \cite{Gehrmann:1999as}. It is obtained by taking derivatives with respect to the three external scalar variables $z_{10}^{2}$, $z_{10}\cdot z_{12}$ and $z_{12}^{2}$. 

Through the use of, e.g., the Magnus--Dyson exponential matrix \cite{Argeri:2014qva}, it is possible to find a dimensionless canonical basis \cite{Henn:2013pwa} $\mathcal{J}$ given by
\begin{equation}\label{eq:canBasis}
    \mathcal{J} = (z_{12}^{2})^{\Dperp-2}\begin{bmatrix}
        \mathcal{T}_1\\
        \mathcal{T}_2\\
        \mathcal{T}_3\\
        \frac{2(\Dperp-3)(\Dperp-4)\mathcal{T}_4+
        z_{10}\cdot z_{20} \mathcal{T}_1+z_{10}\cdot z_{12} \mathcal{T}_2+z_{12}\cdot z_{02}\mathcal{T}_3}{2\kappa}\\
        (\Dperp-2)\mathcal{T}_5\\
        (\Dperp-2)\mathcal{T}_6\\
        (\Dperp-2)\mathcal{T}_7
    \end{bmatrix}\,,
\end{equation}
where we have defined $\kappa=+\sqrt{(z_{10}\cdot z_{12})^{2}-z_{10}^{2}z_{12}^{2}}$. This basis is called “canonical” because it satisfies a system of differential equations in which the dependence on $\varepsilon$ factorizes linearly:
\begin{align}\label{eq:canonical}
    \d\mathcal{J} = \varepsilon\, [\d\Omega]\cdot\mathcal{J}\,.
\end{align}and all non‑trivial entries of the matrix $\Omega$ can be written purely in terms of logarithms. Moreover, $\Omega$ inherits an expected triangular structure from IBPs:
\begin{subequations}
    \begin{align}
        \label{eq:deq_matrix_rv}
    & \hspace{-0.3cm} \Omega = \left[
\begin{array}{ccccccc}
 \log(w_{1}w_{2}) & 0 & 0 & 0 & 0 & 0 & 0 \\
 0 & \log(w_{1}) & 0 & 0 & 0 & 0 & 0 \\
 0 & 0 & \log(w_{2}) & 0 & 0 & 0 & 0 \\
 \frac{1}{2} \log(w_{5}) & \frac{1}{2} \log (w_{4}) & -\frac{1}{2} \log \left(w_{4}w_{5}\right) & 2 \log (w_{3}) & 0 & 0 & 0 \\
 \log \left(\frac{w_{1}}{w_{2}}\right) & \log \left(\frac{w_{1}}{w_{2}}\right) & \log \left(\frac{w_{2}}{w_{1}}\right) & 2 \log (w_{5}) & \log (w_{1} w_{2}) & 0 & 0 \\
 0 & 2 \log (w_{1}) & 0 & 0 & 0 & \log (w_{1}) & 0 \\
 -\log (w_{2}) & \log (w_{2}) & \log (w_{2}) & 2 \log \left(w_{4}w_{5}\right) & 0 & 0 & \log (w_{2}) \\
\end{array}
\right]\,,
\\&
w_{1} = \zeta\overline\zeta, \quad w_{2} = (1-\zeta)(1-\overline\zeta), \quad w_{3} = \frac{\zeta-\overline\zeta}{2}, \quad w_{4} = \frac{\overline\zeta}{2}, \quad w_{5} = \frac{\zeta(1-\overline\zeta)}{\overline\zeta(1-\zeta)}\,.\label{eq:letters}
    \end{align}
\end{subequations}
The five letters in \eqref{eq:letters} are written in terms of dimensionless variables $\zeta\,, \overline\zeta$ to rationalize the square root $\kappa$ and are defined analogously to \eqref{eq:zetadef}
\begin{equation}
   z_{10}^2 = \zeta \overline{\zeta} z_{12}^2\,, \quad z_{10}\cdot z_{12} =\frac{\zeta{+}\overline{\zeta}}{2} z_{12}^2\,, \quad  z_{20}^2 = (1{-}\zeta) (1{-}\overline{\zeta}) z_{12}^2\,, \quad z_{20}\cdot z_{12} =\frac{\zeta{+}\overline{\zeta}{-}2}{2} z_{12}^2\,.
\end{equation}
with the explicit solution branch chosen as
\begin{equation}\label{eq:rat_var_trans_explicit}
    \zeta = \frac{z_{10}\cdot z_{12}+\kappa}{z_{12}^2}\,, \qquad \overline\zeta = \frac{z_{10}\cdot z_{12}-\kappa}{z_{12}^2}\,.
\end{equation}
This choice of inverse is compatible with the condition $(\zeta-\overline\zeta)^2 < 0$, enforcing that $\zeta$ and $\overline\zeta$ are complex conjugates, as well as the Cauchy--Schwarz inequality $(z_{10}\cdot z_{12})^2 \leq z_{10}^{2}z_{12}^{2}$.  In many physical problems (e.g., 2D CFT), one introduces $\zeta$ to effectively describe the two real degrees of freedom—relative length and relative angle—into one complex variable whose magnitude is the length ratio and whose phase is the angle between the vectors.

Upon rationalizing the alphabet, the solution to \eqref{eq:canonical} can systematically be expressed as a path-ordered exponential
\begin{equation}\label{eq:path_exp}
    \mathcal{J} = \mathbb{P}\exp\left[\varepsilon\int_{\Gamma}\d\Omega\right]\cdot\mathcal{J}_{0}\,,
\end{equation}
where $\Gamma$ is a path in the $(\zeta, \overline\zeta, z_{12}^{2})$ space and $\mathcal{J}_0$ is a vector of boundary constants (initial conditions, which we provide explicitly below). Each order in the expansion evaluates to polylogarithmic integrals of uniform weight, defined recursively as
\begin{equation*}
    G(a_1, {\ldots} , a_n;z) = \int_{0}^{z} \frac{\d t}{t-a_1} \, G(a_2, {\ldots}, a_n;t), \qquad G(\underbracket[0.4pt]{0, {\ldots}, 0}_{n \text{ times}};z) = \frac{1}{n!} \, \log^n(z)\, \qquad G(;z)=1\,.
\end{equation*}

The expansion is simplified in terms of classical polylogarithms $\text{Li}_n$ using an internal symbol‐map routine (see \cite{Goncharov:2010jf,Duhr:2011zq}), yielding \eqref{RV amplitude left} once \eqref{eq:form_factors_def} is rewritten in terms of $\mathcal{J}$. These manipulations at the level of $\mathcal{J}$ were numerically cross‑checked with the help of the \texttt{PolyLogTools} package \cite{Duhr:2019tlz}.

\paragraph{More details: boundary conditions used.} In order to fix the boundary vector $\mathcal{J}_0$, an independent evaluation of the master integrals must be calculated at at least one point in the  $(\zeta, \overline\zeta, z_{12}^{2})$ space. The first three master integrals can be evaluated generically using the simple integral
\begin{equation}
    \int\dbar^{\D}k\frac{\e^{ik\cdot z}}{(k^2)^{n}} = \frac{\Gamma\left(1-\varepsilon-n\right)}{2^{2n}\pi^{1-\varepsilon}\Gamma(n)}(z^{2})^{n+\varepsilon-1}\,,
\end{equation}
which immediately leads to
\begin{equation}
    \begin{split}
    \mathcal{T}_1 = \frac{C_\varepsilon^{2}}{16 \varepsilon^2}\left(z_{10}^{2}z_{20}^{2}\right)^{\varepsilon}\,, \qquad
    \mathcal{T}_2 = \frac{C_\varepsilon^{2}}{16 \varepsilon^2}(z_{10}^2z_{12}^2)^{\varepsilon}
    \,, \qquad
    \mathcal{T}_3 = \frac{C_\varepsilon^{2}}{16 \varepsilon^2}\left(z_{12}^{2}z_{20}^{2}\right)^{\varepsilon}\,.
    \end{split}
\end{equation}
For the remaining masters $\mathcal{T}_4,{\ldots}, \mathcal{T}_7$, the integrals become tractable only at specific kinematic points. We choose to work at the collinear point
\begin{equation}\label{eq:p0}
    \mathcal{Q}_{0}: \{z_{10}^{2}=1,z_{12}^{2}=1,z_{10}\cdot z_{12}=1\}\,.
\end{equation}
Using Schwinger parameters we find that at $\mathcal{Q}_{0}$
\begin{equation}
    \mathcal{T}_4|_{\mathcal{Q}_{0}} = -\frac{C_\varepsilon^{2}}{64\varepsilon^2(1+\varepsilon)(1+2\varepsilon)}\,.
\end{equation}
To compute a boundary for $\mathcal{T}_5{\ldots} \mathcal{T}_7$, we this we define two further integrals
\begin{equation}
    A_{q} = \int\dbar^{\D}k_{1}\dbar^{\D}k_{2}\frac{\e^{i(k_{1}\cdot z_{10}+k_{2}\cdot z_{12})}}{(k_{1}^{2})^{1-q}(k_{1}-k_{2})^{2}(k_{2}^{2})^{q}}\,, \qquad
    B_{q} = \int\dbar^{\D}k_{1}\dbar^{\D}k_{2}\frac{\e^{i(k_{1}\cdot z_{10}+k_{2}\cdot z_{12})}}{(k_{1}^{2})^{1+q}(k_{2}^{2})^{1-q}}\,.
\end{equation}
Their relevance can be seen by expanding them around small $q$
\begin{equation}
        A_{q} = \mathcal{T}_3+q\,\mathcal{T}_7+\mathcal{O}(q^{2})\,, \qquad 
        B_{q} = \mathcal{T}_4-q\,\mathcal{T}_6+\mathcal{O}(q^{2})\,.
\end{equation}
Evaluating $A_{q}$ and $B_{q}$ at $\mathcal{Q}_{0}$ using Schwinger parameters and expanding to linear order in $q$ we find
\begin{equation}
        \mathcal{T}_5|_{\mathcal{Q}_{0}}= -\frac{C_\varepsilon^{2}}{32\varepsilon^3}\,,\qquad
        \mathcal{T}_6|_{\mathcal{Q}_{0}}= 0\,,\qquad
        \mathcal{T}_7|_{\mathcal{Q}_{0}}= \frac{C_\varepsilon^{2}}{32\varepsilon^3}\,.
\end{equation}
With a boundary point found for all master integrals in the problem, \eqref{eq:path_exp} can be inverted to fix the boundary constant as
\begin{equation}\label{eq:bc_fix_wrong}
    \mathcal{J}_{0} = \left.\mathbb{P}\exp\left[\varepsilon\int_{\Gamma_{0}}\d\Omega\right]^{-1}\cdot\mathcal{J}\right\rvert_{\mathcal{Q}_{0}}\,.
\end{equation}
This relation is however not well defined, as the integrals $\mathcal{T}_1,{\ldots}, \mathcal{T}_3$ diverge at $\mathcal{Q}_{0}$. This problem can be circumvented by defining the point $\mathcal{Q}(t)$ that depends on a small deviation parameter $t$
\begin{equation}
    \mathcal{Q}(t): \{z_{10}^2(t) = 1, (z_{10}\cdot z_{12})(t) = 1-\frac{t^{2}}{2}, z_{12}^2(t) = 1\}\,,
\end{equation}
and instead recasting \eqref{eq:bc_fix_wrong} as
\begin{equation} \label{eq:bc exp}
    \mathcal{J}_{0} = \lim_{t \to 0}\left[\left.\mathbb{P}\exp\left[\varepsilon\int_{\Gamma_{0}}\d\Omega\right]^{-1}\cdot\mathcal{J}\right\rvert_{\mathcal{Q}(t)}\right].
\end{equation}
The divergence in $\mathcal{T}_1,{\ldots}, \mathcal{T}_3$ in the $t \to 0$ limit cancels with divergences in the inverse kernel, and $\mathcal{J}_0$ is finite. In practice, it suffices to take this limit to high precision numerically, and reconstruct the output using the PSLQ algorithm \cite{Ferguson1998APT}. Upon performing this operation, the boundary vector is obtained:
\begin{equation}
    \mathcal{J}_{0} = \frac{C_\varepsilon^{2}}{16 \varepsilon^2}(
    1,
    1,
    1,
    0,
    1,
    0,
    -1)^\top\,,
\end{equation}
which is \emph{exact} in $\varepsilon$. Clearly, the boundary conditions satisfy all the expected relations between the master basis in \eqref{eq:canBasis} at the point \eqref{eq:p0}.
Crucially, the $1/\varepsilon^2$ and $1/\varepsilon$ poles cancel in expanding \eqref{eq:path_exp}.

\bibliographystyle{JHEP}
\bibliography{refs}

@article{Chirilli:2013kca,
    author = "Chirilli, Giovanni A. and Kovchegov, Yuri V.",
    title = "{Solution of the NLO BFKL Equation and a Strategy for Solving the All-Order BFKL Equation}",
    eprint = "1305.1924",
    archivePrefix = "arXiv",
    primaryClass = "hep-ph",
    doi = "10.1007/JHEP06(2013)055",
    journal = "JHEP",
    volume = "06",
    pages = "055",
    year = "2013"
}

@article{Vladimirov:2016dll,
    author = "Vladimirov, Alexey A.",
    title = "{Correspondence between Soft and Rapidity Anomalous Dimensions}",
    eprint = "1610.05791",
    archivePrefix = "arXiv",
    primaryClass = "hep-ph",
    doi = "10.1103/PhysRevLett.118.062001",
    journal = "Phys. Rev. Lett.",
    volume = "118",
    number = "6",
    pages = "062001",
    year = "2017"
}

@article{Chetyrkin:1981qh,
      author         = "Chetyrkin, K. G. and Tkachov, F. V.",
      title          = "{Integration by Parts: The Algorithm to Calculate beta
                        Functions in 4 Loops}",
      journal        = "Nucl. Phys.",
      volume         = "B192",
      year           = "1981",
      pages          = "159-204",
      doi            = "10.1016/0550-3213(81)90199-1",
      SLACcitation   = "%%CITATION = NUPHA,B192,159;%%"
}

@article{Balitsky:2007feb,
    author = "Balitsky, Ian and Chirilli, Giovanni A.",
    title = "{Next-to-leading order evolution of color dipoles}",
    eprint = "0710.4330",
    archivePrefix = "arXiv",
    primaryClass = "hep-ph",
    reportNumber = "JLAB-THY-07-741",
    doi = "10.1103/PhysRevD.77.014019",
    journal = "Phys. Rev. D",
    volume = "77",
    pages = "014019",
    year = "2008"
}

@article{Caron-Huot:2015bja,
    author = "Caron-Huot, Simon",
    title = "{Resummation of non-global logarithms and the BFKL equation}",
    eprint = "1501.03754",
    archivePrefix = "arXiv",
    primaryClass = "hep-ph",
    doi = "10.1007/JHEP03(2018)036",
    journal = "JHEP",
    number = "1",
    volume = "03",
    pages = "036",
    year = "2018"
}

@article{Balitsky:1995ub,
    author = "Balitsky, I.",
    title = "{Operator expansion for high-energy scattering}",
    eprint = "hep-ph/9509348",
    archivePrefix = "arXiv",
    reportNumber = "MIT-CTP-2470",
    doi = "10.1016/0550-3213(95)00638-9",
    journal = "Nucl. Phys. B",
    volume = "463",
    pages = "99--160",
    year = "1996"
}

@article{Kovchegov:1999yj,
    author = "Kovchegov, Yuri V.",
    title = "{Small-x $F_2$ structure function of a nucleus including multiple pomeron exchanges}",
    eprint = "hep-ph/9901281",
    archivePrefix = "arXiv",
    reportNumber = "NUC-MN-99-1-T, TPI-MINN-99-05",
    doi = "10.1103/PhysRevD.60.034008",
    journal = "Phys. Rev. D",
    volume = "60",
    pages = "034008",
    year = "1999"
}

@article{Caron-Huot:2016tzz,
    author = "Caron-Huot, Simon and Herranen, Matti",
    title = "{High-energy evolution to three loops}",
    eprint = "1604.07417",
    archivePrefix = "arXiv",
    primaryClass = "hep-ph",
    doi = "10.1007/JHEP02(2018)058",
    journal = "JHEP",
    number = "1",
    volume = "02",
    pages = "058",
    year = "2018"
}

@article{Frellesvig:2019uqt,
    author = "Frellesvig, Hjalte and Gasparotto, Federico and Mandal, Manoj K. and Mastrolia, Pierpaolo and Mattiazzi, Luca and Mizera, Sebastian",
    title = "{Vector Space of Feynman Integrals and Multivariate Intersection Numbers}",
    eprint = "1907.02000",
    archivePrefix = "arXiv",
    primaryClass = "hep-th",
    doi = "10.1103/PhysRevLett.123.201602",
    journal = "Phys. Rev. Lett.",
    volume = "123",
    number = "20",
    pages = "201602",
    year = "2019"
}

@article{Frellesvig:2020qot,
    author = "Frellesvig, Hjalte and Gasparotto, Federico and Laporta, Stefano and Mandal, Manoj K. and Mastrolia, Pierpaolo and Mattiazzi, Luca and Mizera, Sebastian",
    title = "{Decomposition of Feynman Integrals by Multivariate Intersection Numbers}",
    eprint = "2008.04823",
    archivePrefix = "arXiv",
    primaryClass = "hep-th",
    doi = "10.1007/JHEP03(2021)027",
    journal = "JHEP",
    number = "1",
    volume = "03",
    pages = "027",
    year = "2021"
}

@article{Frellesvig:2019kgj,
    author = "Frellesvig, Hjalte and Gasparotto, Federico and Laporta, Stefano and Mandal, Manoj K. and Mastrolia, Pierpaolo and Mattiazzi, Luca and Mizera, Sebastian",
    title = "{Decomposition of Feynman Integrals on the Maximal Cut by Intersection Numbers}",
    eprint = "1901.11510",
    archivePrefix = "arXiv",
    primaryClass = "hep-ph",
    doi = "10.1007/JHEP05(2019)153",
    journal = "JHEP",
    number = "1",
    volume = "05",
    pages = "153",
    year = "2019"
}

@article{Chestnov:2022xsy,
    author = "Chestnov, Vsevolod and Frellesvig, Hjalte and Gasparotto, Federico and Mandal, Manoj K. and Mastrolia, Pierpaolo",
    title = "{Intersection numbers from higher-order partial differential equations}",
    eprint = "2209.01997",
    archivePrefix = "arXiv",
    primaryClass = "hep-th",
    doi = "10.1007/JHEP06(2023)131",
    journal = "JHEP",
    number = "1",
    volume = "06",
    pages = "131",
    year = "2023"
}

@article{Fontana:2023amt,
    author = "Fontana, Gaia and Peraro, Tiziano",
    title = "{Reduction to master integrals via intersection numbers and polynomial expansions}",
    eprint = "2304.14336",
    archivePrefix = "arXiv",
    primaryClass = "hep-ph",
    reportNumber = "ZU-TH 19/23",
    doi = "10.1007/JHEP08(2023)175",
    journal = "JHEP",
    number = "1",
    volume = "08",
    pages = "175",
    year = "2023"
}

@article{Mastrolia:2018uzb,
    author = "Mastrolia, Pierpaolo and Mizera, Sebastian",
    title = "{Feynman Integrals and Intersection Theory}",
    eprint = "1810.03818",
    archivePrefix = "arXiv",
    primaryClass = "hep-th",
    doi = "10.1007/JHEP02(2019)139",
    journal = "JHEP",
    number = "1",
    volume = "02",
    pages = "139",
    year = "2019"
}

@article{Weinzierl:2020gda,
    author = "Weinzierl, Stefan",
    title = "{Applications of intersection numbers in physics}",
    eprint = "2011.02865",
    archivePrefix = "arXiv",
    primaryClass = "hep-th",
    doi = "10.22323/1.383.0021",
    journal = "PoS",
    volume = "MA2019",
    pages = "021",
    year = "2022"
}

@article{Henn:2013pwa,
    author = "Henn, Johannes M.",
    title = "{Multiloop integrals in dimensional regularization made simple}",
    eprint = "1304.1806",
    archivePrefix = "arXiv",
    primaryClass = "hep-th",
    doi = "10.1103/PhysRevLett.110.251601",
    journal = "Phys. Rev. Lett.",
    volume = "110",
    pages = "251601",
    year = "2013"
}

@article{Balitsky:2009xg,
    author = "Balitsky, Ian and Chirilli, Giovanni A.",
    title = "{NLO evolution of color dipoles in $\mathcal{N}=4$ SYM}",
    eprint = "0903.5326",
    archivePrefix = "arXiv",
    primaryClass = "hep-ph",
    reportNumber = "JLAB-THY-09-961",
    doi = "10.1016/j.nuclphysb.2009.07.003",
    journal = "Nucl. Phys. B",
    volume = "822",
    pages = "45--87",
    year = "2009"
}

@article{Vladimirov:2017ksc,
    author = "Vladimirov, Alexey",
    title = "{Structure of rapidity divergences in multi-parton scattering soft factors}",
    eprint = "1707.07606",
    archivePrefix = "arXiv",
    primaryClass = "hep-ph",
    doi = "10.1007/JHEP04(2018)045",
    journal = "JHEP",
    number = "1",
    volume = "04",
    pages = "045",
    year = "2018"
}

@article{Hatta:2008st,
    author = "Hatta, Yoshitaka",
    title = "{Relating $\e^+/\e^-$ annihilation to high energy scattering at weak and strong coupling}",
    eprint = "0810.0889",
    archivePrefix = "arXiv",
    primaryClass = "hep-ph",
    doi = "10.1088/1126-6708/2008/11/057",
    journal = "JHEP",
    number = "1",
    volume = "11",
    pages = "057",
    year = "2008"
}

@article{Gelis:2010nm,
    author = "Gelis, Francois and Iancu, Edmond and Jalilian-Marian, Jamal and Venugopalan, Raju",
    title = "{The Color Glass Condensate}",
    eprint = "1002.0333",
    archivePrefix = "arXiv",
    primaryClass = "hep-ph",
    doi = "10.1146/annurev.nucl.010909.083629",
    journal = "Ann. Rev. Nucl. Part. Sci.",
    number = "1",
    volume = "60",
    pages = "463--489",
    year = "2010"
}

@article{Kovchegov:2006vj,
    author = "Kovchegov, Yuri V. and Weigert, Heribert",
    title = "{Triumvirate of Running Couplings in Small-x Evolution}",
    eprint = "hep-ph/0609090",
    archivePrefix = "arXiv",
    doi = "10.1016/j.nuclphysa.2006.10.075",
    journal = "Nucl. Phys. A",
    number = "1",
    volume = "784",
    pages = "188--226",
    year = "2007"
}

@article{Gelis:2012ri,
    author = "Gelis, F.",
    title = "{Color Glass Condensate and Glasma}",
    eprint = "1211.3327",
    archivePrefix = "arXiv",
    primaryClass = "hep-ph",
    reportNumber = "IPHT-T12-121",
    doi = "10.1142/S0217751X13300019",
    journal = "Int. J. Mod. Phys. A",
    number = "1",
    volume = "28",
    pages = "1330001",
    year = "2013"
}

@article{Kotikov:1991pm,
    author = "Kotikov, A. V.",
    title = "{Differential equation method: The Calculation of $n$-point Feynman diagrams}",
    doi = "10.1016/0370-2693(91)90536-Y",
    journal = "Phys. Lett. B",
    number = "1",
    volume = "267",
    pages = "123--127",
    year = "1991",
    note = "[Erratum: Phys.Lett.B 295, 409--409 (1992)]"
}

@inbook{scalar,
author = "Huang, Kerson",
publisher = {John Wiley \& Sons, Ltd},
isbn = {9783527617371},
title = {Scalar Fields},
booktitle = {Quantum Field Theory: From Operators to Path Integrals},
chapter = {2},
pages = {17-39},
doi = {https://doi.org/10.1002/9783527617371.ch2},
year = {1998}
}

@book{Peskin:1995ev,
    author = "Peskin, Michael E. and Schroeder, Daniel V.",
    title = "{An Introduction to quantum field theory}",
    isbn = "978-0-201-50397-5",
    publisher = "Addison-Wesley",
    address = "Reading, USA",
doi= {https://doi.org/10.1201/9780429503559},
    year = "1995"
}

@article{Remiddi:1997ny,
    author = "Remiddi, Ettore",
    title = "{Differential equations for Feynman graph amplitudes}",
    eprint = "hep-th/9711188",
    archivePrefix = "arXiv",
    reportNumber = "DFUB-97-15, DFUB 97-15",
    doi = "10.1007/BF03185566",
    journal = "Nuovo Cim. A",
    volume = "110",
    pages = "1435--1452",
    year = "1997"
}

@article{Laporta:2000dsw,
    author = "Laporta, S.",
    title = "{High precision calculation of multiloop Feynman integrals by difference equations}",
    eprint = "hep-ph/0102033",
    archivePrefix = "arXiv",
    doi = "10.1142/S0217751X00002159",
    journal = "Int. J. Mod. Phys. A",
    volume = "15",
    pages = "5087--5159",
    year = "2000"
}

@article{Peraro:2019svx,
    author = "Peraro, Tiziano",
    title = "{\texttt{FiniteFlow}: multivariate functional reconstruction using finite fields and dataflow graphs}",
    eprint = "1905.08019",
    archivePrefix = "arXiv",
    primaryClass = "hep-ph",
    reportNumber = "ZU-TH 24/19",
    doi = "10.1007/JHEP07(2019)031",
    journal = "JHEP",
    number = "1",
    volume = "07",
    pages = "031",
    year = "2019"
}

@article{Brunello:2023rpq,
    author = "Brunello, Giacomo and Chestnov, Vsevolod and Crisanti, Giulio and Frellesvig, Hjalte and Mandal, Manoj K. and Mastrolia, Pierpaolo",
    title = "{Intersection numbers, polynomial division and relative cohomology}",
    eprint = "2401.01897",
    archivePrefix = "arXiv",
    primaryClass = "hep-th",
    doi = "10.1007/JHEP09(2024)015",
    journal = "JHEP",
    volume = "09",
    pages = "015",
    year = "2024"
}

@article{Accardi:2012qut,
    author = "Accardi, A. and others",
    editor = "Deshpande, A. and Meziani, Z. E. and Qiu, J. W.",
    title = "{Electron Ion Collider: The Next QCD Frontier}: {Understanding the glue that binds us all}",
    eprint = "1212.1701",
    archivePrefix = "arXiv",
    primaryClass = "nucl-ex",
    reportNumber = "BNL-98815-2012-JA, JLAB-PHY-12-1652",
    doi = "10.1140/epja/i2016-16268-9",
    journal = "Eur. Phys. J. A",
    volume = "52",
    number = "9",
    pages = "268",
    year = "2016"
}

@article{Brunello:2023fef,
    author = "Brunello, Giacomo and Crisanti, Giulio and Giroux, Mathieu and Mastrolia, Pierpaolo and Smith, Sid",
    title = "{Fourier calculus from intersection theory}",
    eprint = "2311.14432",
    archivePrefix = "arXiv",
    primaryClass = "hep-th",
    doi = "10.1103/PhysRevD.109.094047",
    journal = "Phys. Rev. D",
    volume = "109",
    number = "9",
    pages = "094047",
    year = "2024"
}

@article{Balitsky:2024xvi,
    author = "Balitsky, Ian and Chirilli, Giovanni A.",
    title = "{Conformal BK equation at QCD Wilson--Fisher point}",
    eprint = "2407.08660",
    archivePrefix = "arXiv",
    primaryClass = "hep-ph",
    reportNumber = "JLAB-THY-24-4106",
    doi = "10.1007/JHEP10(2024)015",
    journal = "JHEP",
    volume = "10",
    pages = "015",
    year = "2024"
}

@article{Balitsky:2006wa,
    author = "Balitsky, Ian",
    title = "{Quark contribution to the small-x evolution of color dipole}",
    eprint = "hep-ph/0609105",
    archivePrefix = "arXiv",
    reportNumber = "JLAB-THY-06-541",
    doi = "10.1103/PhysRevD.75.014001",
    journal = "Phys. Rev. D",
    volume = "75",
    pages = "014001",
    year = "2007"
}

@article{Balitsky:1997mk,
    author = "Balitsky, Ian",
    editor = "Repond, Jos\'e and Krakauer, Daniel",
    title = "{Operator expansion for diffractive high-energy scattering}",
    eprint = "hep-ph/9706411",
    archivePrefix = "arXiv",
    reportNumber = "JLAB-THY-97-22",
    doi = "10.1063/1.53693",
    journal = "AIP Conf. Proc.",
    volume = "407",
    number = "1",
    pages = "953",
    year = "1997"
}

@book{Kovchegov:2012mbw,
    author = "Kovchegov, Yuri V. and Levin, Eugene",
    title = "{Quantum Chromodynamics at High Energy}",
    doi = "10.1017/9781009291446",
    isbn = "978-1-009-29144-6, 978-1-009-29141-5, 978-1-009-29142-2, 978-0-521-11257-4, 978-1-139-55768-9",
    publisher = "Oxford University Press",
    volume = "33",
    year = "2013"
}

@article{Lee:2012cn,
    author = "Lee, R. N.",
    title = "{Presenting \texttt{LiteRed}: a tool for the Loop InTEgrals REDuction}",
    eprint = "1212.2685",
    archivePrefix = "arXiv",
    primaryClass = "hep-ph",
    month = "12",
    year = "2012"
}

@article{Gehrmann:1999as,
    author = "Gehrmann, T. and Remiddi, E.",
    title = "{Differential equations for two-loop four-point functions}",
    eprint = "hep-ph/9912329",
    archivePrefix = "arXiv",
    reportNumber = "TTP-99-49",
    doi = "10.1016/S0550-3213(00)00223-6",
    journal = "Nucl. Phys. B",
    volume = "580",
    pages = "485--518",
    year = "2000"
}

@article{Argeri:2014qva,
    author = "Argeri, Mario and Di Vita, Stefano and Mastrolia, Pierpaolo and Mirabella, Edoardo and Schlenk, Johannes and Schubert, Ulrich and Tancredi, Lorenzo",
    title = "{Magnus and Dyson Series for Master Integrals}",
    eprint = "1401.2979",
    archivePrefix = "arXiv",
    primaryClass = "hep-ph",
    doi = "10.1007/JHEP03(2014)082",
    journal = "JHEP",
    volume = "03",
    pages = "082",
    year = "2014"
}

@article{Duhr:2019tlz,
    author = "Duhr, Claude and Dulat, Falko",
    title = "{\texttt{PolyLogTools} \textemdash{} polylogs for the masses}",
    eprint = "1904.07279",
    archivePrefix = "arXiv",
    primaryClass = "hep-th",
    reportNumber = "CP3-19-17, CERN-TH-2019-045, SLAC-PUB-17423",
    doi = "10.1007/JHEP08(2019)135",
    journal = "JHEP",
    volume = "08",
    pages = "135",
    year = "2019"
}

@book{Ferguson1998APT,
  title="{A Polynomial Time, Numerically Stable Integer Relation Algorithm}",
  author="{Helaman R. P. Ferguson and Daivd H. Bailey and Paul Kutler}",
    publisher = "{}",
  year={1998},
  url={https://api.semanticscholar.org/CorpusID:1024451}
}

@article{Ducloue:2019jmy,
    author = "Duclou\'e, B. and Iancu, E. and Soyez, G. and Triantafyllopoulos, D. N.",
    title = "{HERA data and collinearly-improved BK dynamics}",
    eprint = "1912.09196",
    archivePrefix = "arXiv",
    primaryClass = "hep-ph",
    doi = "10.1016/j.physletb.2020.135305",
    journal = "Phys. Lett. B",
    volume = "803",
    pages = "135305",
    year = "2020"
}

@article{Banfi:2002hw,
    author = "Banfi, A. and Marchesini, G. and Smye, G.",
    title = "{Away from jet energy flow}",
    eprint = "hep-ph/0206076",
    archivePrefix = "arXiv",
    reportNumber = "BICOCCA-FT-02-10",
    doi = "10.1088/1126-6708/2002/08/006",
    journal = "JHEP",
    volume = "08",
    pages = "006",
    year = "2002"
}

@article{Mueller:2018llt,
    author = "Mueller, Alfred H.",
    title = "{Conformal spacelike-timelike correspondence in QCD}",
    eprint = "1804.07249",
    archivePrefix = "arXiv",
    primaryClass = "hep-th",
    doi = "10.1007/JHEP08(2018)139",
    journal = "JHEP",
    volume = "08",
    pages = "139",
    year = "2018"
}

@article{Hofman:2008ar,
    author = "Hofman, Diego M. and Maldacena, Juan",
    title = "{Conformal collider physics: Energy and charge correlations}",
    eprint = "0803.1467",
    archivePrefix = "arXiv",
    primaryClass = "hep-th",
    doi = "10.1088/1126-6708/2008/05/012",
    journal = "JHEP",
    volume = "05",
    pages = "012",
    year = "2008"
}

@article{Alekhin:2012ig,
    author = "Alekhin, S. and Blumlein, J. and Moch, S.",
    title = "{Parton Distribution Functions and Benchmark Cross Sections at NNLO}",
    eprint = "1202.2281",
    archivePrefix = "arXiv",
    primaryClass = "hep-ph",
    reportNumber = "DESY-12-023, DO-TH-11-31, LPN-12-033, SFB-CPP-12-08",
    doi = "10.1103/PhysRevD.86.054009",
    journal = "Phys. Rev. D",
    volume = "86",
    pages = "054009",
    year = "2012"
}

@article{Marchesini:2015ica,
    author = "Marchesini, Giuseppe and Mueller, A. H.",
    title = "{The BMS equation and $ c\overline{c} $ production; a comparison of the BMS and BK equations}",
    eprint = "1510.08763",
    archivePrefix = "arXiv",
    primaryClass = "hep-ph",
    doi = "10.1007/JHEP02(2016)010",
    journal = "JHEP",
    volume = "02",
    pages = "010",
    year = "2016"
}

@article{Lappi:2015fma,
    author = {Lappi, T. and M\"antysaari, H.},
    title = "{Direct numerical solution of the coordinate space Balitsky-Kovchegov equation at next to leading order}",
    eprint = "1502.02400",
    archivePrefix = "arXiv",
    primaryClass = "hep-ph",
    doi = "10.1103/PhysRevD.91.074016",
    journal = "Phys. Rev. D",
    volume = "91",
    number = "7",
    pages = "074016",
    year = "2015"
}

@article{Iancu:2015joa,
    author = "Iancu, E. and Madrigal, J. D. and Mueller, A. H. and Soyez, G. and Triantafyllopoulos, D. N.",
    title = "{Collinearly-improved BK evolution meets the HERA data}",
    eprint = "1507.03651",
    archivePrefix = "arXiv",
    primaryClass = "hep-ph",
    doi = "10.1016/j.physletb.2015.09.071",
    journal = "Phys. Lett. B",
    volume = "750",
    pages = "643--652",
    year = "2015"
}

@article{Kinoshita:1962ur,
    author = "Kinoshita, T.",
    title = "{Mass singularities of Feynman amplitudes}",
    doi = "10.1063/1.1724268",
    journal = "J. Math. Phys.",
    volume = "3",
    pages = "650--677",
    year = "1962"
}

@article{Lee:1964is,
    author = "Lee, T. D. and Nauenberg, M.",
    editor = "Feinberg, G.",
    title = "{Degenerate Systems and Mass Singularities}",
    doi = "10.1103/PhysRev.133.B1549",
    journal = "Phys. Rev.",
    volume = "133",
    pages = "B1549--B1562",
    year = "1964"
}

@article{Dixon:2012yy,
    author = "Dixon, Lance J. and Duhr, Claude and Pennington, Jeffrey",
    title = "{Single-valued harmonic polylogarithms and the multi-Regge limit}",
    eprint = "1207.0186",
    archivePrefix = "arXiv",
    primaryClass = "hep-th",
    reportNumber = "SLAC-PUB-15132",
    doi = "10.1007/JHEP10(2012)074",
    journal = "JHEP",
    volume = "10",
    pages = "074",
    year = "2012"
}

@article{Stockinger:2005gx,
    author = "Stockinger, Dominik",
    title = "{Regularization by dimensional reduction: consistency, quantum action principle, and supersymmetry}",
    eprint = "hep-ph/0503129",
    archivePrefix = "arXiv",
    reportNumber = "IPPP-05-06, DCPT-05-12",
    doi = "10.1088/1126-6708/2005/03/076",
    journal = "JHEP",
    volume = "03",
    pages = "076",
    year = "2005"
}

@article{Siegel:1979wq,
    author = "Siegel, Warren",
    title = "{Supersymmetric Dimensional Regularization via Dimensional Reduction}",
    reportNumber = "HUTP-79/A006",
    doi = "10.1016/0370-2693(79)90282-X",
    journal = "Phys. Lett. B",
    volume = "84",
    pages = "193--196",
    year = "1979"
}

@article{Moch:2004pa,
    author = "Moch, S. and Vermaseren, J. A. M. and Vogt, A.",
    title = "{The Three loop splitting functions in QCD: The Nonsinglet case}",
    eprint = "hep-ph/0403192",
    archivePrefix = "arXiv",
    reportNumber = "DESY-04-047, SFB-CPP-04-09, NIKHEF-04-001",
    doi = "10.1016/j.nuclphysb.2004.03.030",
    journal = "Nucl. Phys. B",
    volume = "688",
    pages = "101--134",
    year = "2004"
}

@article{Jalilian-Marian:1997ubg,
    author = "Jalilian-Marian, Jamal and Kovner, Alex and Weigert, Heribert",
    title = "{The Wilson renormalization group for low x physics: Gluon evolution at finite parton density}",
    eprint = "hep-ph/9709432",
    archivePrefix = "arXiv",
    reportNumber = "TPI-MINN-97-26, NUC-MINN-97-11-T, HEP-MINN-1607, OUTP-97-45-P, CAVENDISH-HEP-97-15",
    doi = "10.1103/PhysRevD.59.014015",
    journal = "Phys. Rev. D",
    volume = "59",
    pages = "014015",
    year = "1998"
}

@article{Jalilian-Marian:1997qno,
    author = "Jalilian-Marian, Jamal and Kovner, Alex and Leonidov, Andrei and Weigert, Heribert",
    title = "{The BFKL equation from the Wilson renormalization group}",
    eprint = "hep-ph/9701284",
    archivePrefix = "arXiv",
    reportNumber = "TPI-MINN-96-28-T, NUC-MINN-96-22-T, HEP-MINN-96-1524",
    doi = "10.1016/S0550-3213(97)00440-9",
    journal = "Nucl. Phys. B",
    volume = "504",
    pages = "415--431",
    year = "1997"
}

@article{Kovner:2014lca,
    author = "Kovner, Alex and Lublinsky, Michael and Mulian, Yair",
    title = "{NLO JIMWLK evolution unabridged}",
    eprint = "1405.0418",
    archivePrefix = "arXiv",
    primaryClass = "hep-ph",
    doi = "10.1007/JHEP08(2014)114",
    journal = "JHEP",
    volume = "08",
    pages = "114",
    year = "2014"
}

@article{Balitsky:2013fea,
    author = "Balitsky, Ian and Chirilli, Giovanni A.",
    title = "{Rapidity evolution of Wilson lines at the next-to-leading order}",
    eprint = "1309.7644",
    archivePrefix = "arXiv",
    primaryClass = "hep-ph",
    reportNumber = "JLAB-THY-13-1806",
    doi = "10.1103/PhysRevD.88.111501",
    journal = "Phys. Rev. D",
    volume = "88",
    pages = "111501",
    year = "2013"
}

@article{AbdulKhalek:2022hcn,
    author = "Abdul Khalek, R. and others",
    title = "{Snowmass 2021 White Paper: Electron Ion Collider for High Energy Physics}",
    eprint = "2203.13199",
    archivePrefix = "arXiv",
    primaryClass = "hep-ph",
    reportNumber = "FERMILAB-PUB-22-125-QIS-SCD-T",
    month = "3",
    year = "2022"
}

@article{Lappi:2016fmu,
    author = {Lappi, T. and M\"antysaari, H.},
    title = "{Next-to-leading order Balitsky-Kovchegov equation with resummation}",
    eprint = "1601.06598",
    archivePrefix = "arXiv",
    primaryClass = "hep-ph",
    doi = "10.1103/PhysRevD.93.094004",
    journal = "Phys. Rev. D",
    volume = "93",
    number = "9",
    pages = "094004",
    year = "2016"
}

@article{Morreale:2021pnn,
    author = "Morreale, Astrid and Salazar, Farid",
    title = "{Mining for Gluon Saturation at Colliders}",
    eprint = "2108.08254",
    archivePrefix = "arXiv",
    primaryClass = "hep-ph",
    reportNumber = "Universe 2021",
    doi = "10.3390/universe7080312",
    journal = "Universe",
    volume = "7",
    number = "8",
    pages = "312",
    year = "2021"
}

@article{Mueller:1993rr,
    author = "Mueller, Alfred H.",
    title = "{Soft gluons in the infinite momentum wave function and the BFKL pomeron}",
    reportNumber = "SLAC-PUB-10047, CU-TP-609",
    doi = "10.1016/0550-3213(94)90116-3",
    journal = "Nucl. Phys. B",
    volume = "415",
    pages = "373--385",
    year = "1994"
}

@article{Mueller:1994gb,
    author = "Mueller, Alfred H.",
    title = "{Unitarity and the BFKL pomeron}",
    eprint = "hep-ph/9408245",
    archivePrefix = "arXiv",
    reportNumber = "CU-TP-640",
    doi = "10.1016/0550-3213(94)00480-3",
    journal = "Nucl. Phys. B",
    volume = "437",
    pages = "107--126",
    year = "1995"
}

@article{Balitsky:2001mr,
    author = "Balitsky, I. I. and Belitsky, Andrei V.",
    title = "{Nonlinear evolution in high density QCD}",
    eprint = "hep-ph/0110158",
    archivePrefix = "arXiv",
    reportNumber = "DOE-ER-40762-008, UMD-PP-02-010, YITP-SB-01-44, JLAB-THY-01-27",
    doi = "10.1016/S0550-3213(02)00149-9",
    journal = "Nucl. Phys. B",
    volume = "629",
    pages = "290--322",
    year = "2002"
}

@article{Grabovsky:2013mba,
    author = "Grabovsky, A. V.",
    title = "{Connected contribution to the kernel of the evolution equation for 3-quark Wilson loop operator}",
    eprint = "1307.5414",
    archivePrefix = "arXiv",
    primaryClass = "hep-ph",
    doi = "10.1007/JHEP09(2013)141",
    journal = "JHEP",
    volume = "09",
    pages = "141",
    year = "2013"
}

@article{Kovner:2013ona,
    author = "Kovner, Alex and Lublinsky, Michael and Mulian, Yair",
    title = "{Jalilian-Marian, Iancu, McLerran, Weigert, Leonidov, Kovner evolution at next to leading order}",
    eprint = "1310.0378",
    archivePrefix = "arXiv",
    primaryClass = "hep-ph",
    doi = "10.1103/PhysRevD.89.061704",
    journal = "Phys. Rev. D",
    volume = "89",
    number = "6",
    pages = "061704",
    year = "2014"
}

@misc{Iancu2019High,
  author       = {Iancu, Edmond},
  title        = "{High energy QCD \& the Color Glass Condensate}",
  howpublished = {Lecture slides from “The Myriad” workshop, ICTS Bangalore,
                  \href{https://www.icts.res.in/sites/default/files/extremeqandg2019-2019-04-11-%20Edmond%20Iancu..1.pdf}
                       {\texttt{https://www.icts.res.in/Edmond\_Iancu.pdf}}},
  year         = {2019},
}

@article{Bern:1991aq,
    author = "Bern, Zvi and Kosower, David A.",
    title = "{The Computation of loop amplitudes in gauge theories}",
    reportNumber = "FERMILAB-PUB-91-111-T, PITT-91-05",
    doi = "10.1016/0550-3213(92)90134-W",
    journal = "Nucl. Phys. B",
    volume = "379",
    pages = "451--561",
    year = "1992"
}

@article{Chiu:2011qc,
    author = "Chiu, Jui-yu and Jain, Ambar and Neill, Duff and Rothstein, Ira Z.",
    title = "{The Rapidity Renormalization Group}",
    eprint = "1104.0881",
    archivePrefix = "arXiv",
    primaryClass = "hep-ph",
    doi = "10.1103/PhysRevLett.108.151601",
    journal = "Phys. Rev. Lett.",
    volume = "108",
    pages = "151601",
    year = "2012"
}

@article{Caucal:2022ulg,
    author = {Caucal, Paul and Salazar, Farid and Schenke, Bj{\"o}rn and Venugopalan, Raju},
    title = "{Back-to-back inclusive dijets in DIS at small x: Sudakov suppression and gluon saturation at NLO}",
    eprint = "2208.13872",
    archivePrefix = "arXiv",
    primaryClass = "hep-ph",
    doi = "10.1007/JHEP11(2022)169",
    journal = "JHEP",
    volume = "11",
    pages = "169",
    year = "2022"
}

@article{Kang:2023oqj,
    author = "Kang, Zhong-Bo and Penttala, Jani and Zhao, Fanyi and Zhou, Yiyu",
    title = "{Transverse energy-energy correlators in the color-glass condensate at the electron-ion collider}",
    eprint = "2311.17142",
    archivePrefix = "arXiv",
    primaryClass = "hep-ph",
    reportNumber = "MIT-CTP/5649",
    doi = "10.1103/PhysRevD.109.094012",
    journal = "Phys. Rev. D",
    volume = "109",
    number = "9",
    pages = "094012",
    year = "2024"
}

@article{Veneziano:1976wm,
    author = "Veneziano, G.",
    title = "{Some Aspects of a Unified Approach to Gauge, Dual and Gribov Theories}",
    reportNumber = "CERN-TH-2200",
    doi = "10.1016/0550-3213(76)90412-0",
    journal = "Nucl. Phys. B",
    volume = "117",
    pages = "519--545",
    year = "1976"
}

@article{tHooft:1973alw,
    author = "'t Hooft, Gerard",
    editor = "Taylor, J. C.",
    title = "{A Planar Diagram Theory for Strong Interactions}",
    reportNumber = "CERN-TH-1786",
    doi = "10.1016/0550-3213(74)90154-0",
    journal = "Nucl. Phys. B",
    volume = "72",
    pages = "461",
    year = "1974"
}

@article{Brower:2006ea,
    author = "Brower, Richard C. and Polchinski, Joseph and Strassler, Matthew J. and Tan, Chung-I",
    title = "{The Pomeron and gauge/string duality}",
    eprint = "hep-th/0603115",
    archivePrefix = "arXiv",
    reportNumber = "BROWN-HET-1462",
    doi = "10.1088/1126-6708/2007/12/005",
    journal = "JHEP",
    volume = "12",
    pages = "005",
    year = "2007"
}

@article{Caron-Huot:2013fea,
    author = "Caron-Huot, Simon",
    title = "{When does the gluon reggeize?}",
    eprint = "1309.6521",
    archivePrefix = "arXiv",
    primaryClass = "hep-th",
    doi = "10.1007/JHEP05(2015)093",
    journal = "JHEP",
    volume = "05",
    pages = "093",
    year = "2015"
}

@article{Brunello:2024tqf,
    author = "Brunello, Giacomo and Chestnov, Vsevolod and Mastrolia, Pierpaolo",
    title = "{Intersection Numbers from Companion Tensor Algebra}",
    eprint = "2408.16668",
    archivePrefix = "arXiv",
    primaryClass = "hep-th",
    month = "8",
    year = "2024"
}

@misc{Mathematica,
  author = {Wolfram Research{,} Inc.},
  title = {Mathematica, {V}ersion 14.2},
  url = {https://www.wolfram.com/mathematica},
  note = {Champaign, IL, 2024}
}

@article{Hatta:2017fwr,
    author = "Hatta, Y. and Iancu, E. and Mueller, A. H. and Triantafyllopoulos, D. N.",
    title = "{Resumming double non-global logarithms in the evolution of a jet}",
    eprint = "1710.06722",
    archivePrefix = "arXiv",
    primaryClass = "hep-ph",
    doi = "10.1007/JHEP02(2018)075",
    journal = "JHEP",
    volume = "02",
    pages = "075",
    year = "2018"
}

@article{Goncharov:2010jf,
    author = "Goncharov, Alexander B. and Spradlin, Marcus and Vergu, C. and Volovich, Anastasia",
    title = "{Classical Polylogarithms for Amplitudes and Wilson Loops}",
    eprint = "1006.5703",
    archivePrefix = "arXiv",
    primaryClass = "hep-th",
    reportNumber = "BROWN-HET-1602",
    doi = "10.1103/PhysRevLett.105.151605",
    journal = "Phys. Rev. Lett.",
    volume = "105",
    pages = "151605",
    year = "2010"
}

@article{Duhr:2011zq,
    author = "Duhr, Claude and Gangl, Herbert and Rhodes, John R.",
    title = "{From polygons and symbols to polylogarithmic functions}",
    eprint = "1110.0458",
    archivePrefix = "arXiv",
    primaryClass = "math-ph",
    reportNumber = "IPPP-11-56, DCPT-11-112",
    doi = "10.1007/JHEP10(2012)075",
    journal = "JHEP",
    volume = "10",
    pages = "075",
    year = "2012"
}

@article{Angeles-Martinez:2015sea,
    author = "Angeles-Martinez, R. and others",
    title = "{Transverse Momentum Dependent (TMD) parton distribution functions: status and prospects}",
    eprint = "1507.05267",
    archivePrefix = "arXiv",
    primaryClass = "hep-ph",
    reportNumber = "DESY-15-111, NIKHEF-2015-023, RAL-P-2015-006, JLAB-THY-15-2020",
    doi = "10.5506/APhysPolB.46.2501",
    journal = "Acta Phys. Polon. B",
    volume = "46",
    number = "12",
    pages = "2501--2534",
    year = "2015"
}

@article{Moult:2018jzp,
    author = "Moult, Ian and Zhu, Hua Xing",
    title = "{Simplicity from Recoil: The Three-Loop Soft Function and Factorization for the Energy-Energy Correlation}",
    eprint = "1801.02627",
    archivePrefix = "arXiv",
    primaryClass = "hep-ph",
    doi = "10.1007/JHEP08(2018)160",
    journal = "JHEP",
    volume = "08",
    pages = "160",
    year = "2018"
}

@article{Yang:2023dwc,
    author = "Yang, Zhong and He, Yayun and Moult, Ian and Wang, Xin-Nian",
    title = "{Probing the Short-Distance Structure of the Quark-Gluon Plasma with Energy Correlators}",
    eprint = "2310.01500",
    archivePrefix = "arXiv",
    primaryClass = "hep-ph",
    doi = "10.1103/PhysRevLett.132.011901",
    journal = "Phys. Rev. Lett.",
    volume = "132",
    number = "1",
    pages = "011901",
    year = "2024"
}

@article{Gromov:2015vua,
    author = "Gromov, Nikolay and Levkovich-Maslyuk, Fedor and Sizov, Grigory",
    title = "{Pomeron Eigenvalue at Three Loops in $\mathcal N=$ 4 Supersymmetric Yang-Mills Theory}",
    eprint = "1507.04010",
    archivePrefix = "arXiv",
    primaryClass = "hep-th",
    doi = "10.1103/PhysRevLett.115.251601",
    journal = "Phys. Rev. Lett.",
    volume = "115",
    number = "25",
    pages = "251601",
    year = "2015"
}

@article{Argeri:2007up,
    author = "Argeri, Mario and Mastrolia, Pierpaolo",
    title = "{Feynman Diagrams and Differential Equations}",
    eprint = "0707.4037",
    archivePrefix = "arXiv",
    primaryClass = "hep-ph",
    reportNumber = "ZU-TH-19-07",
    doi = "10.1142/S0217751X07037147",
    journal = "Int. J. Mod. Phys. A",
    volume = "22",
    pages = "4375--4436",
    year = "2007"
}

@article{Kotikov:1990kg,
    author = "Kotikov, A. V.",
    title = "{Differential equations method: New technique for massive Feynman diagrams calculation}",
    reportNumber = "ITF-90-31E",
    doi = "10.1016/0370-2693(91)90413-K",
    journal = "Phys. Lett. B",
    volume = "254",
    pages = "158--164",
    year = "1991"
}

@book{Smirnov:2012gma,
    author = "Smirnov, Vladimir A.",
    title = "{Analytic tools for Feynman integrals}",
    doi = "10.1007/978-3-642-34886-0",
    volume = "250",
    year = "2012",
    publisher = "Springer"
}

@article{Luo:2002ti,
    author = "Luo, Ming-xing and Wang, Hua-wen and Xiao, Yong",
    title = "{Two loop renormalization group equations in general gauge field theories}",
    eprint = "hep-ph/0211440",
    archivePrefix = "arXiv",
    doi = "10.1103/PhysRevD.67.065019",
    journal = "Phys. Rev. D",
    volume = "67",
    pages = "065019",
    year = "2003"
}

@article{Albacete:2014fwa,
    author = "Albacete, Javier L. and Marquet, Cyrille",
    title = "{Gluon saturation and initial conditions for relativistic heavy ion collisions}",
    eprint = "1401.4866",
    archivePrefix = "arXiv",
    primaryClass = "hep-ph",
    doi = "10.1016/j.ppnp.2014.01.004",
    journal = "Prog. Part. Nucl. Phys.",
    volume = "76",
    pages = "1--42",
    year = "2014"
}

@inbook{Iancu:2003xm,
    author = "Iancu, Edmond and Venugopalan, Raju",
    editor = "Hwa, Rudolph C. and Wang, Xin-Nian",
    title = "{The Color glass condensate and high-energy scattering in QCD}",
    booktitle = "{Quark-gluon plasma 4}",
    eprint = "hep-ph/0303204",
    archivePrefix = "arXiv",
    doi = "10.1142/9789812795533_0005",
    pages = "249--3363",
    month = "3",
    year = "2003"
}

@article{Caola:2010cy,
    author = "Caola, Fabrizio and Forte, Stefano and Rojo, Juan",
    editor = "McLerran, Larry and Dunlop, James and Morrison, Dave and Venugopalan, Raju",
    title = "{HERA data and DGLAP evolution: Theory and phenomenology}",
    eprint = "1007.5405",
    archivePrefix = "arXiv",
    primaryClass = "hep-ph",
    reportNumber = "IFUM-962-FT",
    doi = "10.1016/j.nuclphysa.2010.08.009",
    journal = "Nucl. Phys. A",
    volume = "854",
    pages = "32--44",
    year = "2011"
}

@article{Mantysaari:2018nng,
    author = {M{\"a}ntysaari, Heikki and Zurita, Pia},
    title = "{In depth analysis of the combined HERA data in the dipole models with and without saturation}",
    eprint = "1804.05311",
    archivePrefix = "arXiv",
    primaryClass = "hep-ph",
    doi = "10.1103/PhysRevD.98.036002",
    journal = "Phys. Rev. D",
    volume = "98",
    pages = "036002",
    year = "2018"
}

@article{Fadin:1998py,
    author = "Fadin, Victor S. and Lipatov, L. N.",
    title = "{BFKL pomeron in the next-to-leading approximation}",
    eprint = "hep-ph/9802290",
    archivePrefix = "arXiv",
    reportNumber = "DESY-98-033",
    doi = "10.1016/S0370-2693(98)00473-0",
    journal = "Phys. Lett. B",
    volume = "429",
    pages = "127--134",
    year = "1998"
}

@article{Ciafaloni:1998gs,
    author = "Ciafaloni, Marcello and Camici, Gianni",
    title = "{Energy scale(s) and next-to-leading BFKL equation}",
    eprint = "hep-ph/9803389",
    archivePrefix = "arXiv",
    reportNumber = "DFF-302-03-98",
    doi = "10.1016/S0370-2693(98)00551-6",
    journal = "Phys. Lett. B",
    volume = "430",
    pages = "349--354",
    year = "1998"
}

@article{Salam:1998tj,
    author = "Salam, G. P.",
    title = "{A Resummation of large subleading corrections at small x}",
    eprint = "hep-ph/9806482",
    archivePrefix = "arXiv",
    reportNumber = "IFUM-627-FT",
    doi = "10.1088/1126-6708/1998/07/019",
    journal = "JHEP",
    volume = "07",
    pages = "019",
    year = "1998"
}

@article{BRODSKY1998299,
title = {Quantum chromodynamics and other field theories on the light cone},
journal = {Physics Reports},
volume = {301},
number = {4},
pages = {299-486},
year = {1998},
issn = {0370-1573},
doi = {https://doi.org/10.1016/S0370-1573(97)00089-6},
url = {https://www.sciencedirect.com/science/article/pii/S0370157397000896},
author = {Stanley J. Brodsky and Hans-Christian Pauli and Stephen S. Pinsky}
}

@article{Caron-Huot:2022eqs,
    author = "Caron-Huot, Simon and Kologlu, Murat and Kravchuk, Petr and Meltzer, David and Simmons-Duffin, David",
    title = "{Detectors in weakly-coupled field theories}",
    eprint = "2209.00008",
    archivePrefix = "arXiv",
    primaryClass = "hep-th",
    reportNumber = "CALT-TH 2022-31",
    doi = "10.1007/JHEP04(2023)014",
    journal = "JHEP",
    volume = "04",
    pages = "014",
    year = "2023"
}

@article{Gardi:2006rp,
    author = "Gardi, Einan and Kuokkanen, Janne and Rummukainen, Kari and Weigert, Heribert",
    title = "{Running coupling and power corrections in nonlinear evolution at the high-energy limit}",
    eprint = "hep-ph/0609087",
    archivePrefix = "arXiv",
    reportNumber = "CERN-PH-TH-2006-170, CAVENDISH-HEP-06-21",
    doi = "10.1016/j.nuclphysa.2006.12.004",
    journal = "Nucl. Phys. A",
    volume = "784",
    pages = "282--340",
    year = "2007"
}

@book{Cheng:1984vwu,
    author = "Cheng, Ta-Pei and Li, Ling-Fong",
    title = "{Gauge Theory of Elementary Particle Physics}",
    isbn = "978-0-19-851961-4, 978-0-19-851961-4",
    publisher = "Oxford University Press",
    address = "Oxford, UK",
    year = "1984"
}

@article{Panzer:2014caa,
    author = "Panzer, Erik",
    title = "{Algorithms for the symbolic integration of hyperlogarithms with applications to Feynman integrals}",
    eprint = "1403.3385",
    archivePrefix = "arXiv",
    primaryClass = "hep-th",
    doi = "10.1016/j.cpc.2014.10.019",
    journal = "Comput. Phys. Commun.",
    volume = "188",
    pages = "148--166",
    year = "2015"
}

@article{Brown:2008um,
    author = "Brown, Francis",
    title = "{The Massless higher-loop two-point function}",
    eprint = "0804.1660",
    archivePrefix = "arXiv",
    primaryClass = "math.AG",
    doi = "10.1007/s00220-009-0740-5",
    journal = "Commun. Math. Phys.",
    volume = "287",
    pages = "925--958",
    year = "2009"
}

@article{Blumlein:2021enk,
    author = {Bl{\"u}mlein, J. and Marquard, P. and Schneider, C. and Sch{\"o}nwald, K.},
    title = "{The three-loop unpolarized and polarized non-singlet anomalous dimensions from off shell operator matrix elements}",
    eprint = "2107.06267",
    archivePrefix = "arXiv",
    primaryClass = "hep-ph",
    reportNumber = "DESY-21-104, DO{\textendash}TH 21/23, TTP 21{\textendash}024, RISC Report Series 21{\textendash}13, SAGEX{\textendash}21{\textendash}15, DESY 21--104,DO--TH 21/23,TTP 21--024,RISC Report Series
  21--13,SAGEX--21--15",
    doi = "10.1016/j.nuclphysb.2021.115542",
    journal = "Nucl. Phys. B",
    volume = "971",
    pages = "115542",
    year = "2021"
}

\end{document}